\documentclass[fleqn,usenatbib]{mnras}

\usepackage[T1]{fontenc}
\usepackage{float}
\usepackage{tikz}
\usetikzlibrary{decorations.markings}
\DeclareRobustCommand{\VAN}[3]{#2}
\let\VANthebibliography\thebibliography
\def\thebibliography{\DeclareRobustCommand{\VAN}[3]{##3}\VANthebibliography}

\usepackage{graphicx}	
\usepackage{amsmath}	
\usepackage{subcaption}
\usepackage{tabularx}

\usepackage{newtxtext,newtxmath}

\DeclareMathOperator{\sech}{sech}

\DeclareMathOperator{\arcsinh}{arcsinh}

\usepackage{nccmath}

\newcommand{\lb}{\ell_b} 

\renewcommand{\vec}[1]{\mathbf{#1}}	
\newcommand{\dd}{\mathrm{d}}        

\newcommand{\cm}{\,{\rm cm}}    
\newcommand{\km}{\,{\rm km}}    
\newcommand{\m}{\,{\rm m}}      
\newcommand{\pc}{\,{\rm pc}}     
\newcommand{\kpc}{\,{\rm kpc}}  
\newcommand{\Mpc}{\,{\rm Mpc}}  

\newcommand{\GHz}{\,{\rm GHz}}  
\newcommand{\MHz}{\, {\rm MHz}} 
\newcommand{\s}{\,{\rm s}}      
\newcommand{\yr}{\,{\rm yr}}    

\newcommand{\kms}{\km\s^{-1}}    

\newcommand{\muG}{\,\mu{\rm G}} 

\newcommand{\K}{\,{\rm K}}      

\newcommand{\rad}{\,{\rm rad}} 

\newcommand{\brms}{\,b_{\rm rms}}
\newcommand{\border}{\Bar{B} }

\newcommand{\bpar}{{B}_{0\parallel}}
\renewcommand{\ne}{n_{\rm e}}

\newcommand{\ningal}{N_{\rm ingal}}

\newcommand{\RM}{\text{RM}}

\newcommand{\ipratio}{\,{\rm ip}_{\rm ratio}}
\newcommand{\rgal}{\,{\rm R}_{\rm gal}}
\newcommand{\rhalo}{\,{\rm R}_{\rm halo/CGM}}
\newcommand{\rperp}{{\rm r}_{\rm \perp}}
\newcommand{\cspi}{\,{C}_{\rm spi}}
\newcommand{\cell}{\,{C}_{\rm ell}}
\newcommand{\tspi}{\,{T}_{\rm spi}}
\newcommand{\tell}{\,{T}_{\rm ell}}
\newcommand{\sigmarm}{{\rm \sigma}_{\rm RM}}
\newcommand{\sigmarrm}{{\rm \sigma}_{\rm RRM}}
\newcommand{\sigmadata}{{\rm \sigma}_{\rm data}}
\newcommand{\sigmag}{{\rm \sigma}_{\rm G}}
\newcommand{\sigmash}{{\rm \sigma}_{\rm SH}}
\newcommand{\sigmalb}{{\rm \sigma}_{\rm LB}}

\newcommand{\DKS}{\,D_{\rm KS}}
\newcommand{\pKS}{\,p_{\rm KS}}

\newcommand{\stddev}{\,{\rm \sigma}}
\newcommand{\mean}{\,{\rm \mu}}
\newcommand{\kurtosis}{\,{\rm \mathcal{K}}}
\newcommand{\skewness}{\,{\rm \mathcal{S}}}

\newcommand{\RMsource}{\text{RM}_{\text{source}}}
\newcommand{\RMIGM}{\text{RM}_{\text{IGM}}}
\newcommand{\RMIngal}{\text{RM}_{\text{ingal}}}
\newcommand{\RMMW}{\text{RM}_{\text{MW}}}

\newcommand{\RRM}{\text{RRM}}

\defcitealias{VernstromEA2019}{V19}
\defcitealias{Faraday2020}{Faraday 2020}

\newcommand\Eq[1]{Eq.\,\ref{#1}}
\newcommand\Fig[1]{Fig.~\ref{#1}}
\newcommand\Sec[1]{Sec.~\ref{#1}}
\newcommand\Tab[1]{Table~\ref{#1}}
\graphicspath{{./}{figures/}}
\newcommand\rev[1]{#1}
\newcommand\revb[1]{#1}
\newcommand\revc[1]{#1}

\title[Magnetic fields in elliptical galaxies]{Magnetic fields in elliptical galaxies: Using the Laing-Garrington effect in radio galaxies and polarized emission from background radio sources}

\author[Shah and Seta]{
Hilay Shah$^{1}$\thanks{E-mail: hshah@ph.iitr.ac.in}
and Amit Seta$^{2}$\thanks{E-mail: amit.seta@anu.edu.au}
\\
$^{1}$Department of Physics, Indian Institute of Technology Roorkee, 247667, India\\
$^{2}$Research School of Astronomy and Astrophysics, Australian National University, Canberra, ACT 2611, Australia
}

\date{Accepted XXX. Received YYY; in original form ZZZ}

\pubyear{2020}

\begin{document}
\label{firstpage}
\pagerange{\pageref{firstpage}--\pageref{lastpage}}
\maketitle

\begin{abstract}
   Magnetic fields in elliptical galaxies are poorly constrained due to a lack of significant synchrotron emission from them. This paper explores properties of magnetic fields in ellipticals via two methods. First, we exploit the Laing-Garrington effect (asymmetry in the observed polarization fraction between radio galaxy jets) for 57 galaxies with redshifts up to $0.5$. We use the differences in polarization fraction and rotation measure between the jet and counterjet to estimate the small- and large-scale magnetic fields in and around ellipticals (including their circumgalactic medium). We find that the small-scale field (at scales smaller than the driving scale of turbulence, approximately $300~{\rm pc}$) lies in the range $0.1~\text{--}~2.75~\mu{\rm G}$. The large-scale field (at scales of $100~{\rm kpc}$) is an order of magnitude smaller than the small-scale field. In the second method, we cross-match the Faraday rotation measures (RM) of a few hundred (out of $3098$) extragalactic radio sources with galaxy catalogs to explore the effect of the number and morphology of intervening galaxies on the observed RM distribution. We use both Gaussian and non-Gaussian functions to describe the RM distribution and derive its statistical properties. Finally, using the difference in the observed polarization fraction between the intervening spirals and ellipticals, we estimate the small-scale magnetic fields at the center of ellipticals to be $\sim6~\mu{\rm G}$. Both methods with different observations and analysis techniques give magnetic field strengths consistent with previous studies ($\leq10\mu{\rm G}$), and the results can be used to constrain dynamo theories and galaxy evolution simulations.
\end{abstract}






\section{Introduction}
\label{sec:intro}
\revb{Magnetic fields are ubiquitous in the universe and are a key ingredient in cosmic evolution and structure formation.}  They \revb{can permeate} smaller astrophysical objects ($10^{-10}~\text{--}~10^{-8} \pc$) such as planets and stars and also \revb{objects} on much larger-scales ($10^{2}~\text{--}~10^{4} \pc$) such as \revb{the} interstellar and intergalactic medium (ISM and IGM). Their role in galaxy formation and evolution is not completely known yet, but recent cosmological simulations suggest that they are an important component of galaxy evolution \citep{VoortEA2020}. However, their influence on the physical processes in galaxies such as star formation \citep{BirnboimEA2015, KrumholzF2019}, heating of the ISM \citep{Raymond1992}, launching of galactic winds \citep{EGSFB17,HopkinsEA2018,VoortEA2020}, and propagation of cosmic rays \citep{Cesarsky1980, SSSBW17, SSWBS18} is anything but negligible.

Understanding the impact of magnetic fields on such processes requires details of their origin, strength, spatial scales, and evolution over galactic and cosmic time scales. \revb{Dynamo} theory \revb{\citep{Larmor1919, Herzenberg1958,BS2005,Fed16,Rincon19}}, a process by which the kinetic energy of the medium is converted to magnetic energy, is widely accepted as the magnetic field amplification mechanism in astrophysical objects \citep{Beck1996,BS2005}. Most astrophysical systems are turbulent, and without a self-sustaining dynamo process, \revb{turbulent} magnetic diffusion quickly destroys the seed magnetic fields in a fraction of the galactic lifetime \citep{SS2008}. For example, in spiral galaxies, the decay time scale of the magnetic fields due to the turbulent diffusion is of the order of $10^{8} \yr$, whereas the typical lifetime of a spiral galaxy is around $10^{10} \yr$ \citep[see Sec. 1.5 in][]{Seta2019}. So, the observed astrophysical magnetic fields are \revb{amplified and} maintained by a dynamo \revb{mechanism}.

Most previous studies of galactic magnetic fields deal with spiral galaxies \citep{BeckEA2015}. The magnetic fields in spiral galaxies ($B$) are divided into large-scale or mean \revb{fields} (\revb{$\border$}, ordered over $\kpc$ scales) and small-scale or fluctuating \revb{fields} ($b$, correlation length less than the driving scale of turbulence, $\lesssim 100 \pc$). \revb{In the presence of magnetic fields, cosmic ray electrons, generated by star-formation, supernovae events, active galactic nuclei (AGN), $\gamma$-ray bursts, emit synchrotron radiation which is intrinsically linearly polarized \citep{Duric1988, KleinEA2015}.} The synchrotron intensity traces the total magnetic field perpendicular to the line of sight. The \revb{polarization angle $\psi_{0}$ of a linearly polarized synchrotron radiation wave emitted by a radio source is rotated in the magneto-ionic medium along the line of sight, at a wavelength $\lambda$ \citep[see Sec. 3.3 in][]{KleinEA2015} such that:
\begin{ceqn}\label{eq:rm_angle}
\begin{align}
        \psi_{\rm obs} = \psi_{0} + \RM~\lambda^2,
\end{align}
\end{ceqn}
where $\psi_{\rm obs}$ is the observed polarization angle, and $\RM$ is the Faraday rotation measure. $\RM$ in the observer's frame is related to the properties of the magneto-ionic medium by the equation
\begin{ceqn}
\begin{align}\label{eq:rm_integral}
    \RM = \frac{0.81}{(1+z)^2}~\int_{0}^{L}~\frac{n_e(z, l)}{\cm^{-3}}~\frac{\vec{B}(z,l)}{\muG}~\cdot~\frac{\vec{\dd l}}{\pc}~\rad/\m^{2},
\end{align}
\end{ceqn}
where $z$ is the redshift of the source, $\ne(\cm^{-3}~)$ is the thermal electron density, $\vec{B} (\muG~)$ is the magnetic field, and $\dd l$ is an infinitesimal element along the line of sight (LOS) with $L$ being the total path length. The observed Faraday rotation measure ($\RM$) provides information about the magnetic field component parallel to the line of sight, where the $\RM$ values directly probe the \rev{ordered field, which is the large-scale field component for our study due to substantially large path lengths}.}

\revb{When random fluctuations of magnetic fields (small-scale magnetic fields) are present in a medium, the polarization signal can be reduced due to various depolarization mechanisms, \citep{Sokoloff1998} and it also leads to random fluctuations in $\RM$.} \revb{Thus,} information about the small-scale fields can be obtained from the depolarization measurements \revb{and the fluctuations in $\RM$ (given sufficient spatial resolution).} In the Milky Way and nearby spiral galaxies, depending on the galaxy and the region within it, the total field strength lies in the range $5~\text{--}~25 \muG$ with the small-scale field stronger than the large-scale by a factor of $1~\text{--}~5$ \citep{Fletcher10,BeckEA2015}.

Motivated by observational results, the dynamo theory is also conventionally divided into small- and large-scale \revb{dynamos}. The small-scale dynamo only requires turbulence and a very modest value of the magnetic Reynolds number ($\sim 10^{2}\text{--}10^{3}$) to generate small-scale fields \citep{Kazantsev1968, SCTMM04, BS2005, SetaEA2020}. The large-scale dynamo requires turbulence and large-scale scale properties of galaxies such as differential rotation, shear, and density stratification to generate fields over $\kpc$ scales \citep{RSS88, BlackmanEA1998, BS2005, SS2008}.

Besides the small-scale dynamo, small-scale random magnetic fields in spiral galaxies can also originate from the following two mechanisms: tangling of the large-scale field due to turbulence \citep[Sec. 4.1 in][]{SetaF2020} and amplification of small-scale magnetic fields due to compressive motions \citep{Fed11, Fed14}. In spiral galaxies, turbulence is driven at a range of scales by a variety of phenomena, \revb{like from} molecular cloud scales ($\sim 1\text{--}10\pc$) by stellar winds \citep{MacLowK2004} to galaxy scales ($\sim \kpc$s) by gravitational instabilities \citep{KrumholzEA2018}. However, the most energetic events in the ISM of spiral galaxies are supernova explosions, which drive turbulence at scales of around $100 \pc$ (roughly \revb{the order of} the size of a typical supernova remnant). The characteristic scale of small-scale magnetic fields is around $50 \pc$ \citep{GaenslerEA2005, FletcherEA2011, BeckEA2015}. The present-day telescope resolution \revb{in spiral galaxies is hundreds of $\pc$ in radio bands ($3-21 \cm$) \citep{BeckEA2007, TabatabaeiEA2008, KierdorfEA2020}} \revb{with the Effelsberg telescope and Very Large Array (VLA).} \revb{At such resolutions}, it is very difficult to differentiate between the small-scale field generated by the tangling of the large-scale field, the small-scale dynamo action, and the compression of magnetic fields. Thus, at present, it is not possible to get a pure observational signature of magnetic fields generated by the small-scale dynamo in spiral galaxies.

The observations of magnetic fields in elliptical galaxies can serve as a probe of the small-scale dynamo action since the large-scale dynamo is inactive due to lack of \revb{large-scale differential} rotation \citep{SetaEA2020II}. Type 1a supernova explosions from old stars and stellar winds \citep{MShu96, MathewsEA1997} drive turbulence in elliptical galaxies, which in turn amplifies magnetic fields by the small-scale dynamo action. For a typical elliptical galaxy, \revb{the study of} \citet{SetaEA2020II} estimated the driving scale of turbulence to be of the order of $300 \pc$, the root mean square (rms) turbulent velocity to be of the order of $2.5 \kms$, and the small-scale dynamo amplified magnetic field strength in the range $0.2~\text{--}~1 \muG$. Furthermore, they \revb{estimated} the \revb{small-scale} magnetic fields in a small sample ($30$ sources) of ellipticals using the Laing-Garrington effect (polarization asymmetry in jets \revc{hosted by radio galaxies with a central AGN component}) and \revb{obtained} magnetic fields in the range $0.1~\text{--}~6.0 \muG$. \revb{In this paper, we aim to obtain magnetic fields in ellipticals from a richer \revb{($210$ sources)} and more recent dataset of sources exhibiting the Laing-Garrington effect, which presumably traces the magnetic fields of the depolarizing medium, usually the halo of hot gas around the ellipticals and/or the intracluster medium (ICM) \citep{LaingEA1988, GarringtonEA1991(int)}. Thus,} we also use the statistical properties of polarized emission from \revb{extragalactic} background radio sources when seen through intervening galaxies to estimate magnetic fields in ellipticals, primarily in their \revb{haloes or the circumgalactic medium (CGM), which is the gas around galaxies outside their disks but inside their virial radii. Typical virial radius scales are $\sim 200-250 \kpc$ in spirals \revb{\citep{Putman2012CGM, TumlinsonEA2017}}.}

The structure of the paper is as follows. In \Sec{sec:Maths}, we provide the details of the terminology used throughout the paper and discuss analysis techniques. In \Sec{sec:laing_garrington}, we use the Laing-Garrington effect to estimate magnetic fields in ellipticals (data in \Sec{sec:catalog1}, methodology in \Sec{sec:laing_garrington_methodology}, and results in \Sec{sec:laing_garrington_results}). The magnetic fields using polarized emission from background radio sources is presented in \Sec{sec:polarized_emission} (data in \Sec{sec:rm_catalog} and \Sec{sec:galaxy_catalog}, analysis in \Sec{sec:sdss_analysis}, and results in \Sec{sec:polarized_emission_results}). Finally, we summarize our results in \Sec{sec:conclusions}. \revb{Standard cosmology (flat $\Lambda$CDM model) is used throughout the paper.}

\section{Analysis Techniques}
\label{sec:Maths} 

\subsection{Using the Laing-Garrington effect}
\label{pol}
The observed polarization difference between the radiation from a jet and counterjet (jet facing away from the observer), usually referred to as the Laing-Garrington effect \citep{LaingEA1988, GarringtonEA1988}, can be used to estimate the magnetic field of the host galaxy. \revb{Along the observer's line of sight,} radiation from the counterjet travels a longer distance through the host galaxy's medium than radiation from the jet, leading to a higher depolarization for the counterjet due to the \revb{random fluctuations in the Faraday rotation \revb{measures} ($\sigmarm$) present because of the small-scale} magnetic fields of the galaxy. \revb{Apart from the magnetic fields, the $\sigmarm$, and depolarization in turn, depends on various other parameters such as the electron density ($\ne$), correlation length ($\lb$), path length ($L$), and redshift ($z$) (details in \Sec{sec:laing_garrington_methodology}).} Assuming that the jet and counterjet have similar intrinsic polarization properties, the difference in the observed depolarization (or equivalently polarization) signal can be associated with the magnetic field of the halo or the CGM of the host galaxy \revb{in the regions along the line of sight.} \rev{\citep[e.g.,~][]{SetaEA2020II}.}

\rev{The synchrotron emission in ellipticals is from the central active galactic nucleus (AGN) and the ISM (due to cosmic ray electrons and magnetic fields) \citep{Jiang_2010, CrockerEA2011}. \revb{The study of} \citet{Nyland2017} shows that \revb{$1.4 \GHz$} radio emission \revb{in the host galaxy} extends to $18 \kpc$ in extended radio galaxies (see their Sec.~7). \revb{In the majority of their galaxies, the radio emission is associated with star formation (SF) activity, and in some rare cases, the extended jets of an active nucleus. \revb{The radio galaxy (RG) catalog we adopt consists of radio observations from AGN jets, lobes, and hotspots that extend much beyond the host galaxy's environment \citep{VernstromEA2019}, having impact parameters mostly in hundreds of $\kpc$.} Thus, in extended radio galaxies (ERGs), the extent of SF activity is negligible near their usual halo extent of a few hundred $\kpc$, as well as for the large} impact parameters we encounter in our sample of RGs. \revb{Ignoring the internal Faraday depolarization of the polarized emission from the AGN jets (or lobes, hotspots) in our sample of radio galaxies}, it would be reasonable to assume that the polarization properties \revb{of this emission} are not affected much by \revb{any other} synchrotron emission \revb{(due to lack star formation at such great distances beyond the host galaxy)} along the line of sight in the \revb{ERGs}. \revb{For giant radio galaxies (GRGs) extending to a few $\Mpc$, a very small fraction (12/820) show the potential for star formation \citep{DabhadeEA2020}.}} \revb{For this reason, polarized emission from radio galaxies do not have deep Faraday screens in general, unless the polarization is seen from the radio core associated with the host galaxy. Thus, depolarization in our sample of RGs mainly occurs in the} medium in which the plane of polarization is only rotated by Faraday rotation due to thermal electrons and magnetic fields \revb{, and for this,} the degree of polarization $p$ depends on the  observing wavelength ($\lambda$) and the standard deviation of rotation measure fluctuations ($\sigmarm$) as \citep{BurnEA1966, Sokoloff1998}

\begin{ceqn}
\begin{align} \label{eq:pol}
    p = p_0~\exp(-2 \sigmarm^2 \lambda^4),
\end{align}
\end{ceqn}
where $p_0$ is the \revb{intrinsic} degree of polarization. 

Assuming that the intrinsic polarization properties of the jet and counterjet in an elliptical are \revb{similar}, the ratio of the observed degree of jet and counterjet polarization, $p_{\rm j}$ and $p_{\rm cj}$, at the observing wavelength $\lambda$ is expressed as
\begin{ceqn}
\begin{align} \label{eq:pol_ratio}
    \frac{p_{\rm j}}{p_{\rm cj}} = \exp(-2\lambda^4({\sigmarm}_{\rm j}^2-{\sigmarm}_{\rm cj}^2)),
\end{align}
\end{ceqn}
where ${\sigmarm}_{\rm j}$ and ${\sigmarm}_{\rm cj}$ are the standard \revb{deviations} of the rotation measure fluctuations for the jet and counterjet, respectively. \revb{For smaller physical separation between the lobes, the contribution from the galactic foreground (including the IGM) is similar \citep{VernstromEA2019}, and most likely cancels out on division.} Using \Eq{eq:pol_ratio}, we can estimate the standard deviation of the rotation measure fluctuations in the halo (or the circumgalactic medium) of the elliptical galaxy (also accounting for the redshift of the source \revb{now}), ${\sigmarm}_{\rm ell}$, as
\begin{ceqn}
\begin{align}\label{eq:sigma_rme}
    {\sigmarm}_{\rm ell} = \frac{\left({\sigmarm}_{\rm cj}^2-{\sigmarm}_{\rm j}^2~\right)^{1/2}}{(1+z)^2}.
\end{align}
\end{ceqn}
We use this analysis\footnote{\revb{The study of} \citet{SetaEA2020II} uses the ratio of observed depolarization fraction for the jet and counterjet instead of the degree of polarization, which we use.} for the data classified as physical pairs ($210$ sources) in \citet{VernstromEA2019} \rev{(\citetalias{VernstromEA2019})} to compute ${\sigmarm}_{\rm ell}$ for those host \revb{galaxies' halo or CGM}. \revb{Then,} assuming two different thermal electron density distributions, \revb{we} estimate magnetic field strengths from the computed ${\sigmarm}_{\rm ell}$ values (\Sec{sec:laing_garrington}).

\subsection{Using the Faraday Rotation Measure ($\RM$) from background radio sources}
\label{sec:rm}

The observed RM \revb{from an extragalactic radio source} is actually a sum of contributions from all the intervening magneto-ionic regions between the source and us; thus, $\RM$ can be expressed as
\begin{ceqn}
\begin{align}\label{eq:rm_contri}
    \RM = \RMsource + \RMIGM + \RMIngal +\RMMW,
\end{align}
\end{ceqn}
where $\RMsource$ is the \revb{$\RM$ arising due to the background polarized source (for example, radio lobes)}, $\RMIGM$ is the contribution from the foreground intergalactic medium, $\RMIngal$ is the contribution from intervening galaxies (\revb{including the host galaxy of the background polarized source and }excluding the Milky Way), \revb{and} $\RMMW$ is the contribution from the Milky Way.

Magnetic fields in the IGM are of the order of $10^{-3}~\muG$ \citep{GrassoEA2001, ElisabeteEA2006}, whereas the observed galactic magnetic fields are in the range $1~\text{--}~10\muG$ \citep{BeckEA2015}. The intergalactic thermal electron density \citep[$\sim 1.6 \times 10^{-7} \cm^{-3}$,][]{YaoEA2017} is also much smaller than average thermal electron density in spiral galaxies \citep[$\sim 0.02 \cm^{-3}$,][]{BerkhuijsenM2008,GaenslerEA2008,YaoEA2017}. Thus, even though the path length of the light traversing through the IGM is about a thousand times higher than that for the galaxies (approximately $\kpc$ for galaxies and $\Mpc$ for the IGM), the contribution of the IGM to the observed $\RM$, $\RMIGM$ in \Eq{eq:rm_contri}, can be safely neglected in comparison to the other terms. \rev{The Milky Way contribution, $\RMMW$, can be modeled using the Faraday rotation map of the Galactic sky, such as the Oppermann map \citep{OppermannEA2015}, and the map by \citep{Faraday2020} (referred as \citetalias{Faraday2020} hereon with further details given in \Sec{sec:rm_catalog})}. \revb{Thus,} the Milky Way part can be subtracted from the RM to obtain the residual $\RM$ contribution, $\RRM = \RM - \RMMW$, which would ideally have contributions from the source and the intervening galaxies only. 

Given a distribution for $\RMsource$, the $\RRM$ distribution would depend on the number and morphology of the intervening galaxies. \rev{For a fixed number of intervening galaxies, considering that the spiral galaxies have stronger magnetic fields and higher thermal electron density \citep{TaylorEA1993} than the ellipticals \citep{MathewsEA2003}}, we expect that the standard deviation of $\RRM$ would be higher for intervening spirals than for intervening ellipticals. Furthermore, for a fixed morphology, we would expect that the standard deviation of $\RRM$ would increase with the number of intervening galaxies. We derive the statistical properties of $\RRM$ distribution, computed using the $\RM$ of background sources from the catalog compiled by \revb{\citet{FarnesEA2014}}) and the Milky Way $\RM$ model \revb{from} \rev{\citetalias{Faraday2020}} for different morphologies \citep[decided based on the Galaxy Zoo data,][]{LintottEA2008}, to study their dependence on the number of intervening galaxies and morphology. \revb{From this, we} finally estimate magnetic fields in elliptical galaxies (\Sec{sec:polarized_emission}). The estimated magnetic field in galaxies \revb{from} both \revb{methods} is primarily from the halo or circumgalactic medium. 

\subsection{Terminology}
In the list below, we define common terms which will be used throughout the paper.
\begin{itemize}
    \item $\mean_{\rm sample}$ : mean of a sample dataset
    \item $\stddev_{\rm sample}$ : standard deviation of a sample dataset
    \item $\skewness_{\rm sample}$ : skewness of a sample dataset
    \item $\kurtosis_{\rm sample}$ : kurtosis of a sample dataset
    \item $\sigmarrm$ : standard deviation of the residual $\RM$s, i.e., the $\RM$s from \textbf{multiple} background sources with the Milky Way $\RM$ contribution removed
    \item $\sigmarm$ : standard deviation of fluctuation of rotation measures in a \textbf{single} host galaxy
    \item $\rgal$ : radius of the galaxy (assumed to be spherical)
    \item $\rhalo$ : radius of the galaxy's halo or CGM (10$\rgal$) (assumed to be spherical)
    \item $\rperp$ : impact parameter (distance from galaxy's center at which the light from the background source passes)
    \item $\ipratio$ : $\rperp/\rgal$
    \item $\ningal$ : number of intervening galaxies through which the light from the background source passes before reaching the observer.
\end{itemize}

\section{Magnetic fields in elliptical galaxies using the Laing-Garrington effect}
\label{sec:laing_garrington}
\subsection{Radio Galaxy (RG) Catalog}
\label{sec:catalog1}
Recently, \citet{VernstromEA2019} \rev{(referred as \citetalias{VernstromEA2019} hereon)} presented a study of the classification of \revb{RM sources from the} NVSS catalog \citep{TaylorEA2009} into 317 physical pairs and 5111 random pairs, \revb{by} making use of infrared images from Wide-Field Infrared Survey Explorer (WISE, \citet{WrightEA2010}), optical images from Sloan Digital Sky Survey (SDSS) data release 9 \citep{AhnEA2012}, and the 1.4 GHz \revb{images} from the Faint Images of the Radio Sky at Twenty-cm (FIRST, \citet{BeckerEA1995}). Physical pairs are tantamount to the lobes \revb{or hotspots} of giant radio galaxies (GRGs) or extended radio galaxies (ERGs), where the lobes extend well beyond the host galaxy's local environment into the CGM or surrounding IGM/ICM. The two lobes are probably jet (closer to us) and counterjet (facing away from us) side. \revc{However, a physical pair could also be composed of radio hotspots instead of radio lobes.}

The angular separation between the lobes, polarization values \revb{and} off-axis leakage polarization for each lobe, the redshift of the sources, and the difference in the $\RM$ of two lobes, $\Delta \RM$, with the corresponding error, $\Delta \RM_{\rm error}$, were accessed and utilized for our analysis \footnote{The data is available at \url{https://iopscience.iop.org/0004-637X/878/2/92/suppdata/apjab1f83t1_mrt.txt}.}. We only retain 210 physical pairs for which \revb{the} redshifts are available. Together with their angular separation and \revb{the redshift $z$}, we determine the distance between the \rev{polarized components of the} lobes \revb{(physical separation)} \rev{for which the $\RM$ and polarization values are available}\footnote{\href{https://docs.astropy.org/en/stable/api/astropy.cosmology.LambdaCDM.html}{Astropy.cosmology.LambdaCDM} was used to calculate the angular diameter distance to a body using its redshift.}. \rev{Our RG catalog consists of \revb{$91$ RGs with \revb{physical separation} less than $500 \kpc$}, \revb{$85$} RGs between $500 \kpc~\text{--}~1 \Mpc$, $27$ RGs between $1~\text{--}~2 \Mpc$, and $7$ RGs between $2~\text{--}~4 \Mpc$. Due to the lack of inclination angle data, we do not account for the bent tail type radio galaxies and assume that the angle between the lobes is $180^\circ$. Additionally, the angle between the lobes' orientation and the line of sight is not known from the data. Thus, the distance between the lobes is the projection of distance in the observer's plane. This might lead to inaccurate measurements at the extremities (\revb{infrequent} possibility) in lobe orientation, i.e., lobes oriented $\perp$ or $\parallel$ to the observer's plane.}

The polarization values in \rev{\citetalias{VernstromEA2019}} are taken from the NVSS catalog, and percent polarization is the ratio of average peak polarized intensity to \revb{the peak integrated} \rev{S}tokes I \citep{TaylorEA2009}, \revb{which gives the lower limit to the true observed polarization}. The average and integration are done with images at frequencies 1365 and 1435 $\MHz$. Thus, the equivalent frequency for the average polarization values is 1400 $\MHz$ (to be used for \Eq{eq:pol} and \Eq{eq:pol_ratio}). 

\subsection{Methodology to extract small- and large- scale magnetic field strengths using the RG catalog}
\label{sec:laing_garrington_methodology}
Here, we use the techniques described in \Sec{pol} and the observed polarization properties of physical pairs with redshifts from the dataset in \rev{\citetalias{VernstromEA2019}} to estimate magnetic fields in those radio galaxies. Radio galaxies are usually associated with ellipticals, and \revb{several} observational results suggest that the majority of giant radio galaxies are \rev{giant ellipticals \citep{LillyEA1987, OwenEA1989, HamiltonEA200, HoEA2009, Saripalli_2012, MalareckiEA2013, BanfieldEA2015}.} Furthermore, \citet{GarringtonEA1991(int)} attributed the asymmetry in depolarization of two jets to an external X-ray halo of ionized gas, which is often found in clusters. \revb{To ensure that the depolarization occurs inside the galaxy medium rather than the cluster, we cross-match the physical pairs of \citetalias{VernstromEA2019} with the SDSS catalog (\Sec{sec:galaxy_catalog}), using the methodology described in \Sec{sec:cross-match}, to find the host galaxies of the jets. We find 40 matches in the SDSS catalog, for which the diameter of the halo of the galaxy is considered ten times the diameter of the galaxy (see the last paragraph of \Sec{sec:galaxy_catalog}). The depolarization is assumed to be from the galaxy medium if the halo diameter is greater than the lobes' physical separation; 26/40 RGs satisfy this condition. The average physical separation between the lobes for these $26$ RGs is $\sim 300 \kpc$ and is $\sim 600 \kpc$ for the rest $14$ RGs. Based on this, we now consider only RGs with physical separation less than $500 \kpc$ for a greater probability of depolarization occurring inside the galaxy's medium. Typical CGM diameter in spirals is $\sim 400-500 \kpc$ \citep{Putman2012CGM, TumlinsonEA2017, PakmorEA2020}. Assuming a similar CGM diameter in ellipticals, the physical separation cap at $500 \kpc$ is consistent with CGM/halo scales \citep{Gerhard_2010}.} Hence, we assume that the asymmetry in the depolarization arises mainly from the X-ray halo and CGM of an elliptical galaxy, \rev{as the available data has path lengths in hundreds of $\kpc$} \citep{TumlinsonEA2017} from the galaxy's center, beyond which the depolarization effects can be safely neglected due to negligible thermal electron density and magnetic fields.

 Using \Eq{eq:pol_ratio} and \Eq{eq:sigma_rme} for the data of physical pairs in \rev{\citetalias{VernstromEA2019}}, we obtain the standard deviation of rotation measure fluctuations in the host elliptical galaxies, ${\sigmarm}_{\rm ell}$, for \revc{$84$} sources \revc{(after imposing $500 \kpc$ physical separation cap and removing galaxies with SDSS halo diameters smaller than the physical separation)}. ${\sigmarm}_{\rm ell}$, assuming a profile for thermal electron density, can be used to compute the root mean square (rms) strength of the small-scale random magnetic field in the elliptical galaxy. For thermal electron density distribution, we use two profiles: a uniform density \revb{profile for simplicity \citep{Seta2019}} and a King profile \revb{\citep{SarazinEA1988_King, FeltenEA1996, MathewsEA2003}}. 

 For a uniform thermal electron density (\revb{$\overline{\ne}$ = constant}), the ${\sigmarm}_{\rm ell}$ can be expressed as 
 \revb{
 \begin{ceqn}
 \begin{align}\label{eq:sigma_rm}
    \frac{{\sigmarm}_{\rm ell}}{{(1 + z)^2}} = \frac{(2\pi)^{1/4}}{3^{1/2}}~\K~\overline{\ne} \brms (L \lb)^{1/2},
 \end{align}
 \end{ceqn}
 }
where $z$ is the redshift, $\K = 0.81 \muG^{-1}\cm^3\pc^{-1}\rad/\m^{2}$ is a constant, \revb{and} $L$ is the path length in $\pc$ (distance between the lobes, see \Sec{sec:catalog1}). Assuming the correlation length of small-scale magnetic fields $\lb$ to be $100 \pc$ \citep[see Sec. 5.3.1 in][]{Seta2019} and the average thermal electron density in the halo/CGM of an elliptical galaxy \revb{$\overline{\ne}$} to be $0.01~\cm^{-3}$, $\brms$ can be  estimated. \revc{The correlation length of small-scale magnetic field $\lb$ (assumed to be $100 \pc$) is $3-4$ times smaller \citep[see Tab 2.1 in][]{Seta2019} than the driving scale of turbulence, $300 \pc$ \citep[see Sec. 5.2 in][for full derivation]{Seta2019}.} \revc{Here, the turbulence is assumed to be driven by Type 1a supernovae explosions. This is evident from the observations of iron abundances \citep{MathewsEA2003} and simulations of multiphase gas in elliptical galaxies \citep{Li2020_SN, Li2020_SN2}.}

For a King profile, $\ne=\ne(0)(1+(r/a)^2)^{-3 \beta/2}$, where $\ne(0)$ is the thermal electron density at $r=0$, and $a$ is the core radius\revb{, the} ${\sigmarm}_{\rm ell}$ close to the center ($r=0$) is expressed as
\begin{ceqn}
    \begin{align}\label{eq:king_sigma}
        \frac{{\sigmarm}_{\rm ell} (0)} {(1 + z)^2} = \frac{(2\pi)^{1/4}}{3^{1/2}}~\K~\ne (0) \brms~(a\lb)^{1/2}
   \left(\frac{\Gamma \left(\frac{3}{2}(\gamma+1)-0.5\right)}{\Gamma \left(\frac{3}{2}(\gamma+1)\right)} \right)^{1/2},
\end{align}
\end{ceqn}
where $\beta$ is assumed to be $1/2$ \citep[following the gas distribution, see][]{MatthaeusEA2003}, \revb{and $\gamma=2/3$ (inspired by the flux freezing or cooling flow constraint, which gives $\brms \propto \rho^{2/3}$). Now,} the ${\sigmarm}_{\rm ell}$ at any radius $r$ (on the plane of the sky) can be estimated as 
\begin{ceqn}
\begin{align}\label{eq:king_sigma0}
    {\sigmarm}_{\rm ell} (r)=\frac{{\sigmarm}_{\rm ell} (0)}{(1+z)^2}\left[1+\left(\frac{r}{a}\right)^2 \right]^{-
    \left(\frac{1}{4}\Gamma \left(\frac{3}{2}(\gamma+1)-1\right)\right)}.
\end{align}
\end{ceqn}
\revb{\Eq{eq:king_sigma} contains the contribution of varying $\ne$ and $\brms$, derived via integration of $\sigmarm$ (see \citet{FeltenEA1996} and Appendix A in \citet{BhatEA2013})}. To estimate small-scale rms magnetic field strengths, $\brms$, using the King profile, we assume $\ne(0)=0.1~\cm^{-3}$ \citep{MathewsEA2003}, $\lb=100~\pc$ (as in the uniform thermal electron density case), and $a=3~\kpc$ \citep{GarringtonEA1991(int)}. We consider the calculated ${\sigmarm}_{\rm ell}$ from \Eq{eq:sigma_rme} to be equal to ${\sigmarm}_{\rm ell}(L/2)$ from \Eq{eq:king_sigma0}, where $L/2$ is the approximate distance from the center \revb{in the plane of observation where the extra depolarized hotspot is assumed to be located} (since $L$ is the physical separation \revb{in the plane of observation}). \revb{As \Eq{eq:king_sigma0} predicts the ${\sigmarm}_{\rm ell}$ at a distance from the center in the plane of observation along the line of sight, it is applicable at $L/2$, the location of the extra depolarized hotspot (presumably counterjet).}

Besides using the standard deviation of rotation measure fluctuations, ${\sigmarm}_{\rm ell}$, to estimate the rms strength of small-scale magnetic fields (using \Eq{eq:sigma_rm} or \Eq{eq:king_sigma} and \Eq{eq:king_sigma0}), we also use the observed $\RM$ values from the physical pairs data in \rev{\citetalias{VernstromEA2019}} to estimate the strength of large-scale magnetic fields.

$\RM$ (\Eq{eq:rm_integral}) can be approximated in terms of mean quantities as
\revb{
\begin{ceqn}
\begin{align}\label{eq:rm_average}
 \RM_{\rm j, cj} \approx \frac{\K~\overline{\ne}~\overline{\bpar}~L_{\rm j,cj}}{(1+z)^2},
\end{align}
\end{ceqn}
}
where \revb{$\overline{\bpar}$} is the \revb{ordered} parallel component of the large-scale magnetic field, and $L_{\rm j,cj}$ is the path length of jet and counterjet, respectively. \revb{Note that we assume the elliptical galaxy to be a single and homogeneous Faraday rotating medium to calculate the large-scale fields (\revb{$\overline{\bpar}$}) along the line of sight.} We assume that the difference in the $\RM$ values from the two lobes (jet,  $\RM_{\rm j}$, and counterjet, $\RM_{\rm cj}$) of the radio galaxy, $\Delta \RM$, is primarily due to different distances traversed by the radiation and can be expressed as

\revb{
\begin{ceqn}
\begin{align}\label{eq:rm_average_diff}
    \Delta\RM \approx \frac{\K~\overline{\ne}~\overline{\bpar}~ L}{(1+z)^2},
\end{align}
\end{ceqn}
}
where \revb{$L$} is the \revb{physical} separation between the two lobes in $\pc$. Assuming \revb{$\overline{\ne} = 0.01 \cm^{-3}$} and using $\Delta \RM$ and $L$ from the data, \revb{$\overline{\bpar}$} can be estimated. 
 
It is important to reiterate that $L$ in \Eq{eq:sigma_rm} and \Eq{eq:rm_average_diff} is taken to be equal to the distance \revb{or physical separation} between the two radio lobes \revb{or hotspots} in $\pc$. This means that the \revb{$ \overline{\bpar}$} estimated from \Eq{eq:rm_average_diff} will be averaged over the path length (hundreds of $\kpc$) scales. Hence, the large-scale \revb{fields} extracted from this analysis will be ordered over similar scales, i.e., hundreds of $\kpc$. This also ensures that the effect of small-scale random magnetic fields on $\RM$ cancels out on averaging over such large-length scales since the correlation length of the small-scale field is comparatively quite small, \revb{i.e.,} of the order of $100 \pc$.


\subsection{Results from the analysis of the RG catalog using the Laing-Garrington effect}
\label{sec:laing_garrington_results}
\begin{figure}
	\includegraphics[width=1\columnwidth]{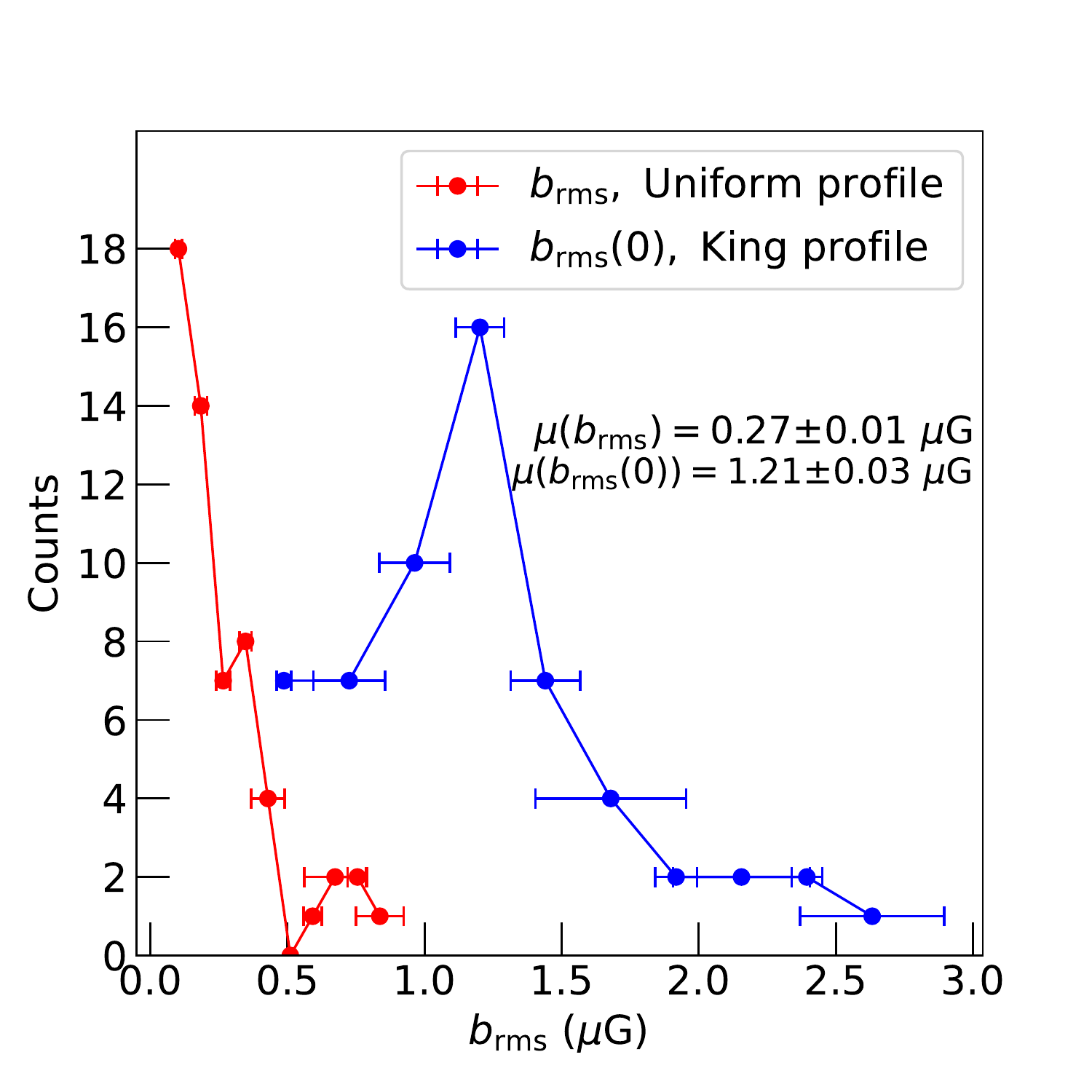}
    \caption{Histogram of rms strengths of small-scale magnetic fields ($\brms$ for the uniform profile and $\brms(0)$ for the King profile) estimated using the Laing-Garrington effect in the RG sample (\revb{$\sim 57$ sources}). The errors of the bins are standard deviations of the variation of the mean of every point in a bin calculated by the bootstrapping method with 500 iterations. The magnetic field strength for both electron density profiles lies in the range \revb{$0.06~\text{--}~2.75 \muG$}.}
    \label{fig:brms}
\end{figure}

\begin{figure*}
	\includegraphics[width=2\columnwidth]{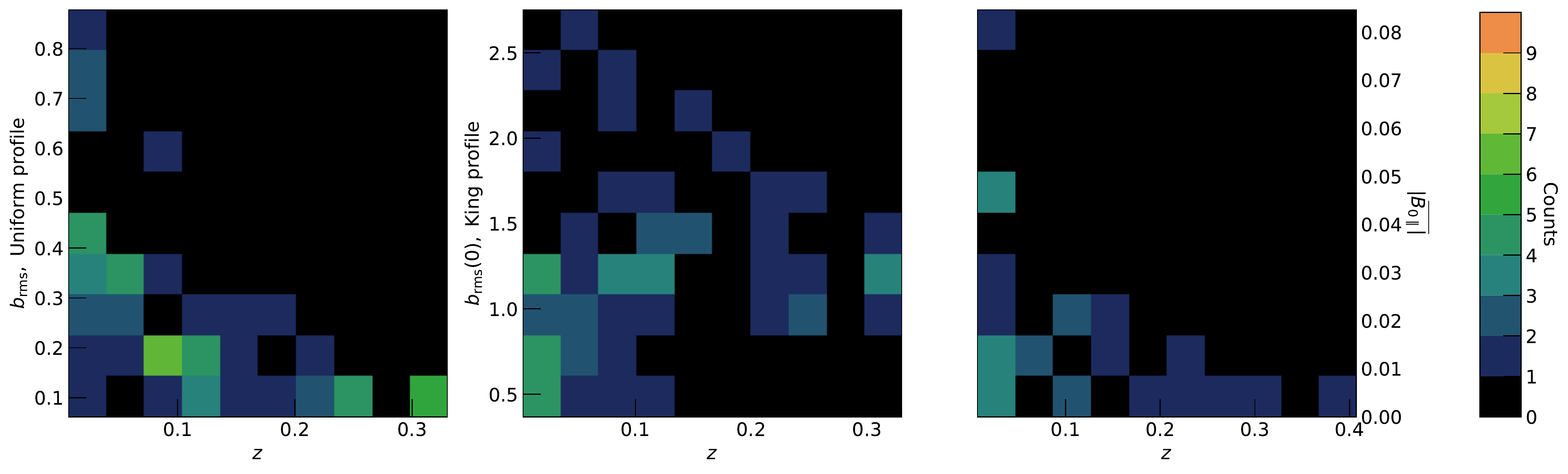}
    \caption{Two-dimensional histogram of the rms small-scale magnetic field strengths \rev{(left, middle plots), and large-scale magnetic field strengths (right plot), as a function of} redshift with colors showing the counts \revb{for all the} cases. \rev{The unit of magnetic fields is $\muG$.} The counts reduce with redshift in both the \rev{rms small-scale field plots}; however, the $\brms(0)$ (King profile) seems to have an increasing trend with the redshift as opposed to $\brms$ (uniform profile). \rev{The $\brms(0)$ somewhat exhibits the trend expected for small-scale magnetic fields in cosmological simulations shown in Fig. 1 of \citet{SetaEA2020II}. Older galaxies (high $z$) tend to have more gas and a more turbulent medium than the younger ones (low $z$), \revb{implying} that their magnetic field strengths will be higher due to the small-scale dynamo action. The large-scale field is an order of magnitude lower than the small-scale field and \revb{fluctuates} around zero with redshift \revb{(we plot only its absolute values)}. However, the dispersion of large-scale field (right plot) seems to reduce with redshift. This \revb{might} be \revb{possible} \revb{due to} a low number of RGs at higher $z$.}}
    \label{fig:z_hist}
\end{figure*}

\begin{figure}
	\includegraphics[width=1\columnwidth]{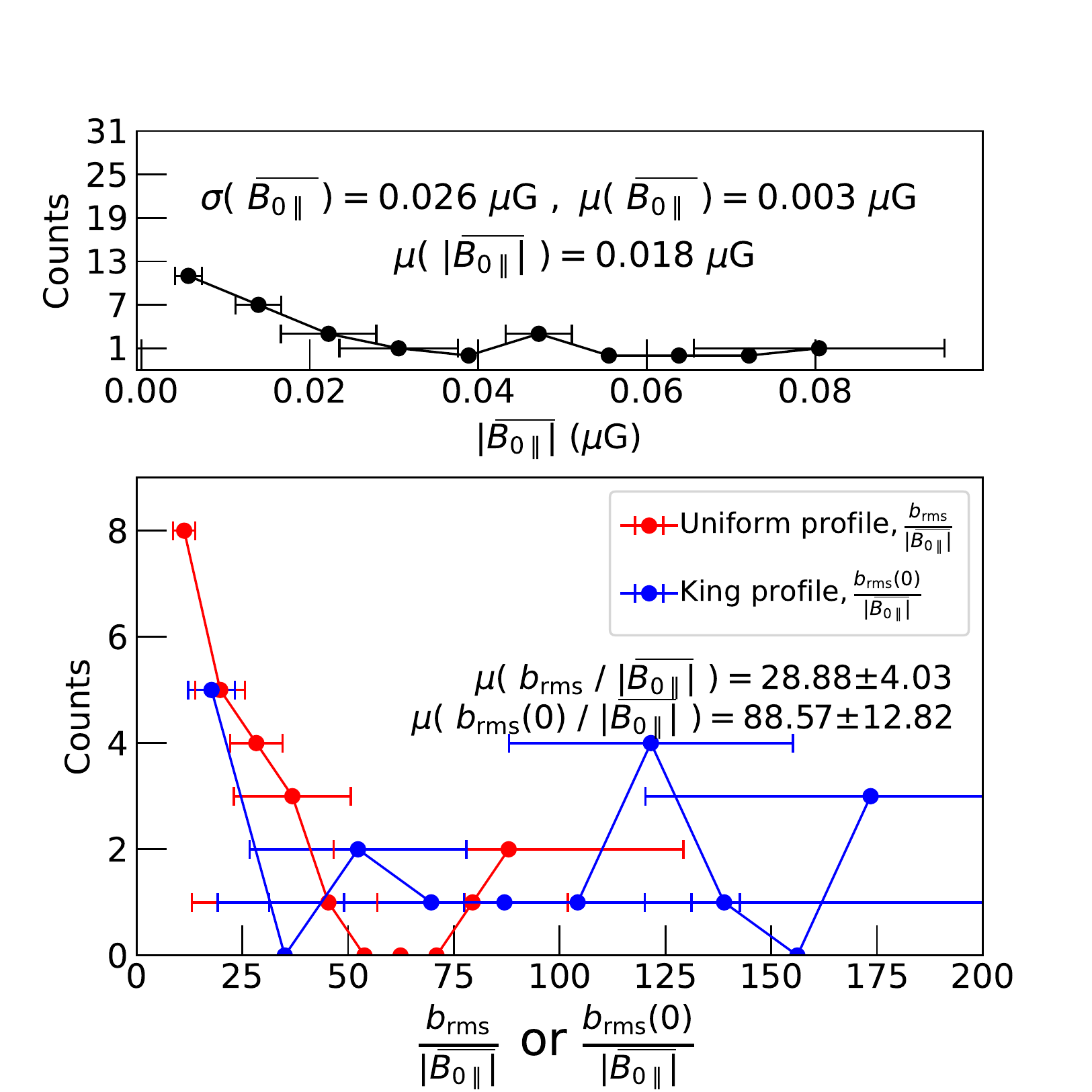}
    \caption{Top: Histogram for the estimated large-scale magnetic field strengths, \revb{$\overline{\bpar}$}, for \revb{$26$} sources. The errors of the bins are standard deviations of the variation of the mean of every point in a bin calculated by the bootstrapping method with 500 iterations. The mean and standard deviation are much smaller than the average $\brms$ or $\brms(0)$ (see \Fig{fig:brms}). This shows that the small-scale magnetic fields in ellipticals are much stronger than the large-scale fields. Bottom: Histogram of the ratio of the small-to-large scale field strengths \revb{for $\sim 20$ RGs in the sample}. Errors calculated are high at higher ratio values due to relatively large \revb{$\overline{\bpar}$} errors at \revb{such} magnitudes. The small-scale magnetic field strengths can be as high as \revb{$25~ \text{--}~90$} times stronger than the large-scale magnetic fields.}
    \label{fig:mean_ratio}
\end{figure}

\revb{After including the RGs with physical separation between the lobes less than $500 \kpc$, and removing the galaxies with SDSS halo diameters smaller than the physical separation ($< 500 \kpc$, SDSS detected), we retain 84 RGs for which depolarization probably occurs in the galaxy's medium. }
The $\brms$ (assuming uniform thermal electron density, \Eq{eq:sigma_rm}), $\brms(0)$ (assuming a King profile for the thermal electron density, \Eq{eq:king_sigma}), and \revb{$\overline{\bpar}$} (\Eq{eq:rm_average_diff}) are calculated for \revb{$84/210$} RGs available from the physical pairs in \rev{\citetalias{VernstromEA2019}}. However, after considering errors in $\brms$ values due to off-axis leakage polarization \citep[see][for further details]{MaEA2019}, \rev{\revb{roughly $57$ RGs} are retained with the magnitude of $\brms~\text{and}~\brms(0)$ being twice their respective errors.} \Fig{fig:brms} shows the histogram of $\brms$ and $\brms(0)$ for the RG sample and the rms field strengths \revb{lies} in the range \revb{$0.06~\text{--}~2.75 \muG$}. The mean of $\brms$ is \revb{$0.27\pm 0.01\muG$} and that of $\brms(0)$ (naturally expected to be slightly higher because \revb{it} only probes the magnetic field strength in the core of the elliptical galaxy) is \revb{$1.21 \pm 0.03 \muG$}. For the whole sample, the estimated rms small-scale magnetic field strength in ellipticals \revb{($\lesssim 3 \muG$)} is much smaller than that of the Milky Way and nearby spiral galaxies, which is observed to be in the range $5~\text{--}~10 \muG$ \citep{BeckEA2015}.

\begin{table*}
\caption{\revb{This table is a representative of the given and extracted values from the data of \citetalias{VernstromEA2019} for depolarization by galaxies (physical separation between the lobes less than $500 \kpc$). The columns are as follows: column1: \citetalias{VernstromEA2019} source identifier number, column2: redshift of the source, column3: path length or projected physical separation between the lobes, column4: diameter of the halo of the host galaxy of the radio lobes (of \citetalias{VernstromEA2019}) identified in SDSS by cross-matching, column5: ratio of the polarization values of the jet and counterjet, column6: calculated standard deviation of rotation measure in elliptical's CGM/halo, column7: small-scale magnetic fields with their one-sigma uncertainties for the uniform $n_{\rm e}$ case, column8: small-scale magnetic fields with their one-sigma uncertainties at the center of ellipticals for the King profile $n_{\rm e}$ case, column9: large-scale magnetic fields, column10: ratio of small-to-large scale magnetic fields for the uniform $n_{\rm e}$ case, column11: ratio of small-to-large scale magnetic fields for the King profile case. The dashes represent the lack of values for the corresponding parameter due to a failure in meeting the error conditions. The complete table is available in the supplementary text.}}
\begin{tabular}{|c|c|c|c|c|c|c|c|c|c|c|c|}
\hline
V19 no. & $z$ & \multicolumn{1}{p{0.6cm}}{\centering $L$ \\ $\rm (kpc)$} & \multicolumn{1}{|p{1cm}|}{\centering $\rm SDSS$ \\ $\rm identified$ \\ $\rm host~size$ $\rm (kpc)$} & $p_{\rm j}/p_{\rm cj}$ & \multicolumn{1}{p{1cm}}{\centering $\sigma_{\rm RM}$ \\ $\rm (rad/m^2)$} & \multicolumn{1}{|p{1.8cm}}{\centering $b_{\rm rms}$ \\ $(\mu{\rm G})$} & \multicolumn{1}{|p{1.8cm}}{\centering $b_{\rm rms}(0)$ \\ $(\mu{\rm G})$} & \multicolumn{1}{|p{2cm}|}{\centering $\overline{B_{0 \parallel}}$ \\ $(\mu{\rm G})$} & $b_{\rm rms} / |\overline{B_{0 \parallel}}|$ & $b_{\rm rms}(0) / |\overline{B_{0 \parallel}}|$ \\
\hline
38 & 0.097 & 174 & - & 1.29 & 6.53 & 0.18 $\pm$ 0.03 & 0.82 $\pm$ 0.15 & - & - & - \\
142 & 0.098 & 208 & 293 & 1.53 & 8.34 & 0.2 $\pm$ 0.02 & 1.14 $\pm$ 0.11 & - & - & - \\
438 & 0.03 & 384 & - & 1.18 & 5.9 & 0.12 $\pm$ 0.02 & 1.06 $\pm$ 0.19 & 0.0061 $\pm$ 0.0007 & 19.97 $\pm$ 5.99 & 174.67 $\pm$ 52.39 \\
521 & 0.3 & 431 & - & 1.04 & 1.82 & - & - & - & - & - \\
1654 & 0.133 & 257 & 2000 & 1.34 & 6.51 & 0.14 $\pm$ 0.02 & 0.98 $\pm$ 0.13 & - & - & - \\
2143 & 0.128 & 235 & 321 & 1.22 & 5.37 & - & - & - & - & - \\
3378 & 0.013 & 183 & - & 2.64 & 14.83 & 0.46 $\pm$ 0.03 & 1.92 $\pm$ 0.12 & 0.0177 $\pm$ 0.0031 & 25.73 $\pm$ 6.08 & 108.01 $\pm$ 25.53 \\
3458 & 0.129 & 209 & - & 1.21 & 5.27 & - & - & - & - & - \\
3892 & 0.095 & 160 & - & 1.43 & 7.74 & 0.22 $\pm$ 0.1 & 0.94 $\pm$ 0.42 & - & - & - \\
4451 & 0.242 & 369 & - & 1.66 & 7.12 & 0.1 $\pm$ 0.02 & 1.26 $\pm$ 0.22 & -0.0111 $\pm$ 0.0046 & 9.26 $\pm$ 5.51 & 113.44 $\pm$ 67.47 \\

\hline
\end{tabular}
\end{table*}

In \Fig{fig:z_hist}, we show the two-dimensional histogram of the rms small-scale magnetic field strengths with redshift (in the range \revb{$0.01~\text{--}~0.46$}) for both the uniform and King thermal electron density profiles\rev{, \revb{and} the large-scale magnetic field strengths}. For the uniform thermal electron density profile (left-panel of \Fig{fig:z_hist}), as the redshift increases, the number of elliptical galaxies decreases, and the magnetic field strength does not seem to vary much beyond redshifts of $0.1$. However, for the King profile case (right-panel of \Fig{fig:z_hist}), the field strength \revb{seemingly} increases with redshift even though the number of ellipticals decreases. The trend with redshift seen for the King profile case \rev{roughly} agrees with that \rev{obtained from the} the cosmological simulations \citep[see Fig. 1 in][]{SetaEA2020II}. \rev{As the redshift increases, the turbulence and amount of gas both increase, enhancing small-scale magnetic fields. \revb{However, a sample with inadequacies in data at higher redshifts could also generate a similar trend; hence, more robust datasets are required to confirm this trend.} We see that the large-scale fields fluctuate roughly around zero, and the dispersion increases as the redshift decreases. This might be because of a different number of sources at different redshifts.}

\revb{$\overline{\bpar}$}, \rev{the large-scale field component of the radio galaxies,} is estimated using \Eq{eq:rm_average_diff} with corresponding errors calculated by propagating $\RM$ uncertainties\revb{. $26$} RGs are retained for which the error in \revb{$\overline{\bpar}$} is less than half of the value. The top panel of \Fig{fig:mean_ratio} shows the histogram of \revb{$|\overline{\bpar}|$} and the computed \revb{absolute} mean and standard deviation of the \revb{$\overline{\bpar}$} distribution (\revb{$0.018 \muG$ and $0.026 \muG$}, respectively) are much smaller than small-scale field strengths (\Fig{fig:brms}). The insignificant value of the mean of \revb{$\overline{\bpar}$} ($0.003 \muG$) might be because of averaging over $\kpc$ scales and/or changing orientation (positive or negative) of the line of sight of the large-scale field. However, the standard deviation \revb{and the absolute mean} of the \revb{$\overline{\bpar}$} distribution \revb{are} also very small compared to the rms small-scale field strengths. 

Furthermore, \rev{for approximately \revb{$20$} radio galaxies (after removing the ratios for which the error is greater than the value)}, for which both the small-scale rms field strengths and \revb{$\overline{\bpar}$} values were available, the bottom panel of \Fig{fig:mean_ratio} shows the ratio of small-to-large scale field strengths (limited to a maximum value of $200$ \revb{to prevent extremely high ratios due to negligible large-scale field strengths;  only $2$ points were removed as a result of this condition}). \revb{All the reported errors were calculated by propagating the errors obtained from observations.} The rms small-scale field strengths, for both the uniform and King thermal electron density profile, are significantly higher (mean of the ratios are \revb{$28.88 \pm 4.03~\text{and}~88.57 \pm 12.82$} for uniform and King \revb{profile} cases, respectively) than that of the large-scale fields. \revb{The same analysis, when performed with three-sigma accuracy, yields very similar results with $\brms$ in the range $0.06 \, \text{--} \, 2.75 \muG$ and mean of the small-to-large scale field ratios in the range $25 \, \text{--} \, 70$. } These ratios are much larger in ellipticals as compared to spirals, where the ratio of small to large scale field strengths lie in the range $1~\text{--}~5$ \citep{FletcherEA2011,BeckEA2015}. Thus, the small-scale field generation mechanisms are probably much more efficient \rev{in our sample of ellipticals} as compared to the large-scale dynamo. \rev{This is probably because the large-scale field dynamo \revb{is inactive} due to a lack of significant rotation in the ellipticals \citep{Beck1996, MShu96}. However, due to the small number of available radio galaxies for which both the small- and large-scale fields can be estimated (\revb{$\sim 20$} out of $210$ galaxies), and large error bars in the bottom plot of \Fig{fig:mean_ratio}, better statistics are needed to confirm that the small-scale field generation mechanisms are dominant.} 
 
\section{Using background polarized radio sources}
\label{sec:polarized_emission}
\subsection{RM catalog}
\label{sec:rm_catalog}
\begin{figure} 
	\includegraphics[width=\columnwidth]{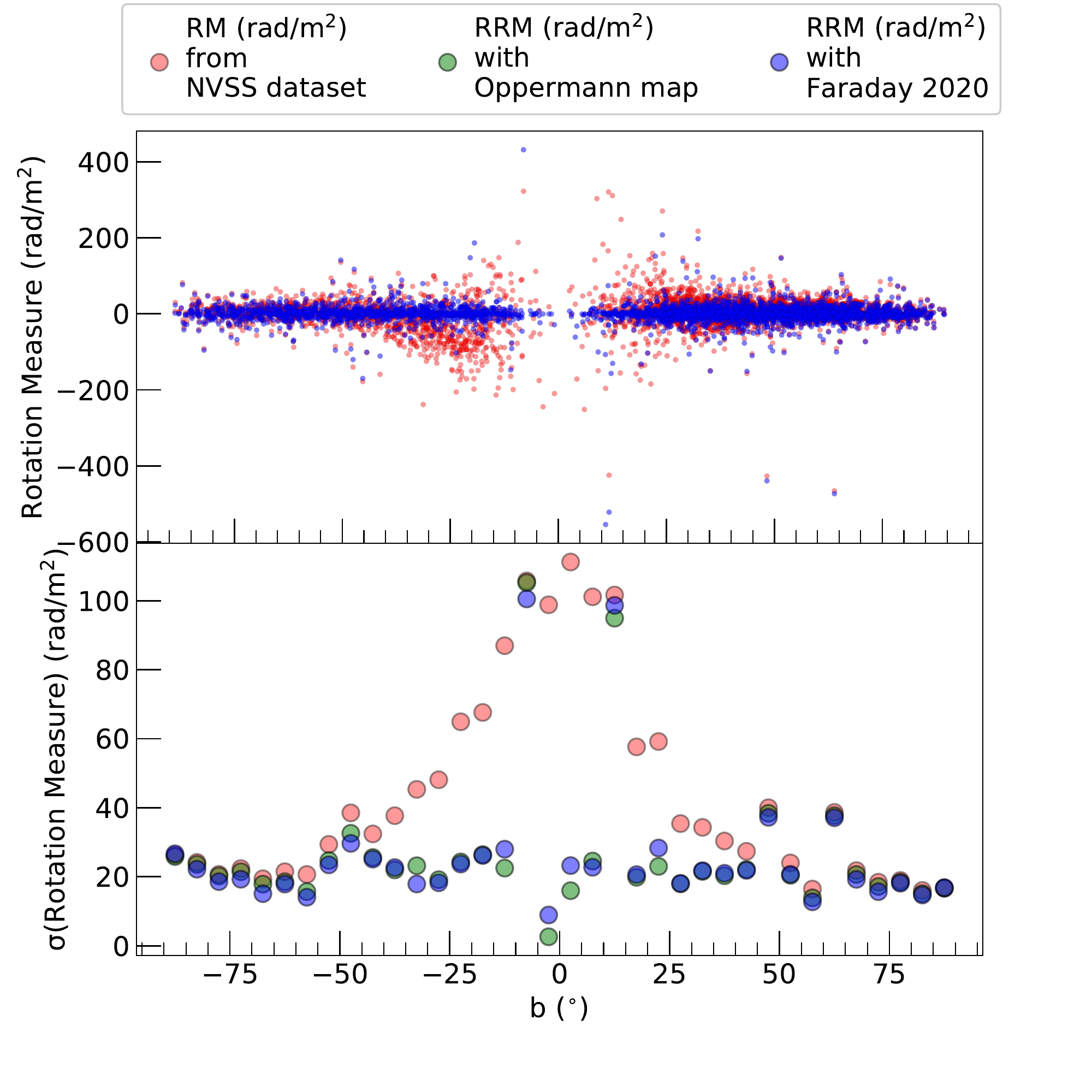}
    \caption{Top: $\RM$ (from \citet{FarnesEA2014}) and $\RRM$ (from \citet{FarnesEA2014} with the Milky Way contribution removed using \rev{\citetalias{Faraday2020}}) as a function of galactic latitude, $\rm b$. Bottom: Dispersions of \revb{rotation measures} when binned into $36$ equidistant bins as a function of Galactic latitude. \rev{$\stddev$ of $\RRM$ with Milky Way subtraction from Oppermann map \citep[][ green dots]{OppermannEA2015}, and \citetalias{Faraday2020} \citep[] [blue dots]{Faraday2020} is shown}. Due to a low number of $\RM$ and $\RRM$ values available in the bins near $b=0^\circ$, the $\sigma$ (Rotation Measure) takes unpredictable values, and we denote the dip at the center of this graph for the corrected case to a small number of $\RRM$ values in that bin. Since the $\RRM$ ideally represents only the \revb{extragalactic} background source distribution, we would expect the standard deviation of $\RRM$ to be flat with the Galactic latitude. However, $\sigma (\RRM)$ is seen to be high in some bins within $-30^\circ < {\rm b} < 30^\circ$ and this might be due to incomplete removal of the Milky Way contribution (especially the large-scale gradients). Thus, we only select sources beyond $\pm 30^{\circ}$ of the Galactic latitude. \revb{For all the subsequent analyses in the study, we adopt \citetalias{Faraday2020} to estimate the Milky Way $\RM$ contribution and finally calculate the residual rotation measure ($\RRM$).}}    
    \label{fig:RRM_GRM_visualizer}
\end{figure}

We use redshift, $\RM$, and polarization data for all $4003$ background polarized sources from \citet{FarnesEA2014} for our study (they obtained the $\RM$ values from \citet{TaylorEA2009} and redshifts by cross-matching the original NVSS data set in \citet{CondonEA1998} to a catalog of NVSS RM redshifts in \citet{HammondEA2012}). The dataset is expected to \revb{contain} mostly \revb{the jet-dominated sources in} classical radio galaxies and quasars \citep[see Sec. 7 of][]{FarnesEA2014}. These $\RM$s also have Milky Way contribution, $\RMMW$, which we remove using the data in \citet{OppermannEA2015} \rev{or \citetalias{Faraday2020} \footnote{The Milky Way $\RM$ data is available at \url{https://wwwmpa.mpa-garching.mpg.de/ift/faraday/2014/index.html} and  \url{https://wwwmpa.mpa-garching.mpg.de/~ensslin/research/data/faraday2020.html}}. Throughout the paper (\revb{except} in  \Fig{fig:RRM_GRM_visualizer} \revb{where} we show a comparison between both the maps), we use \citetalias{Faraday2020}.} We use RING-ordered \href{https://healpix.jpl.nasa.gov/}{HEALPix} maps at resolution \rev{$N_{\rm side}=512$} to extract $\RMMW$ for 3145728 pixels across the whole sky. The angular resolution of each pixel is around \rev{$0.1~\deg$} \footnote{\url{https://lambda.gsfc.nasa.gov/toolbox/tb_pixelcoords.cfm}}. Based on the coordinates of the background $\RM$ source, the closest pixel number from the HEALPix map was calculated, and the \revb{Galactic rotation measure} ($\RMMW$) for that pixel was subtracted from the background \revb{(extragalactic)} $\RM$ data to obtain the residual rotation measure, $\RRM$.

The top panel of \Fig{fig:RRM_GRM_visualizer} shows $\RM$ with and without subtracting the Milky Way contribution as a function of the Galactic latitude, $\rm b$, and the bottom panel shows their standard deviation over Galactic \revb{latitude} bins of size $5^{\circ}$. We would expect the standard deviation of residual contributions to remain nearly uniform throughout the sky as they only represent the background source distribution. However, there is some dependence of the standard deviation of $\RRM$ on the Galactic latitude, especially in the range  $-30^\circ < {\rm b} < 30^\circ$. This is probably due to incomplete Milky Way subtraction, despite correcting with the model from \citet{OppermannEA2015} \rev{and \citetalias{Faraday2020}}. Thus, for our study, to avoid any unintended Milky Way contribution, all the background $\RM$ sources appearing inside the Galactic latitude \revb{range} $-30^\circ < {\rm b} < 30^\circ$ were removed. As a result, only 3098 of 4003 points were preserved outside the $\pm 30^\circ$ of the Galactic plane.

\rev{The NVSS rms resolution for the coordinates of the faintest radio sources (flux
densities $<$ 15 mJy, only $227~\text{out of a total of}~37000$ sources) is $7~\rm{arcsec}$\footnote{ \url{https://www.cv.nrao.edu/nvss/}}. This is about $60 \kpc$ at $z=1$, significantly lesser than the average $\rhalo$ of the galaxies in our sample (about $500 \kpc$, see \Sec{sec:sigma_ip}). Additionally, for the rest of the majority of the sources, rms resolution is less than $1~\rm{arcsec}$. This is equivalent to a distance of less than $10 \kpc$ at $z=1$, which is small enough to prevent any false detection of the intervening galaxies.
}

\subsection{Galaxy Catalog}
\label{sec:galaxy_catalog}

To explore the effects of intervening galaxies on \revb{RM from the} background \revb{polarized} sources, we use the galaxy catalog provided by the 16th data release from the SDSS \citep{AhumadaEA2020}. It is the fourth data release from SDSS-IV (following DR13, DR14, DR15). DR16 provides spectra along with \revb{spectroscopic}
redshifts for around \revb{1} million unique galaxies, with the help of the eBOSS \rev{\citep[Extended Baryon Oscillation Spectroscopic Survey,][]{DawsonEA2016}} ELG \rev{(Emission Line Galaxies)} program and the data of APOGEE-2 South \rev{\citep[Apache Point Observatory Galactic Evolution Experiment,][]{MajewskiEA2017}}. The SDSS catalogs were found and queried on the Catalog Archive Server \citep{ThakarEA2008} \footnote{\url{https://skyserver.sdss.org/casjobs/}}. These catalogs contain photometric and spectroscopic properties, along with their derived object parameters. 

The following parameters were extracted for galaxies from the CasJobs server: Galaxy Coordinates, Petrosian radius in u, g, r, i, and z bands \footnote{Details of Petrosian radius can be found at \url{https://www.sdss.org/dr12/algorithms/magnitudes/}.}, spectroscopic redshifts \revb{(robust upto $z=0.7$)}\revb{.} Morphology classifications of galaxies with/without spectra from the Galaxy Zoo \citep{LintottEA2008} were \revb{also} obtained. An object is classified as a galaxy only if both the spectroscopic and photometric information classifies it as one. The coordinates and the Petrosian radii are obtained from the photometric observations, whereas redshifts are obtained from the spectroscopic data. Photometric, spectroscopic data of the SDSS and the Galaxy Zoo were connected using a common unique SDSS identifier, specobjid.

Galaxy Zoo 1 project \citep{LintottEA2011} categorized a total of $738,175$ galaxies from the SDSS DR6, $92 \%$ of which have spectra available in SDSS DR7. Votes from more than $100000$ volunteers, corrected for weightage of every user, were used to report $p_{\rm ell}$ and $p_{\rm spi}$, which are the probabilities of a galaxy being an elliptical or a spiral, respectively. As this is a statistical study, it would be useful to have most of the galaxies classified, so we take $p=0.65$ to be the threshold to categorize a galaxy into a morphological type (probability at which the Galaxy Zoo project claims to have less than $10\%$ classification error). A galaxy in the sample is categorized into one of the following four categories: `Elliptical' (if $p_{\rm ell}>0.65$ and $p_{\rm spi}<0.35$), `Spiral' (if $p_{\rm spi}>0.65$ and $p_{\rm ell}<0.35$), `Not Sure' (if both the conditions above are not satisfied), and `Not Available' (if the probabilities for a galaxy type are unavailable in the Zoo dataset). 

The maximum of the five Petrosian radii in u, g, r, i, and z bands is selected as the galaxy's radius, referred to as $\rgal$ throughout the text. The galaxy's contribution to $\RM$ is considered up to $10$ times the galaxy radius, which \revb{likely traces} the halo or CGM ($\rhalo=10\rgal$ with a maximum of $1 \Mpc$). Giant elliptical haloes can extend up to $1 \Mpc$ \citep{UsonEA1990} and the CGM up to a few hundred $\kpc$ \citep{TumlinsonEA2017}. Thus, $\rhalo$ can be the radius of the halo \revb{or the} CGM.

\subsection{Cross-matching the catalogs}
\label{sec:cross-match}
\begin{figure*}
    \begin{subfigure}{1\textwidth}
    \centering
    \includegraphics[width=1\columnwidth]{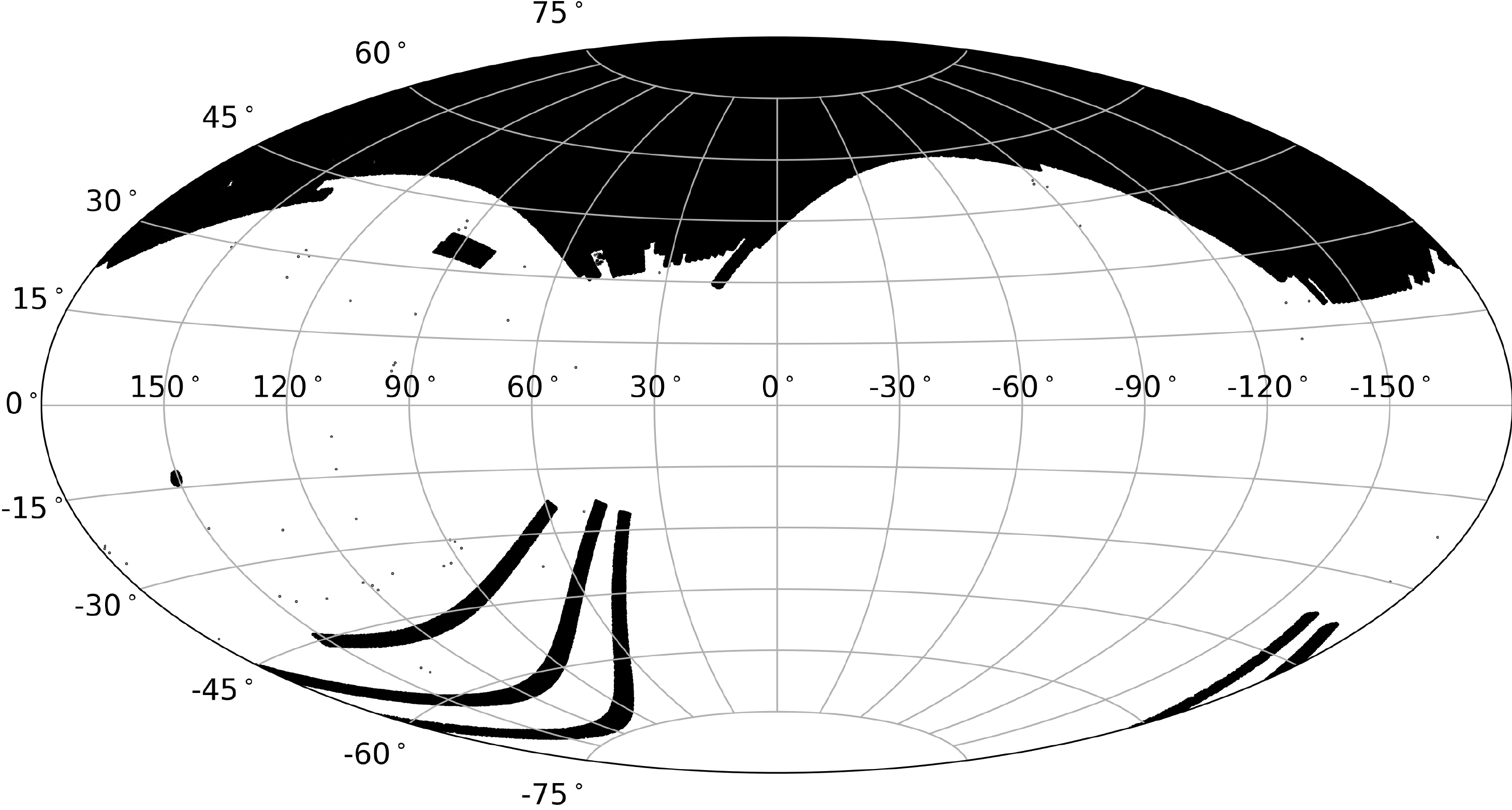} 
    \caption{Shaded parts are the regions covered by galaxies from the SDSS DR16 catalog. The figure contains approximately \revb{0.9} million galaxies. }
    \end{subfigure}
    \begin{subfigure}{1\textwidth}
    \includegraphics[width=1\columnwidth]{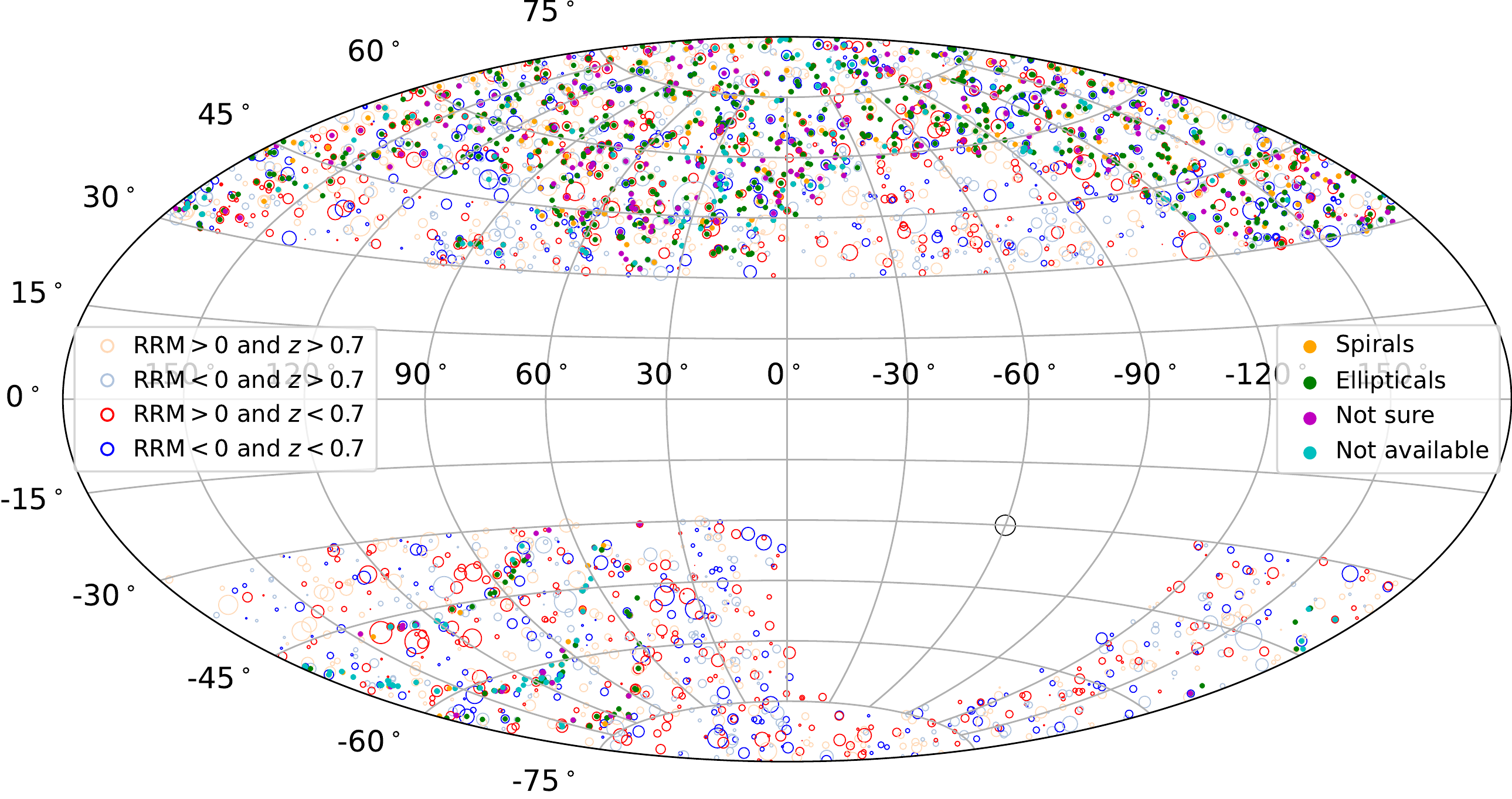} 
    \caption{\revb{Unfilled circles} are the RM sources with positive and negative $\RRM$ values with \revb{$z<0.7$ or $z>0.7$}. The size of the circles is scaled to the magnitude of the $\RRM$ of each source. The black circle at coordinates ($\rm -60^\circ,~-30^\circ$) represents an absolute $\RRM$ value of $100~\rad/\m^2$. Approximately 3000 \revb{background $\RM$} sources \revb{(available after imposing latitude and redshift availability conditions)} are plotted in this figure. Solid dots depict \revb{only} the intervening galaxies of the four types mentioned in the legend.}
    \end{subfigure}    
    \caption{\rev{All the coordinates in these plots are in the galactic frame.} (a) All-sky maps of available galaxies from the SDSS DR16. (b) Available $\RM$ sources and known morphology of galaxies in the same region.}
    \label{fig:money_plot}
\end{figure*}

\Fig{fig:money_plot}(a) shows the available galaxies from SDSS DR16. \Fig{fig:money_plot}(b) shows the $\RM$ data sources and the known morphology of galaxies according to the Galaxy Zoo project. In \Fig{fig:money_plot}(b), visually, there is \revb{a significant} overlap between the intervening galaxies and $\RM$ sources.

We use coordinates of \rev{background polarized} sources \rev{with available RM values from the NVSS catalog} and galaxies from the galaxy catalog \rev{(SDSS DR16 catalog)}, respectively, to cross-match and detect the intervening galaxies. \revb{These background polarized sources are mostly jet-dominated sources due to their steep spectrum \citep[see Fig. 5 and Sec. 7 of][]{FarnesEA2014} and their contribution is considered part of the $\RMsource$ term in \Eq{eq:rm_contri}.} For a particular source, the DR16 catalog is scanned to identify the \rev{intervening galaxies} that contain the source within its $\rhalo$ ($10 \rgal$, where $\rgal$ varies with the galaxy). \rev{Also, see the left panel of Fig. 11 in \citetalias{VernstromEA2019} for a schematic}. To ensure that the galaxies intervene, only those with redshifts \revb{(minus two-sigma error)} smaller than that of the RM source \revb{(plus two-sigma error)} are chosen\revb{, with an additional relaxation of $z=0.0005$ which corresponds to $\sim 2 \Mpc$ for nominal redshifts.} \revb{Thus, the redshift condition for the intervention is $z_{\rm gal}-2 {\rm error}(z_{\rm gal})<z_{\rm RM}+2 {\rm error}(z_{\rm RM})+0.0005$. This condition ensures that the host galaxies of the background sources are also identified as intervening galaxies, and the additional relaxation covers the possible difference between the redshifts of the host galaxy's center and the background polarized source jet. Galaxies from the SDSS catalog have robust spectroscopic redshifts upto $z=0.7$. Thus, when the redshift of the RM source is $>0.7$, there could be undetected intervening galaxies in its line of sight.} Impact parameter values (distance between the \rev{intervening} galaxy's center and the point where radiation from the background source penetrates \rev{that} galaxy) are also calculated using the coordinates.

\subsection{Analysis of $\RRM$s via probability distribution functions (PDFs)}
\label{sec:sdss_analysis}
After obtaining the number and morphology of intervening galaxies for $\RM$ sources by cross-matching galaxies of known morphology with the $\RRM$ sample, we aim to understand the contribution of the $\RM$ from the intervening galaxies to the observed $\RRM$. We would naively expect that on increasing the number of galaxies, the standard deviation of $\RRM$ distribution due to background \revb{extragalactic} polarized sources would increase. Similarly, we would expect that the intervening spiral galaxies would affect the standard deviation more than the elliptical galaxies since the spirals have significantly stronger fields than ellipticals (\Sec{sec:laing_garrington} and \citet{SetaEA2020II}).

\subsubsection{Gaussian and non-Gaussian $\RRM$ PDFs}
\label{sec:GnG}

The \rev{probability density function (PDF) of $\RM$ values} of the background polarized sources is \rev{not completely known} and thus is \rev{commonly} assumed to be a Gaussian distribution\revb{,} largely described by the mean and standard deviation \rev{(see Fig. 1 of \citet{SaralaEA2001}, \revb{and Fig. 3 and} Fig. 7 of \citet{BasuEA2018} for some Gaussian and non-Gaussian \revb{fits and background $\RM$ distributions}).} In this study, we explore the possibility of the \rev{PDF to be both a Gaussian and non-Gaussian. Non-Gaussian PDFs, because of the long heavy tail at higher values of $|\RM|$, can better capture the effects of the outliers. Furthermore, the $\RM$ distributions from magnetohydrodynamic turbulent simulations and observations can be non-Gaussian \citep[e.g.~see Fig. 3a, 6(a), and A1 in][]{SetaF2021}.} Thus, in addition to computing the \revb{statistical parameters} of $\RRM$ \revb{distribution} using the data, we fit the $\RRM$ probability distribution function with the following three (Gaussian and non-Gaussian distributions with controllable parameters) distributions \revb{to obtain statistical parameters of RRM distribution.}

\begin{itemize}
    \item Gaussian distribution (G)
        \begin{ceqn}
        \begin{align}\label{eq:g_dist}
             \rm g(\RRM) = \frac{1}{(2\pi)^{1/2}\sigmarrm}~exp\left(\frac{-(\RRM- \mean( \RRM ))^2}{2\sigmarrm^2}\right),
        \end{align}
        \end{ceqn}
     where $\mean( \RRM )$ and $\sigma_{\RRM}$ represent the mean and standard deviation of the $\RRM$ distribution.


    \item Gaussian distribution with sinh-arcsinh transform (SH)
    \begin{subequations}
        \begin{ceqn}
        \begin{align} \label{eq:sh_dist}
            {\rm sh}(y) = \frac{A}{(2\pi)^{1/2} \sigma_y}~\exp\left(\frac{-(y-y_0)^2}{2\sigma_y^2}\right)
        \end{align}
        \end{ceqn}
        with
         \begin{ceqn}
         \begin{align} 
        y = \sinh(\delta \arcsinh(\RRM)-\epsilon), \, \text{and} \\  A = \frac{\delta \cosh(\delta \arcsinh(\RRM)-\epsilon)}{(1+\RRM^2)^{1/2}}. 
    \end{align}
    \end{ceqn}
    \end{subequations}
    Here, the parameters $\delta$ and $\epsilon$ control the $\kurtosis$ (tailedness) and the $\skewness$ (asymmetry) of the non-Gaussian distribution. When $\delta=1~\text{and} ~\epsilon=0$, ${\rm sh}(y)$ turns into a Gaussian distribution. $y_0$ and $\sigma_y$ are the mean and standard deviation of the distribution, respectively. In a general case, the standard deviation of ${\rm sh}(y)$ is not the same as the standard deviation of $\RRM$ distribution.
  
    
    \item Modified line-broadening distribution (LB) 
    \begin{subequations}
        \begin{ceqn}
        \begin{align} \label{eq:lb_dist}
            {\rm lb}(\RRM) = \frac{A}{(\RRM-u_0)^2+w^2}\exp\left(\frac{-\RRM^2}{2\Sigma^2}\right)
        \end{align}
        \end{ceqn}
        with 
        \begin{ceqn}
        \begin{align}
            A = \left(\frac{\sqrt{2\pi}\Sigma}{w^2}.{\rm voigt}\left(\frac{u_0}{w} , \frac{\Sigma^2}{2w^2}\right)\right)^{-1}, 
        \end{align}
        \end{ceqn}
        where ${\rm voigt}$ function is defined by,
        \begin{ceqn}
        \begin{align}
            {\rm voigt}(x,a)=\frac{1}{\sqrt{4 \pi a}}.\int_{-\infty}^{\infty} \frac{\exp(\frac{-(x-y)^2}{4a})}{1+y^2} \,dy. 
        \end{align}
        \end{ceqn}
    \end{subequations}
    Here, $w$ and $u_0$ are the parameters that control the $\kurtosis$ and $\skewness$ of the distribution, respectively. $\Sigma$ controls the standard deviation of the distribution, and it should be noted that the normalizing parameter $(A)$ is a function of the Voigt function. Also, the parameter $\Sigma$ is not necessarily the same as the standard deviation of the $\RRM$ distribution.
\end{itemize}

\begin{figure}
    \begin{subfigure}{0.5\textwidth}
    \includegraphics[width=1\columnwidth]{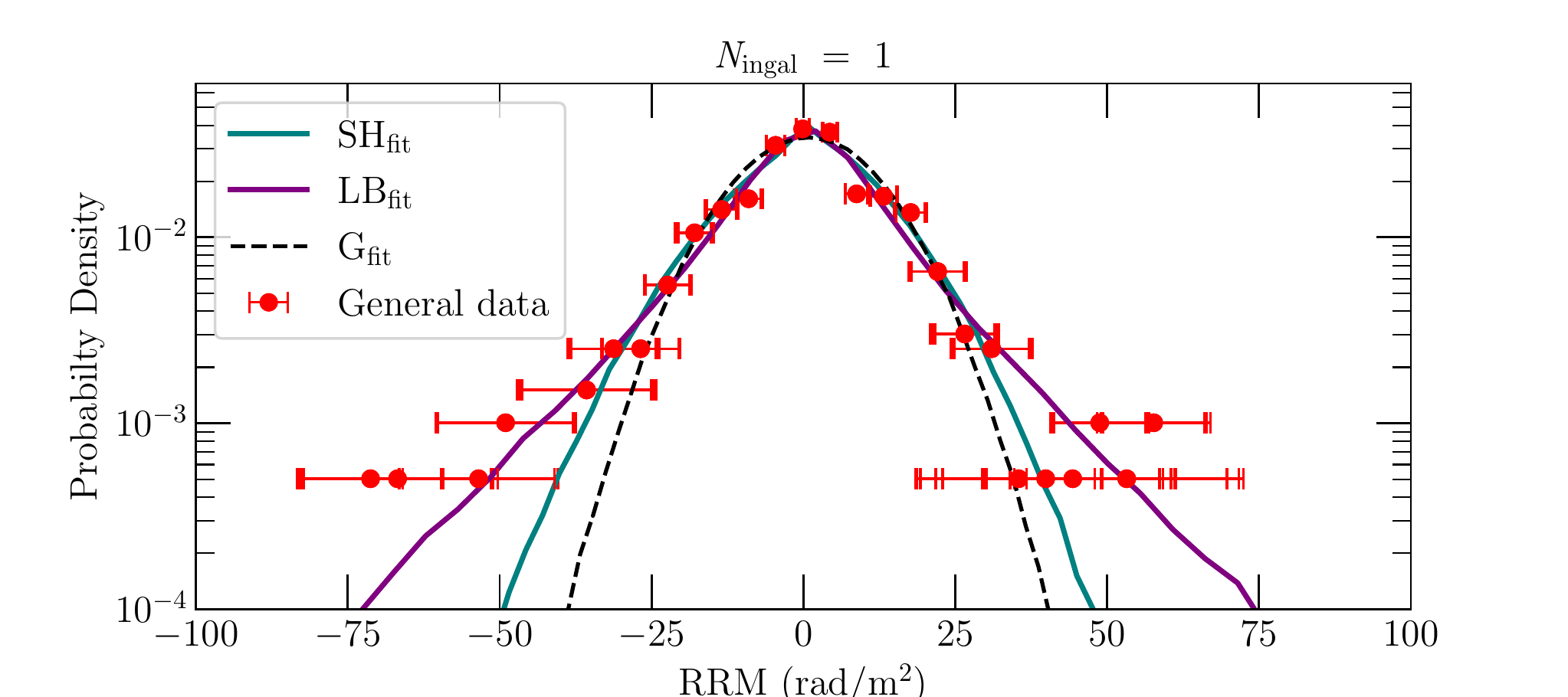} 
    \caption{\revb{The} red curve represents the PDF (with a total of $30$ bins) of $\RRM~(\rad/\m^2)$ for the background RM sources whose radiation passes through exactly \revb{one} galaxy inside $\rhalo$.}
    \label{fig:gen}
    \end{subfigure}
    \begin{subfigure}{0.5\textwidth}
    \includegraphics[width=1\columnwidth]{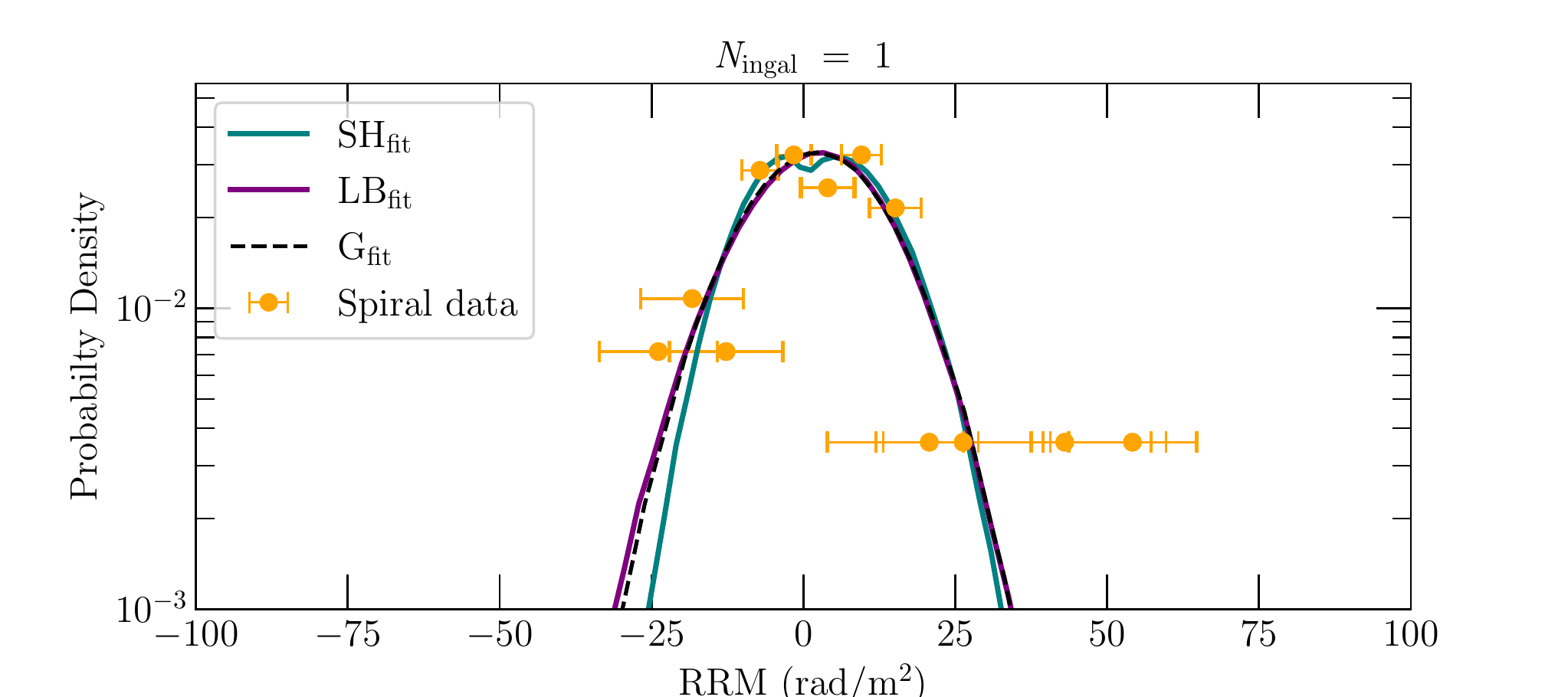}
    \caption{\revb{The} orange curve represents the PDF (with a total of $15$ bins) of $\RRM~(\rad/\m^2)$ for the background RM sources whose radiation passes through exactly \revb{one} spiral inside $\rhalo$. Comparatively shorter tails are observed in this particular PDF\revb{,} suggesting a lack of outliers.}
    \label{fig:spi}
    \end{subfigure}
    \begin{subfigure}{0.5\textwidth}
    \includegraphics[width=1\columnwidth]{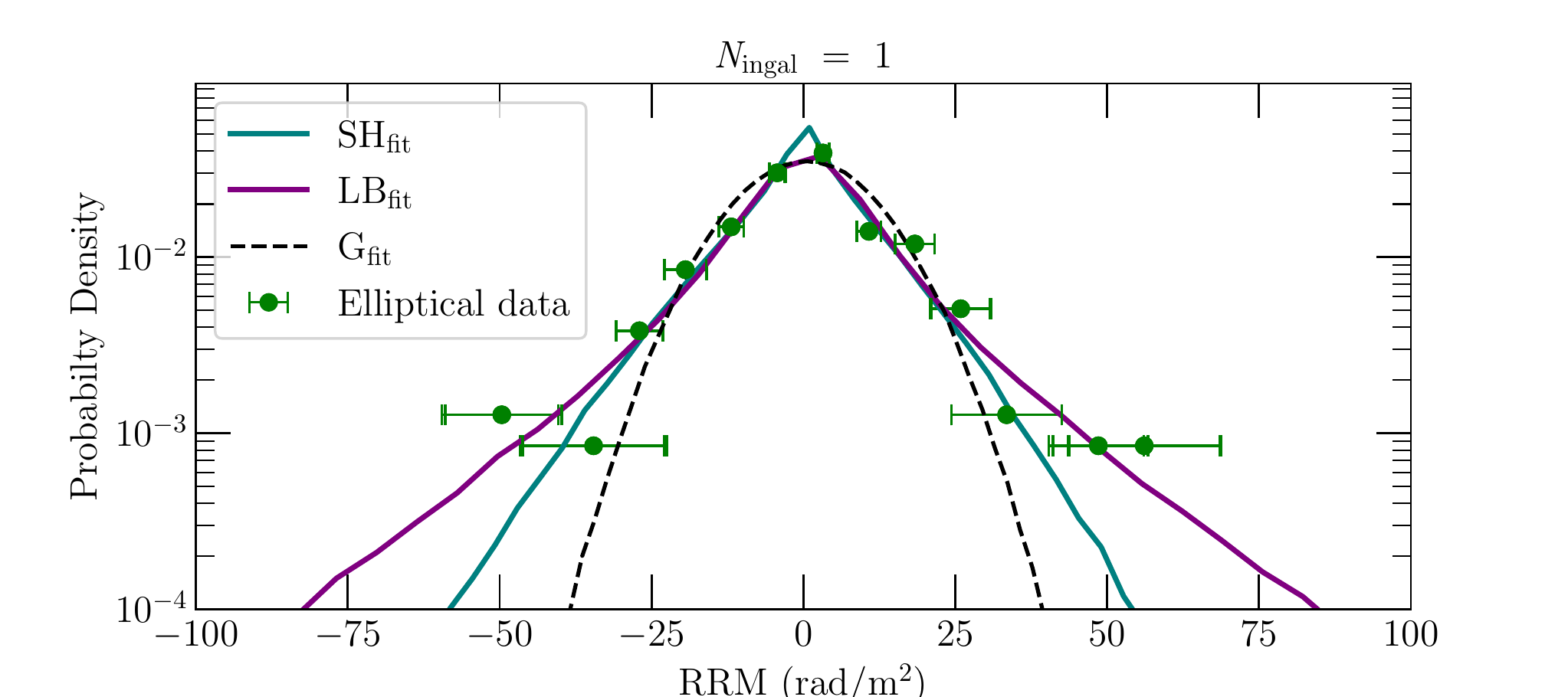}
    \caption{\revb{The} green curve represents the PDF (with a total of $15$ bins) of $\RRM~(\rad/\m^2)$ for the background RM sources whose radiation passes through exactly \revb{one} elliptical inside $\rhalo$. }
    \label{fig:ell}
    \end{subfigure}
    \caption{PDFs of $\RRM$ for sources with one intervening galaxy ($\ningal=1$) along the path length \revb{and $z<0.7$} for three different cases depending on the morphology: (a) General (all galaxies), (b) spiral, and (c) elliptical. For each case, three different fits using three different fitting functions (\Eq{eq:g_dist} for G$_{\rm fit}$, \Eq{eq:sh_dist} for SH$_{\rm fit}$, and \Eq{eq:lb_dist} for LB$_{\rm fit}$) are created using the  dataset of $\RRM$ points. All the parameters extracted from the data and fits are provided in \Tab{tab:fitparams}.}
    \label{fig:three_cases}
    \end{figure}

In \Fig{fig:three_cases}, we show the data and the fits using these functions with one intervening galaxy, $\ningal=1$, and for general data (data of all galaxies in (a)), spiral data (data of only one intervening spiral galaxy in (b)), and elliptical data (data of only \revb{one} elliptical galaxy in (c)), respectively\revb{, for RM sources with $z<0.7$}. \revc{We ignore the absolute $\RRM$ values greater than $100 \rad/\m^{2}$ to avoid \revc{the sensitivity of our results to a few outliers \citep[also see Sec. 2.2.2 in][]{LanP2020}}\revb{,} \rev{just $18$ out of $3098$ available sources are classified as outliers with this condition (see the top part of \Fig{fig:RRM_GRM_visualizer})}.} The data is represented with error bars (errors are standard deviations of the variation of mean of every point in a bin calculated by the bootstrapping method with 500 iterations), and \revb{the} three curves (\Eq{eq:g_dist} for G$_{\rm fit}$, \Eq{eq:sh_dist} for SH$_{\rm fit}$, and \Eq{eq:lb_dist} for LB$_{\rm fit}$) are fitted to the PDFs of the data. Using the fitted probability density functions (G, SH, LB), a sample of $10^{5}~\RRM$ points is generated according to its respective probability distribution from the fit for all the three functions. These generated samples can now be utilized to extract various statistical \revb{parameters} like $\sigma, \skewness,$ and $\kurtosis$ (the mean is always very close to zero) of the $\RRM$ distribution from the fits. These properties for the data and fits are given in \Tab{tab:fitparams}. The reported kurtosis is such that the kurtosis of a Gaussian distribution is zero. We also perform \revb{a} \revc{KS test\footnote{The KS test is performed by calculating the maximum absolute difference between the CDFs of the data and fits.}} to compare the $\RRM$ distribution obtained from the data with each fitted function. \revb{Its test statistic values ($\DKS$) and p-values ($\pKS$)} are given in the last two columns of \Tab{tab:fitparams}\revb{, respectively}. 

\begin{table}
    \caption{Statistical properties of the $\RRM$ distribution of the data and the fitted functions for each of the three cases in \Fig{fig:three_cases} with one intervening galaxy. The columns are as follows: 1) the morphology of the intervening galaxy: General (all galaxies, \Fig{fig:three_cases}(a)), spiral (\Fig{fig:three_cases}(b)), and elliptical (\Fig{fig:three_cases}(c)), 2) data or the fitted function, 3) the computed standard deviation, $\sigma$ in $\rad/\m^{2}$, 4) the computed skewness, $\mathcal{S}$, 5) the computed kurtosis, $\mathcal{K}$, such that the kurtosis of a Gaussian distribution is zero, \revb{6) the D-statistic when KS test} is performed to compare the $\RRM$ data distribution with the fitted function, and \revb{7) the KS test $p$-value.}}
    \begin{tabular}{ccccccc}
    \hline
    Morphology & Case & $\sigma$ & $\mathcal{S}$ & $\mathcal{K}$ & $D_{\rm KS}$ & $p_{\rm KS}$ \\
    \hline
    
    $\rm General$ & $\rm Data$ & 15.21 & -0.22 & 3.49 & - & -\\
    $\rm General$ & $\rm G$ & 11.62 & -0.01 & -0.02 & 0.1 & 0.74\\
    $\rm General$ & $\rm SH$ & 13.19 & -0.12 & 0.57 & 0.09 & 0.78 \\
    $\rm General$ & $\rm LB$ & 16.33 & -0.07 & 3.82 & 0.08 & 0.82 \\

    \hline
    
    $\rm Spirals$ & $\rm Data$ & 15.14 & 1.01 & 2.57& - \\
    $\rm Spirals$ & $\rm G$ & 12.15 & -0.01 & -0.02 & 0.1 & 0.86\\
    $\rm Spirals$ & $\rm SH$ & 11.35 & 0.09 & -0.34 & 0.1 & 0.86\\
    $\rm Spirals$ & $\rm LB$ & 12.36 & -0.08 & 0.1 & 0.1 & 0.86\\

    \hline

    $\rm Ellipticals$ & $\rm Data$ & 14.57 & 0.13 & 2.36 & - \\
    $\rm Ellipticals$ & $\rm G$ & 11.41 & -0.01 & -0.02 & 0.15 & 0.71 \\
    $\rm Ellipticals$ & $\rm SH$ & 13.15 & -0.19 & 2.25 & 0.15 & 0.71\\
    $\rm Ellipticals$ & $\rm LB$ & 17.77 & -0.08 & 5.73 & 0.14 & 0.74\\

    \hline
    \end{tabular}
    \label{tab:fitparams}
\end{table}

The kurtosis of the $\RRM$ distribution computed from data is far from zero for all three cases, and thus $\RRM$ distribution (and by extension, the $\RM$ source distribution) can very well be non-Gaussian. \revb{By choosing a confidence level of $95\%$ (or $p$-value $>0.05$)}, this can also be seen via the last two columns in \Tab{tab:fitparams}, which show via \revb{KS test} that the non-Gaussian distributions (SH and LB) can represent the data \revb{equally well as} the Gaussian distribution\revb{, since the $\pKS > 0.05$ for all the distributions}. Also, note that the standard deviation of the data is always much larger than the fitted Gaussian because the higher values of $\RRM$ are not considered with a Gaussian distribution. So, considering just a Gaussian distribution for rotation measure distribution of background sources is probably \revb{incomplete}. Thus, we study the dependence of the statistical properties (primarily the standard deviation of the $\RRM$ distribution) for all four cases (data, G, SH, and LB) on the number and morphology of intervening galaxies.

It is important to classify the morphology of galaxies which contribute \revb{the} most. \revb{There can be contribution from numerous intervening galaxies}, not all of which might contribute significantly. If for a particular background source, the number of spirals, $N_{\rm spi}$, is greater than \revb{twice} the number of ellipticals, $N_{\rm ell}$ and \revb{twice} the number of not sure ones\revb{,} $N_{\rm NS}$, the contribution is assumed to be primarily from the spirals. If $N_{\rm ell} \neq 0~\text{and}~N_{\rm spi}=N_{\rm NS}=0$ then the contribution is assumed to be from the ellipticals. If $N_{\rm NS}>2(N_{\rm spi}+N_{\rm ell})$ then the contribution is assumed to be from the lenticulars. Only those $\RM$ sources with $N_{\rm NA}=0$ are considered. Here, $N_{\rm spi}$, $N_{\rm ell}$, $N_{\rm NS}$, and $N_{\rm NA}$ refer to the number of intervening `Spirals', `Ellipticals', `Not sure', and `Not available' galaxies, respectively. \revc{All the results containing a morphology classification of the intervening galaxies follow the aforementioned conditions.}

\subsection{Results from $\RRM$ PDFs constructed using the background radio sources}
\revb{In this section, results obtained from cross-matching the NVSS and SDSS catalogs (\Sec{sec:cross-match}) are \revc{presented\footnote{For all the subsequent figures apart from \Fig{fig:gen_line}, we constrain our analysis to RM sources with $z<0.7$, for robust results (see \Sec{sec:cross-match}).}. } }
\label{sec:polarized_emission_results}
\subsubsection{Dependence of $\sigmarrm$ on $\ningal$ }
\label{sec:sigma_ningal}

\begin{figure}
    \begin{subfigure}{0.45\textwidth}
    \includegraphics[width=1\columnwidth]{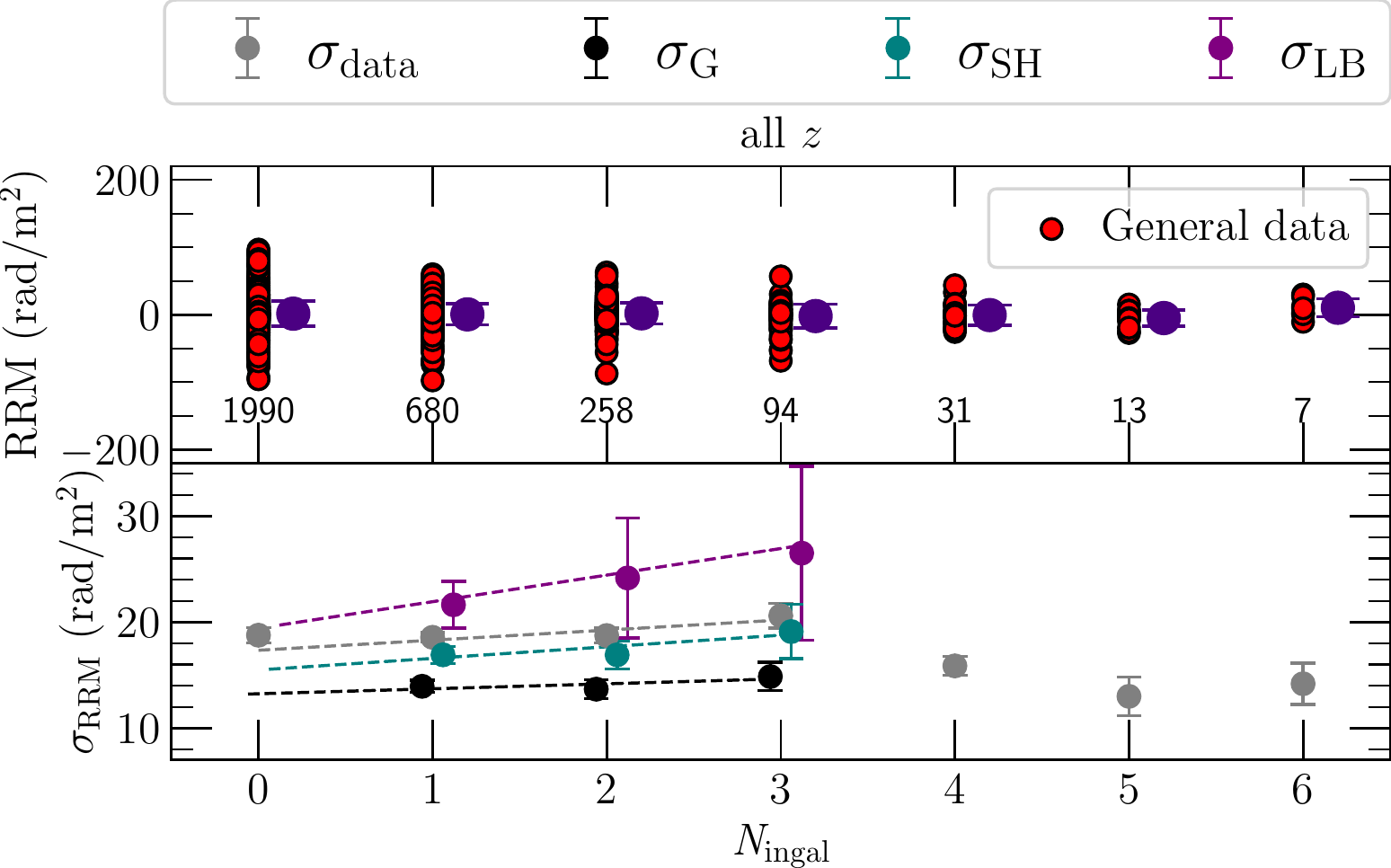} 
    \caption{$\sigmarrm$ vs. $\ningal$ for the general (all morphologies included, \revb{all RM source redshifts}) case. \rev{The curve-fitting functions are used for \revb{just} $\ningal=1-3$, due to lack of datapoints at higher $\ningal$. \revb{The slope values for data, G, SH, \revb{ and} LB cases are $0.97\pm1.23, 0.41\pm1.5, 1.1\pm2.66, \text{ and } 2.65\pm8.61$ respectively, and their corresponding intercepts are $17.32\pm1.97, 13.25\pm2.51, 15.45\pm4.1, \text{ and } 18.93\pm13.12$.}}}
    \label{fig:gen_line}
    \end{subfigure}
    \begin{subfigure}{0.45\textwidth}
    \includegraphics[width=1\columnwidth]{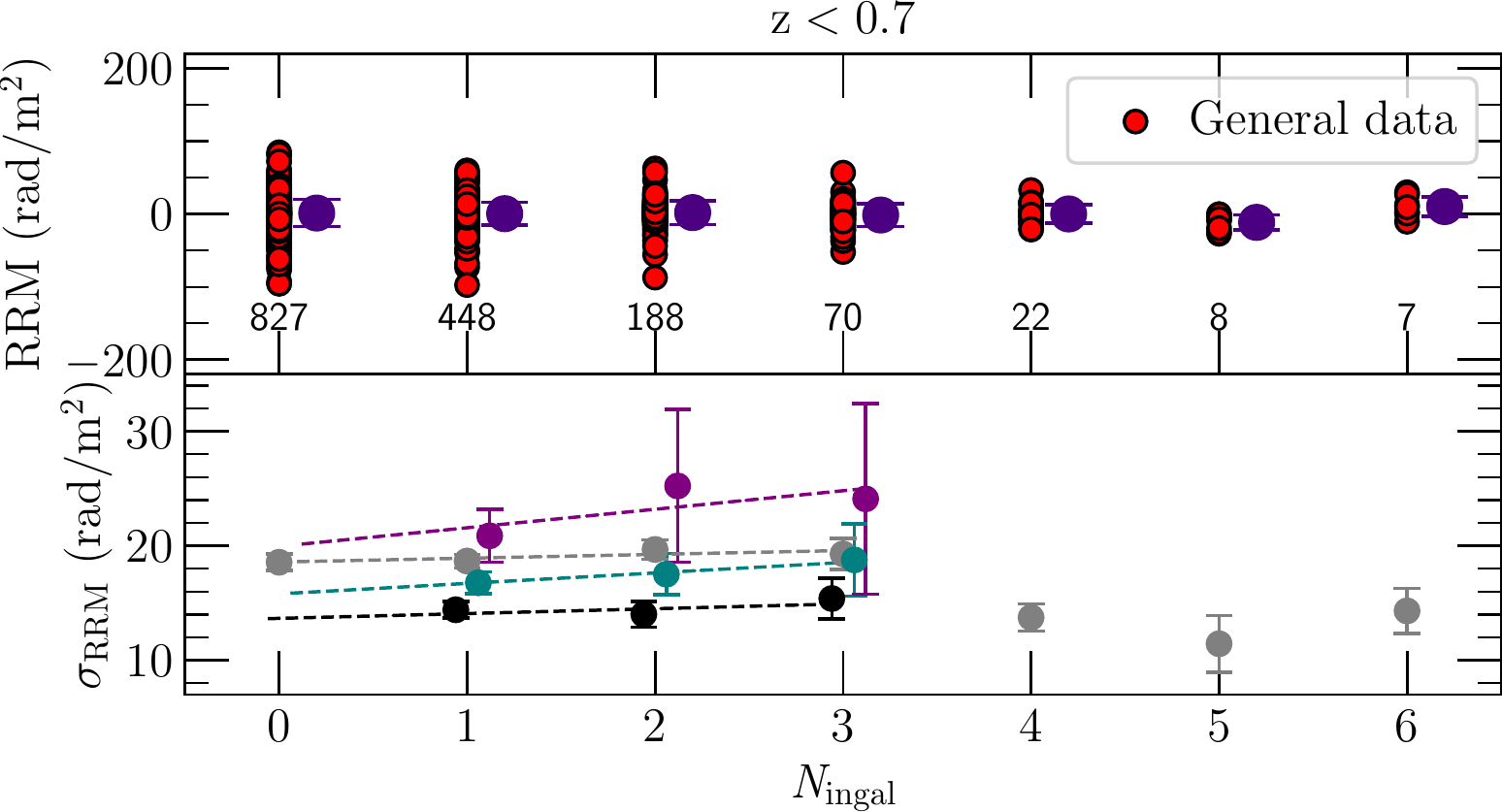} 
    \caption{$\sigmarrm$ vs. $\ningal$ \revb{for the general (with RM source $z<0.7$) case. \rev{The slope values for data, G, SH, LB cases are $0.31\pm1.48 , 0.57\pm1.98 , 1.05\pm3.3 ,~\text{and}~ 1.24\pm8.34 $, and their corresponding intercepts are $18.57\pm2.49,  13.56\pm3.26, 15.57\pm5.14,\text{ and } 20.9\pm13.02$. }}}
    \label{fig:spi_line}
    \end{subfigure}
    \begin{subfigure}{0.45\textwidth}
    \includegraphics[width=1\columnwidth]{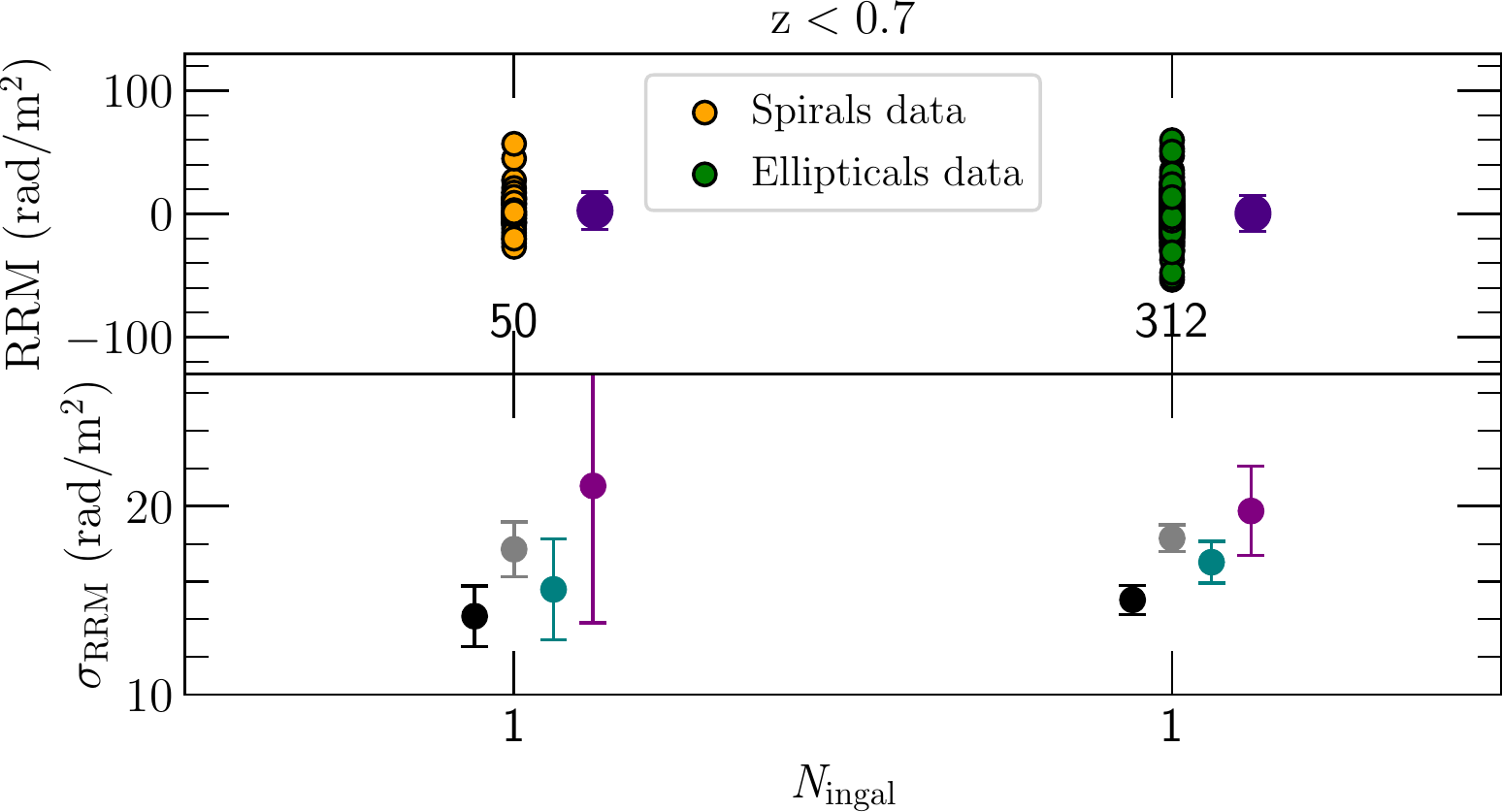} 
    \caption{$\sigmarrm$ vs. $\ningal$ \revb{for just intervening spirals and ellipticals (with RM source $z<0.7$, $\ningal=1$) case. }}
    \label{fig:ell_line}
    \end{subfigure}
    \caption{The \revb{dependence of} $\RRM$ (top panels in each \revb{subfigure}) and $\sigmarrm$ (bottom panels in each \revb{subfigure}) on $\ningal$ for three different intervening galaxy samples based on morphology: all galaxies\revb{,} (a) \revb{and (b) with all RM source $z$ and $z<0.7$, respectively, and} spirals and ellipticals\revb{,} (c). All the top panels show $\RRM$ scatter with the violet point showing \revb{$\mean_{\rm RRM}$ and its errorbar showing} $\sigmarrm$ for each case. The number of sources in each sample is also given in the plot. The bottom panels show $\sigmarrm$ as a function of $\ningal$ for data and three different fitting distributions (described in \Sec{sec:GnG}). \rev{The offsets in $\ningal$ for different cases of $\stddev$ are introduced to improve the legibility of the plot.} The error in $\sigmarrm$ is computed using the $\RM$ uncertainty from the NVSS \citep{FarnesEA2014}. For most cases, $\sigmarrm$ increases with an increase in the number of intervening galaxies. However, the range of $\ningal$ is quite small (upto $6$ for the general case).}
    \label{fig:line_fits}
 \end{figure}
 
In this section, we show the dependence of the standard deviation of the $\RRM$ distribution, $\sigmarrm$, on the number of intervening galaxies, $\ningal$, for three different morphological cases, i.e., general (containing all types of galaxies), spirals, and ellipticals. As described in \Sec{sec:GnG}, we compute four different $\sigmarrm$, one directly from the data ($\sigmadata$) and other three by fitting the following distributions: Gaussian distribution ($\sigmag$), Gaussian distribution with sinh-arcsinh transform ($\sigmash$), and modified line-broadening distribution ($\sigmalb$). \revb{To calculate one-sigma errors in the statistical parameters, $500$ random RRM samples are generated based on one-sigma errors of individual RRM points from the data. The fits are generated for these samples, and the best $250$ (least $\DKS$) statistical parameters extracted from the fits are considered valid, whose mean and standard deviation are reported as the value and one-sigma fluctuation in the value.}

We expect the standard deviation of the $\RRM$ distribution ($\sigmarrm$) to increase with the number of intervening galaxies. With the available data, we explore a linear dependence (see \Sec{sec:bg_sim} for a numerical demonstration) of the form
\begin{ceqn}
\begin{align}\label{eq:linear_fit}
    \sigmarrm (\ningal) = A \ningal + B,
\end{align}
\end{ceqn}
where $A$ is the slope and $B$ is the intercept. \Fig{fig:line_fits} shows the $\RRM$ and $\sigmarrm$ as a function of number of intervening galaxies for all three cases: general (all morphologies included), spirals, and ellipticals.  We also provide the slope and intercept of fitted lines \revb{for the general cases} assuming a linear dependence of $\sigmarrm$ (for both data and fits) on $\ningal$ in \rev{the captions of \Fig{fig:gen_line} \revc{(containing all RM source redshifts)} \revc{and} \Fig{fig:spi_line} \revc{(containing RM source redshifts $<0.7$)}. \revc{We do not use the $\ningal=0$ case for fitting the linear slope because many of those sources might not have any \revb{reported} intervening galaxies \revb{due to} the lack of coverage in SDSS-DR16 data (see \Fig{fig:money_plot}).}}

In \Fig{fig:gen_line} \revb{and \Fig{fig:spi_line}}, we do not overlap our data with results from fitting the distribution beyond $\ningal=3$, as the number of $\RRM$ values are quite low and the fitting procedure does not \revb{give statistically significant results}. The slope tends to be positive (correlation of $\sigmarrm$ with $\ningal$ is consistent with results using MgII absorbers in \citet{FarnesEA2014mg2,MalikEA2020}) for the data and all three distributions \revb{of the general case (\Fig{fig:gen_line}, \Fig{fig:spi_line}),} but it is difficult to conclude that robustly because of large errors in the estimated values. This is probably due to a smaller number of galaxies for each $\ningal$\revb{,} fewer number of $\ningal$ values ($3$ for the general case)\revb{, and large errors in the observed $\RRM$ values}.

The linear dependence of $\sigmarrm$ with $\ningal$ is different from what \revb{the study of} \citet{LanP2020} assumed, i.e., \revb{$\sigmarrm \propto \sqrt{\ningal}$}. The difference arises because they assume a random walk model of $\RRM$ even at a smaller number of intervening galaxies, which we think is probably applicable only at a very high number of intervening galaxies. They also find a higher number of intervening galaxies ($\sim 10$, they do not differentiate between the morphology of intervening galaxies) from their sample (see Fig. 2 in their paper). However, we find only up to six intervening galaxies for our general case. Moreover, they do not observe any correlation of $\sigmarrm$ with $\ningal$. \rev{\citet{AmaralEA2021} performs a similar analysis with no significant correlation detected as well.} According to our fitting functions ($\rm LB$ function in particular), there can be a positive correlation between the two \rev{(see the slope in the captions of \Fig{fig:gen_line}, \Fig{fig:spi_line}}\revb{)}, however, \rev{due to large error in slope values} it might also be uncorrelated. \revb{Thus,} we need further data to conclude this concretely. The differences in the results might also be because of the different galaxy surveys (we use SDDS DR 16, and they use DESI Legacy Imaging \revc{Surveys)\footnote{\revb{A possible cause of differences could be the redshift range, which is $0< z< 0.7$ for the SDSS dataset (spectroscopic), and $0.3<z<1.0$ for the Legacy survey (photometric).}}.} 

\rev{We also classify the SDSS catalog based on galaxy's morphology using \revb{the} Galaxy Zoo data as we expect significant differences between thermal electron density and magnetic fields in spiral and elliptical galaxies.} \revb{In \Fig{fig:ell_line}, we show the same analysis for just intervening spirals and ellipticals ($\ningal=1$), with no notable differences seen between the $\sigmarrm$ of the two. The number of ellipticals ($312$) is much higher than the number of spirals ($50$) because the hosts of the RM sources (mainly ellipticals) are also included as intervening galaxies.}

\rev{Even though magnetic fields cannot be accurately derived using this section's data due to large errors, and lack of data points, we consider a Gaussian and two non-Gaussian distributions for the $\RRM$ PDFs. From non-Gaussian PDFs, we observe that the outliers can play a role in determining the $\sigmarrm$. This can be seen in \Fig{fig:line_fits} as $\sigmalb>\sigmash>\sigmag$ \revb{for} nearly all the cases. Also, from \Tab{tab:fitparams} LB function tends to have higher kurtosis and lower \revb{$\DKS$} values, meaning it fits the outliers well. We believe that this is the reason for higher values of $\sigmalb$. However, more data with higher sensitivity is required to confirm these claims.}

\subsubsection{Dependence of $\sigmarrm$ on the impact parameter ($\rperp$)}
\label{sec:sigma_ip}

\begin{figure}
    \includegraphics[width=1\columnwidth]{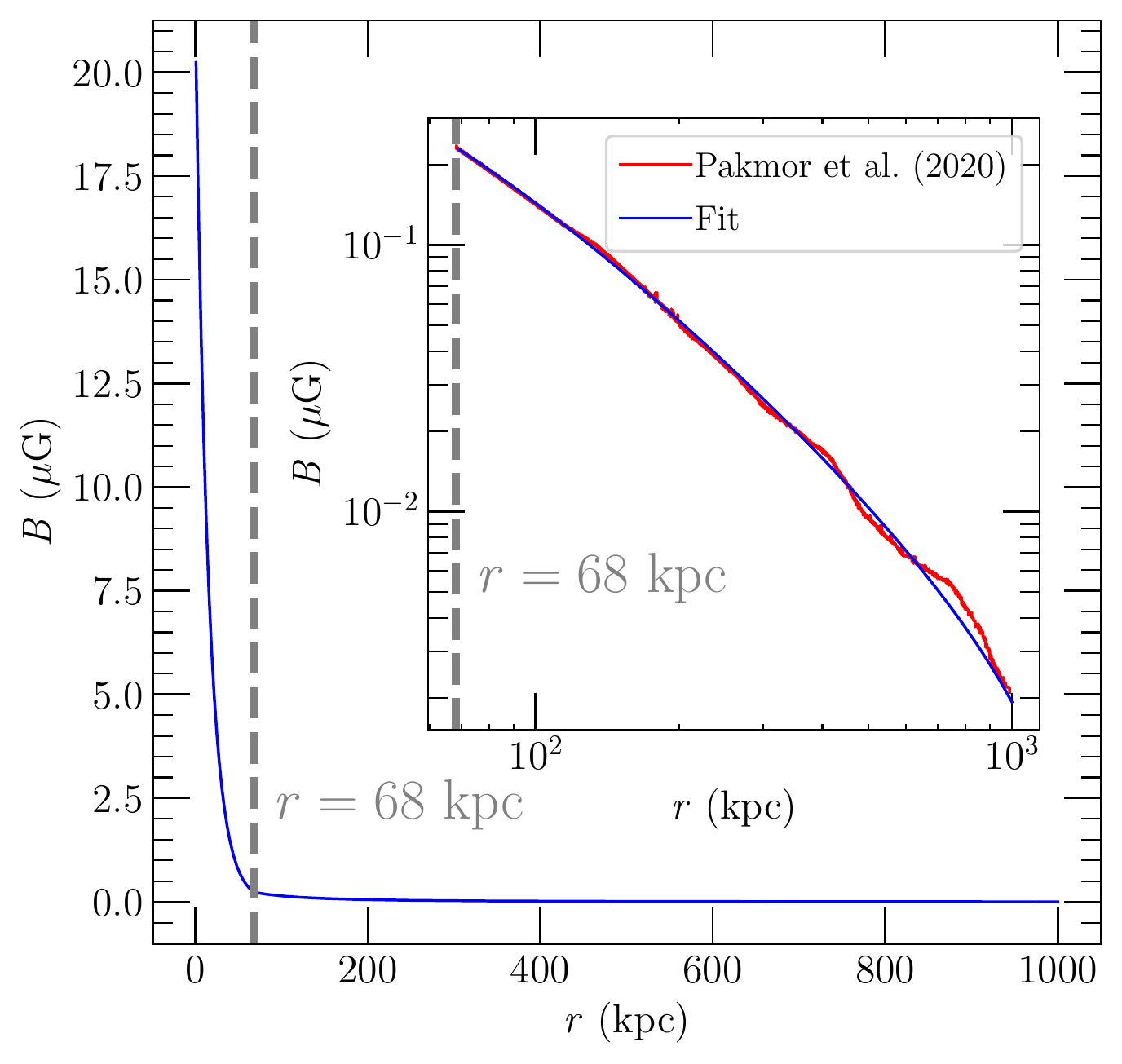} 
    \caption{The adopted large-scale field ($B$) profile for spirals. Below $r=68 \kpc$ an exponentially decaying field is taken. Above $r=68 \kpc$, profile from \citet{PakmorEA2020} (for circumgalactic medium) is obtained and fitted, as shown in the inset. The fitted function $0.592 \exp(-\log(3.03~(r (\kpc)/r_{\rm vir})+0.617)/0.54)-0.00163$ is used as the profile above $r=68 \kpc$.}
    \label{fig:pakmor}
\end{figure}

\begin{figure}
    \begin{tikzpicture}[scale=0.8]
    \draw [color=green!60, fill=gray!5, ultra thick](0,0) circle (5cm);
    \draw [very thick](2.5,-5) -- (2.5,5);
    \filldraw [gray] (0,0) circle (2pt);
    \draw [thick](2.5,2.5) -- (0,0);
    \draw [thick](2.5,3.2) -- (0,0);
    \draw [thin](2.5,0) -- (0,0);
    \draw (2,2) arc (45:54:2cm);
    \draw (2.1,2.34) node {d$\rm \theta$};
    \draw (0.5,0) arc (0:45:0.5cm);
    \draw (0.65,0.24) node [scale=1.5]{$\rm \theta$};
    \draw [thick](2.5,-4.33) -- (0,0);
    \draw (0.5,0) arc (0:-60:0.5cm);
    \draw (0.68,-0.4) node [scale=1.5]{$\rm \alpha_0$};
    \draw (1.5,0.2) node [scale=1.5]{$\rperp$};
    \draw (1.4,1.1) node [scale=1.5]{$r$};
    \draw [thin](2.7,2.5) -- (2.7,3.2);
    \draw[thin] (2.6,3.2) -- (2.8,3.2);
    \draw[thin] (2.6,2.5) -- (2.8,2.5);
    \draw (3.4,2.85) node [scale=1.2]{$\rm \mathit{r}\frac{d\theta}{cos\theta}$};
    \draw[thin] (0,0) -- (-5,0);
    \draw (-2.5,0.3) node [scale=1.5]{$\rhalo$};
    \begin{scope}[very thick,decoration={
        markings,
        mark=at position 0.5 with {\arrow{>}}}
        ] 
        \draw[postaction={decorate}] (2.5,4.5) -- (2.5,4.4);
        \draw[postaction={decorate}] (2.5,-4.6) -- (2.5,-4.7);
    \end{scope}
    \draw (0.8,4.6) node [scale=1.2]{$\rm From~\RM~source$};
    \draw (1.3,-4.6) node [scale=1.2]{$\rm To~observer$};
    
    \end{tikzpicture}
    \caption{The green circle represents the bounds of an intervening galaxy. In the figure, $\rhalo$ and $\rperp$ represent the radius of the galaxy's halo (10 $\rgal$) and the impact parameter, respectively. The radial distance $r$ changes as the radiation traverses through the medium, and $2\alpha_0$ is the angle subtended by the path length (vertical line) at the center. The differential element along the path length is $1/\cos\theta$ times the differential arc at $r$, i.e., $r \dd \theta$.}  
    \label{fig:geometry}
    \end{figure}
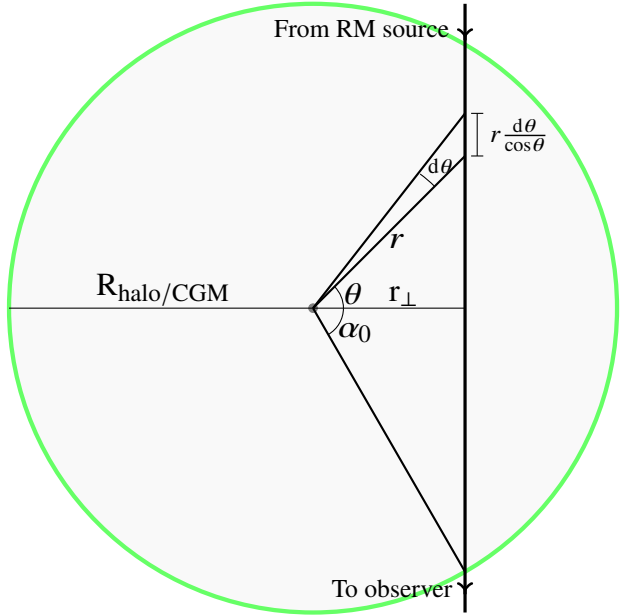
    
\begin{figure*}
    \includegraphics[width=2\columnwidth]{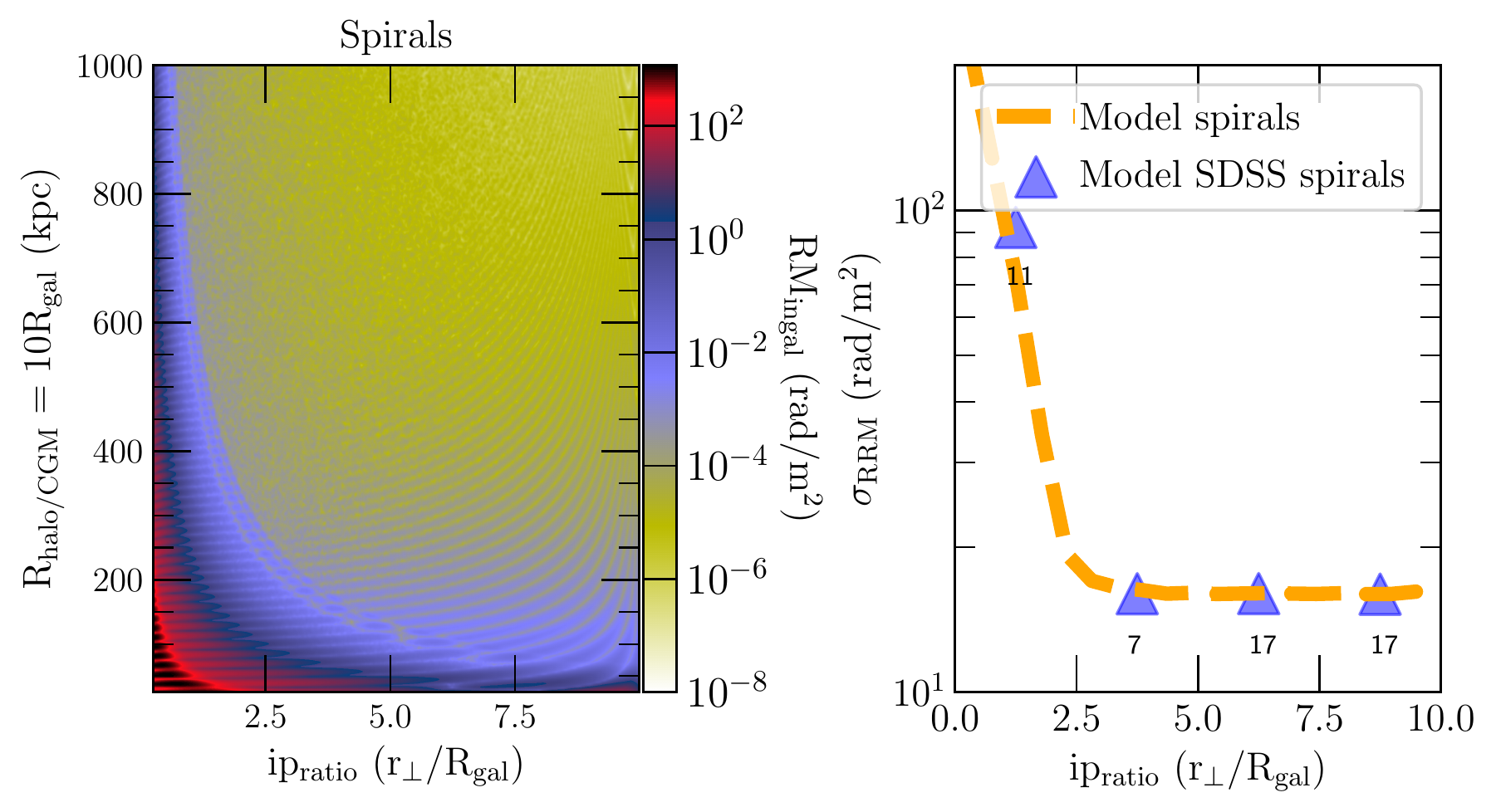} 
    \caption{Left: The histogram represents predictions of $|\RMIngal|$ with the model with varying $\rhalo$ and $\ipratio$, thermal densities, and large-scale fields. $\rhalo$ varies from $25-1000 \kpc$. Even at high $\ipratio$ values, galaxies with a small $\rhalo$ can contribute significantly towards $\RMIngal$, and contribution below $10^{-8}~\rad/\m^2$ is considered as zero. The highest $\RMIngal$ is \revb{$\approx 1000 \rad/\m^2$}, occurring near the origin of the figure. Right: The variation of $\sigmarrm$ with $\ipratio$, assuming a Gaussian background source distribution of $\RM$ with a standard deviation of $16~\rad/\m^{2}$ for all galaxies with $\rhalo$ from $25 \kpc$ to $1000 \kpc$\revb{,} is shown with the orange curve. Then, $\ipratio$ varying from $0-10$ \revb{is} divided into \revb{four} equally sized bins. The dependence of $\sigmarrm$ on $\ipratio$ for SDSS spirals is shown \revb{with} blue triangles, which varies \revb{similarly as the model} spirals in each bin. \revb{However, after removing one galaxy with $\RMIngal=300 \rad/\m^2$, the $\sigmarrm$ of model SDSS spirals varies only slightly.} This explains the uncertainty in the positive slope correlation in \Fig{fig:spi_line}.}
    \label{fig:ip_spi}
\end{figure*}
    
Here, we explore the dependence of $\sigmarrm$ on the impact parameter, $\rperp$ (distance from the center of the galaxy to a region where the line of sight from the sources crosses the galaxy). We assume $\rhalo$ as the extent to which a galaxy can contribute, which is fixed to a maximum of $1~\Mpc$ for both ellipticals and spirals. $\rhalo$ is taken as \revb{ten} times the Petrosian radius ($\rgal$), which varies with the galaxy. In our sample, the minimum $\rhalo$ for spirals and ellipticals is approximately \revb{$46~\kpc$ and $31~\kpc$}, respectively. The mean and standard deviations of $\rhalo$ are \revb{$580~\kpc$ and $360~\kpc$} for spirals, and \revb{$447~\kpc$ and $334~\kpc$ for ellipticals}. \revb{The study of} \citet{LanP2020} \revb{contains} a similar analysis but for a fixed $\rhalo$ ($200~\kpc$) for all galaxies (see their Fig. 4), whereas we vary the upper limit of the impact parameter depending on the size of the individual galaxy. Thus, we normalize $\rperp$ by the Petrosian radius, $\rgal$ \revb{($\ipratio$)}.

We also use simple analytical radial ($r$) profiles of the thermal electron density and magnetic fields in spirals to estimate \revb{the standard deviation of the} $\RRM$ \revb{distribution} as a function of the impact parameter. For spiral galaxies, we adopt the following thermal electron density profile for the disk \citep{TaylorEA1993} and \revb{the} halo \revb{or} CGM ($r$ in $\kpc$)
\begin{ceqn}
\begin{align}
    \ne(r) = &~0.0223 \cm^{-3} \left(\sech\left(\frac{r}{20 \kpc}\right)\right)^2  \nonumber \\ 
    + & ~ 0.1 \exp \left(\left(-\frac{r-3.5 \kpc}{1.8 \kpc}\right)^2\right), ~r \le 68 \kpc,  \label{eq:taylor_text} \\
    \ne(r)  =  & ~ 0.01 \cm^{-3} \left( 1+\left(\frac{r}{3 \kpc}\right)^2 \right)^{-3/4} ,~r > 68\kpc, \label{eq:taylor_king_text}
\end{align}
\end{ceqn}
respectively. According to \revb{the profile from \citet{TaylorEA1993}}, the disk $\ne$ component falls to $10^{-4} \cm^{-3}$ at $r=68 \kpc$ and thus we assume a king profile for the halo/CGM after that limit (since the features of spirals become more homogeneous). The constant $0.01 \cm^{-3}$ in \Eq{eq:taylor_king_text} is obtained by demanding continuity in the overall $\ne$ profile around $r=68 \kpc$. For large-scale magnetic field strengths in spiral galaxies, we use an exponentially decaying profile \citep{BasuEA2018}, i.e. $B_0 = 21.6 \muG \exp(-r/15 \kpc)$ up to $r=68 \kpc$ with \revb{field} reversals at every $15 \kpc$. Beyond $68 \kpc$, we adopt a fit to \revb{the} profile of CGM from \citet{PakmorEA2020} in simulated Milky Way type spiral galaxies (see the top panel of their Fig. 13) \footnote{We use the WebPlotDigitizer, available at \url{https://automeris.io/WebPlotDigitizer/}, to extract data from \citet{PakmorEA2020}.}. The constant $21.6 \muG$ is obtained by demanding continuity in the overall $B$ profile, and is consistent with the observationally derived magnetic field strengths in Fig. 3 of \citet{BasuRoy2013}. \Fig{fig:pakmor} shows the two profiles adopted for the core and halo/CGM of spirals separated by the grey line. The fitting function (see inset of \Fig{fig:pakmor}) obtained is $0.592 \exp(-\log(3.03~(r (\kpc)/r_{\rm vir})+0.617)/0.54)-0.00163$, where $r_{\rm vir}=200 \kpc$ is the virial radius of the galaxy \citep{PakmorEA2020}. This analysis is not performed for ellipticals as their large-scale field profiles are not very well-known, and we obtain extremely low magnitudes for large-scale field strengths from the Laing-Garrington analysis (see \Fig{fig:mean_ratio}).

Using these models in \Eq{eq:rm_integral} and the geometry in \Fig{fig:geometry}, we compute the $\RM$ contribution of the intervening galaxy as
\begin{ceqn}
\begin{align}\label{eq:rm_integral_angle}
  \frac{\RMIngal}{\rad/\m^{2}} = \frac{811.9}{(1+z)^2}\int_{-\alpha_0}^{\alpha_0}\frac{n_e(\rperp/\cos\theta)}{\cm^{-3}} \frac{B(\rperp/\cos\theta)/\rev{\sqrt{3}}}{\muG} \frac{2\rperp/\pi}{\kpc}\frac{\dd \theta}{\cos^2\theta},
\end{align}
\end{ceqn}

In this equation, we divide the large-scale field magnitudes by \revb{$\sqrt{3}$} to determine the component parallel to the line of sight, assuming an isotropic field \revb{in all three directions}. The impact parameter is multiplied with a factor of $2/\pi$ to account for the galaxies' random alignment with the line of sight \citep{BasuEA2018}. The integration limits $\rev{-}\alpha_0$ to $\alpha_0$ correspond to the entire path length as can be seen in \Fig{fig:geometry}. $z$ is taken as 1 to generate the model in \Fig{fig:ip_spi}. The maximum computed value for the spirals goes up to \revb{$\approx 1000~\rad/\m^2$} with a very low probability \revb{of occurrence ($\sim 0.001$)}, which is consistent with \revb{the predictions of} \citet{BasuEA2018} (see the left panel of their Fig. 7). The left panel of \Fig{fig:ip_spi} shows the variation of \revc{this} modeled $\RMIngal$ with $\ipratio$ and $\rhalo$.

We assume a Gaussian distribution of rotation measures from the \revb{background} sources with a standard deviation of $16~\rad/\m^2$ (selected as this is the standard deviation of $\RRM$ at very large impact parameters of the order of $500 \kpc$) and randomly impart our modeled $\RMIngal$ (using \Eq{eq:rm_integral_angle}) along varying $\ipratio$ by considering the presence of all the $\rhalo$ from our \revb{model} sample in the range 25 $\kpc$ to $1000~\kpc$ (according to the left panel of \Fig{fig:ip_spi}). \revc{These are called model spirals}. Then we select the spiral galaxies \revb{from} the SDSS dataset and predict their $\RMIngal$ from the model \revc{(using \Eq{eq:rm_integral_angle})} with the help of their extracted sizes and impact parameters from the observational dataset. \revc{These are called model SDSS spirals}. The right panel of \Fig{fig:ip_spi} shows the predictions \revb{for} both \revc{of these}. \revb{The match between the two is significant.} \revb{However, one galaxy with $\RMIngal=300 \rad/\m^2$ causes the rise in $\sigmarrm$ in the first bin of the modeled SDSS spirals case. After removing that galaxy,} the maximum $\sigmarrm$ of the modeled SDSS spirals is \revb{$17.2 \rad/\m^2$,} which is too sensitive to be observed (errors in \Fig{fig:line_fits} are greater than \revb{$\sim 2.5 \rad/\m^2$}). 

\revc{Additionally,} the scarcity of SDSS \revb{spirals} in the former bins explains the low values of predicted $\sigmarrm$ \revb{(without considering the galaxy with $\RMIngal=300 \rad/\m^2$)}\revc{\footnote{Former bins with lower $\ipratio$ can contribute the highest to $\sigmarrm$.}.} As most of the galaxies are situated in the later bins (where $\RMIngal$ is low), the overall contribution of SDSS spirals \revb{in $\sigmarrm$} is not observable. This explains the \revb{large uncertainties} in the positive slope correlation (see \Fig{fig:spi_line}) observed in \Sec{sec:sigma_ningal}. The ambiguity can be resolved with deeper observations with a sufficient number of galaxies present in each $\ipratio$ bin.


\subsubsection{Influence of intervening galaxies on the polarization fraction}
\label{sec:results_bg_pol}

\begin{figure}
    \includegraphics[width=1\columnwidth]{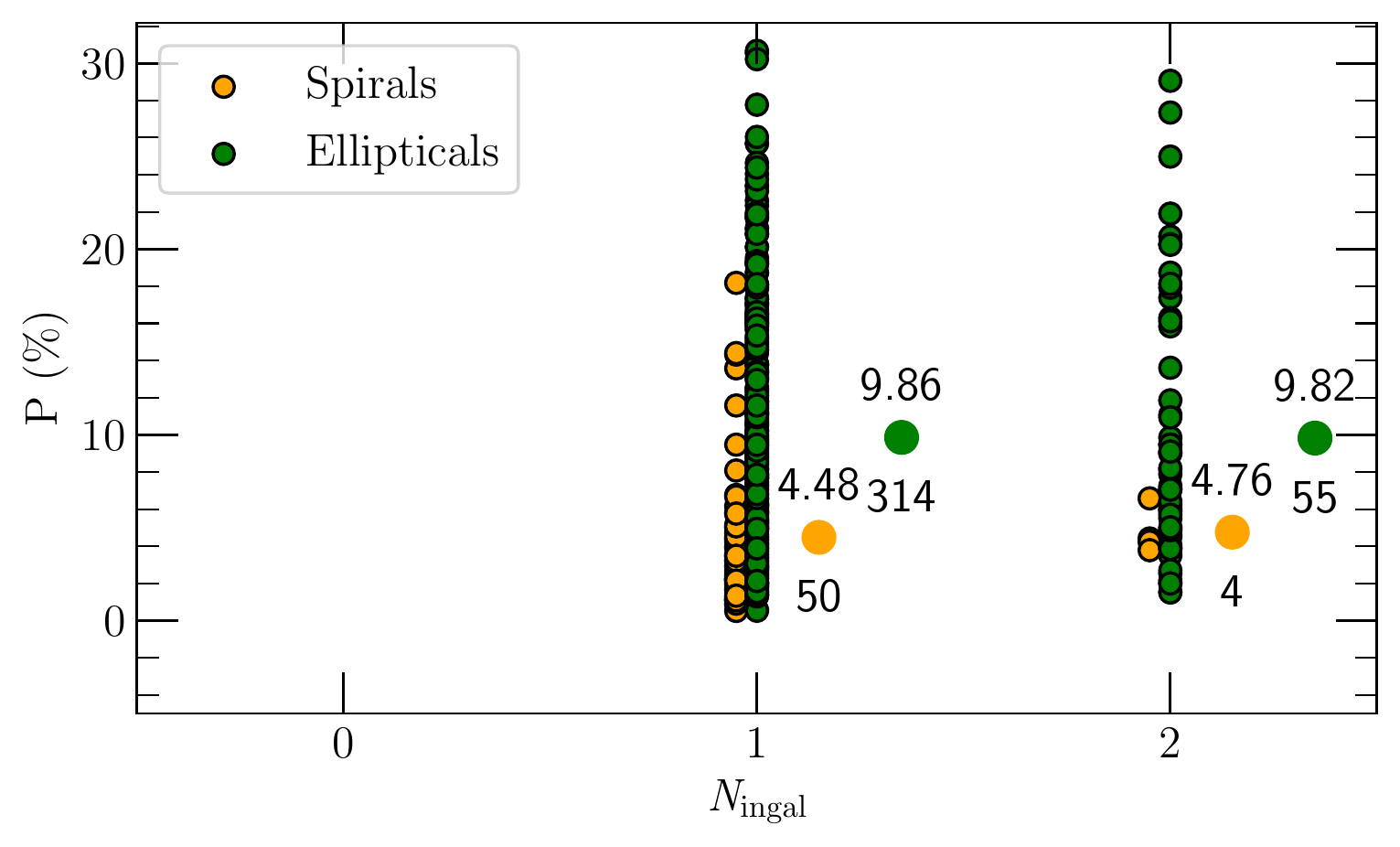} 
    \caption{Polarization fraction, ${\rm P}~(\%)$, as a function of the number of intervening galaxies, $\ningal$, for spirals (orange) and ellipticals (green). For each case, the mean is plotted with dots and their magnitude above it. The numbers below that point show the sample size in each case. Ellipticals tend to have a high polarization fraction than spirals. This implies, statistically, spirals have stronger magnetic fields and higher thermal electron densities.}
    \label{fig:pol_sdss}
\end{figure}

In addition to rotation measure values, we also use the available observed polarization fractions in the \citet{FarnesEA2014} dataset. We use the morphologically classified galaxy catalog to explore the influence of intervening spirals and ellipticals on the polarization fractions of the background sources. \Fig{fig:pol_sdss} shows a scatter plot of the distribution of polarization fraction values of radiation traversing through spirals and ellipticals with different $\ningal$\revc{, and the} mean of the distributions \revb{are} shown \revc{above the points}. The mean polarization fraction values for spirals are lower than that of the ellipticals, \revb{implying} stronger magnetic fields and higher thermal electron density in spirals (making them strong depolarizers). \revc{We exploit} the difference in the values of polarization fractions between spirals and ellipticals \revc{to} estimate the ratio of magnetic field strengths between the two samples.

The thermal electron density profile for spirals is modeled by \Eq{eq:taylor_text} and \Eq{eq:taylor_king_text} \citep{TaylorEA1993}, and for elliptical galaxies, we use the \revb{King} profile
\begin{ceqn}
\begin{align}\label{eq:king}
    \ne (r) = 0.1 \cm^{-3}\left( 1+\left(\frac{r}{3 \kpc}\right)^2 \right)^{-3/4},
\end{align}
\end{ceqn}
where the constant $0.1 \cm^{-3}$ is consistent with \citet{MathewsEA2003}. For spirals, we adopt the flux-freezing condition with $\brms \propto \ne^{2/3}$ and limit the maximum $\brms$ to $10 \muG$ to calculate the proportionality constant. Then we use these models to estimate the $\sigmarm$ using single-scale model of magnetic field structures \citep{BurnEA1966, FeltenEA1996, LaingEA2008}. The $\sigmarm^2$ (according to the geometry in \Fig{fig:geometry}) is 

\begin{ceqn}
\begin{align}\label{eq:sigma_rm_angle_final_text}
    \sigmarm^2=\frac{811.9^2 \lb C^2}{3(1+z)^2} \int_{-\alpha_0}^{\alpha_0}\ne^{2(1+2/3)}(\rperp/{\rm cos\theta})~\rperp\frac{\dd \theta}{{\rm cos^2\theta}},
\end{align}
\end{ceqn}
where $\rperp$ and $\lb$ are in $\kpc$, $\lb$ is the magnetic field correlation length, and $C$ is the proportionality constant in the equation $\brms \propto \ne^{2/3}$ profiles ($\cspi$ for spirals and $\cell$ for ellipticals). All the other units are the standard Gaussian units as used throughout the paper. We represent \Eq{eq:sigma_rm_angle_final_text} as $\sigmarm^2 = TC^2$ to separate the proportionality constant and \revc{other} galaxy dependent terms ($\tspi$ for spirals and $\tell$ for ellipticals). \revb{Since $\sigmarm^2$ is an additive quantity \citep{FeltenEA1996, Laing2008}, we half the value predicted by \Eq{eq:sigma_rm_angle_final_text} for the identified \revc{hosts\footnote{The intervening galaxies in our study also contain the host galaxies. To identify them, we use the condition $\Delta z=|z_{\rm RM}-z_{\rm gal}|<0.001$. If multiple hosts are identified for one RM source, the galaxy with the lowest $\Delta z$ is identified as the host.}} as, on average, the polarized emission emerges halfway in the plane of observation in the sample of host galaxies.}

Then we take an average of logarithms of polarization values (see \Eq{eq:pol}) passing through spirals and ellipticals and subtract the two to remove any contribution from the IGM and source (which are statistically similar for both samples). The only unknown in the subtracted equations is the proportionality constant ($\cell$) of $\brms$ of the ellipticals. The other parameters are known from the NVSS-SDSS cross-matching, correlation lengths ($\lb=30~\pc$ for spirals, $\lb=100~\pc$ for ellipticals), and the assumed $\brms$ profile of the spirals. From this analysis, $\cell$ is

\begin{ceqn}
\begin{align}\label{eq:pol_sdss_analysis_cell_4}
    \cell^2 = \frac{\langle \cspi^2  \sum_{k=1}^{\ningal} {{\tspi}_i}_k \rangle+\frac{\langle \log({P_{\rm spi}}_i) \rangle-\langle \log({P_{\rm ell}}_j) \rangle}{2\lambda^4}}{\langle  \sum_{k=1}^{\ningal} {{\tell}_j}_k \rangle},
\end{align}
\end{ceqn}

where $i$ and $j$ denote the background sources with radiation passing \revb{through} spirals and ellipticals, respectively. $k$ sums over $1$ or $2$ intervening galaxies (see \Fig{fig:pol_sdss}) for a particular source $i$ or $j$. This equation automatically accounts for galaxies with high $\ipratio$ which contribute less as their calculated $\sigmarm^2$ contribution will be low (according to \Eq{eq:sigma_rm_angle_final_text}).

\rev{We extract $\brms$ profile for ellipticals from the calculated \revb{$\cell$ ($\sim 29.51$)} as ${\brms}_{\rm ell} = \cell \ne^{2/3}$, which turns out to be ${\brms}_{\rm ell}(r) = 6.36\muG~(1+(r/3)^2)^{-1/2}$ after assuming a King profile for their $\ne$ (see \Eq{eq:king}). The extracted $\brms(0)$ for ellipticals is approximately $~6.36~\muG$ which is in agreement with the results of \revb{previous studies like \citet{MathewsEA2003} and \citet{SetaEA2020II}.} }

\section{Conclusions}
\label{sec:conclusions}
We aim to estimate magnetic fields in elliptical galaxies from observations, primarily in their haloes and circumgalactic medium (CGM), using the Laing-Garrington effect and polarized emission from background radio sources.

For the Laing-Garrington effect (\Sec{sec:laing_garrington}), we find that the estimated small-scale magnetic field strengths in the sample of galaxies with redshifts ranging from \revb{$0.01$ to $0.46$} lie in the range \revb{$0.06~\text{--}~2.75 \muG$}. The magnetic field strengths in ellipticals are an order of magnitude smaller than that in the Milky Way and nearby spiral galaxies. We find that the large-scale field, albeit over larger path lengths of the order of $100 \kpc$, is an order of magnitude smaller than the random magnetic fields (\Fig{fig:mean_ratio}). A larger sample of Laing-Garrington sources with a higher signal-to-noise ratio of the polarized intensity is required to study these magnetic fields in much greater detail.

With observations of polarized background radio sources, we aim to explore the \revb{effects} of morphology and number of intervening galaxies on the residual rotation measure, $\RRM$ (the observed $\RM$ after subtracting the modeled Milky Way contribution). Most previous studies assume that the $\RM$ distribution of background sources is Gaussian, and so we first explore the possibility of non-Gaussian distributions. We show that non-Gaussian $\RM$ distributions might be better (\Fig{fig:three_cases}) as \revb{they} can account for large rotation measure values. We propose that the standard deviation of the $\RRM$ distribution has a linear dependence on the number of intervening galaxies, $\ningal$, for a lower number of intervening galaxies but goes as $\sigmarrm \propto \sqrt{\ningal}$ for a very high $\ningal$ \rev{(according to results from \Sec{sec:bg_sim})}. Then we find that $\sigmarrm$ values (computed directly from the data, fitting a Gaussian distribution, and two different non-Gaussian distributions) tend to increase with $\ningal$ \rev{(aligned with the results from MgII absorber studies in \citet{Bernet08, FarnesEA2014mg2, MalikEA2020})}. But since the errors in the estimated slopes (\Fig{fig:line_fits}) are high, $\sigmarrm$ can as well be independent of $\ningal$ \rev{(as seen in \citet{LanP2020, AmaralEA2021})}. \revb{We then use simple radial profiles of magnetic fields and thermal electron density for spirals in the data to show that the contribution of the rotation measure of the intervening galaxy to $\sigmarrm$ is negligible. This is probably because the impact parameter is very large for most of the spirals in our sample.} However, we do use the difference in the observed polarization fractions between intervening spirals and ellipticals to obtain a field \rev{strength of \revb{$\sim 6.3 \muG$} at the center of ellipticals} \revb{(which is consistent with the previous studies \citep{MathewsEA2003, SetaEA2020II})}.  A much larger sample with lesser uncertainty in the rotation measure observations is required to study the probability distribution function of $\RRM$ and compute higher-order statistical properties of the distribution. This would robustly confirm the presence of non-Gaussianity, which in turn would also help us study magnetic fields in galaxies in more detail.

\revb{Both the methods have their pros and cons. They both probe the small-scale magnetic fields via their depolarization signatures; however, the Laing-Garrington method cannot individually (for every source) distinguish whether the radiation is depolarized due to the galactic medium or the ICM/IGM surrounding the galaxies. On the contrary, radiation from the background polarized sources passing through intervening galaxy is individually constrained to be inside the galactic halo/CGM scales. Although, the Laing-Garrington method works without assuming any small-scale magnetic field profile. It is, however, necessary to assume the flux freezing condition ($\brms \propto \ne^{\gamma}$) to derive results from the observations of polarized sources. }

Both methods utilize very different datasets and analysis techniques but give similar answers for the strength of magnetic fields in ellipticals, which is of the order of \revb{$1-10 \muG$}. The results can be used to further constrain fluctuation dynamo theories \citep{Rincon19} and also parameters of the magnetohydrodynamic simulations of galaxy formation and evolution \citep[such as][]{PakmorEA2020,NelsonEA2020}.

\section*{Acknowledgements}
\rev{We thank \revb{both anonymous reviewers} for their helpful feedback. We also \revb{thank} Sui Ann Mao and Aritra Basu for \revb{valuable} discussions.} We acknowledge the use of high-performance computing resources provided by the Research School of Astronomy and Astrophysics at the Mount Stromlo Observatory.

\section*{Data availability}
Radio galaxies' data used for the Laing-Garrington analysis (\Sec{sec:catalog1}) has been derived from \citet{VernstromEA2019}. The data is available at \url{https://iopscience.iop.org/0004-637X/878/2/92/suppdata/apjab1f83t1_mrt.txt}. For the background polarized emission analysis, there are three datasets: $\RM$ catalog (\Sec{sec:rm_catalog}), galaxy catalog (\Sec{sec:galaxy_catalog}), and morphology classification of galaxies' catalog (Galaxy Zoo). The $\RM$ catalog is derived from \citet{FarnesEA2014} (NVSS data), galaxy catalog from \citet{AhumadaEA2020} (SDSS DR16 data), and galaxy zoo data from \citet{LintottEA2008}. \rev{To remove the Milky Way $\RM$ contribution ($\RMMW$), we adopted Oppermann's maps \citep[][data available at \url{https://wwwmpa.mpa-garching.mpg.de/ift/faraday/2014/index.html}]{OppermannEA2012} and \citetalias{Faraday2020} (data available at \url{https://wwwmpa.mpa-garching.mpg.de/~ensslin/research/data/faraday2020.html}).} The SDSS catalogs along with the Galaxy Zoo data were found and queried on the Catalog Archive Server \citep{ThakarEA2008} at \url{https://skyserver.sdss.org/casjobs/}. The analyzed data is available upon a reasonable request to the author, Amit Seta (\href{mailto:amit.seta@anu.adu.au}{amit.seta@anu.adu.au}).



\bibliographystyle{mnras}
\bibliography{bellip} 

\begin{thebibliography}{}
\makeatletter
\relax
\def\mn@urlcharsother{\let\do\@makeother \do\$\do\&\do\#\do\^\do\_\do\%\do\~}
\def\mn@doi{\begingroup\mn@urlcharsother \@ifnextchar [ {\mn@doi@}
  {\mn@doi@[]}}
\def\mn@doi@[#1]#2{\def\@tempa{#1}\ifx\@tempa\@empty \href
  {http://dx.doi.org/#2} {doi:#2}\else \href {http://dx.doi.org/#2} {#1}\fi
  \endgroup}
\def\mn@eprint#1#2{\mn@eprint@#1:#2::\@nil}
\def\mn@eprint@arXiv#1{\href {http://arxiv.org/abs/#1} {{\tt arXiv:#1}}}
\def\mn@eprint@dblp#1{\href {http://dblp.uni-trier.de/rec/bibtex/#1.xml}
  {dblp:#1}}
\def\mn@eprint@#1:#2:#3:#4\@nil{\def\@tempa {#1}\def\@tempb {#2}\def\@tempc
  {#3}\ifx \@tempc \@empty \let \@tempc \@tempb \let \@tempb \@tempa \fi \ifx
  \@tempb \@empty \def\@tempb {arXiv}\fi \@ifundefined
  {mn@eprint@\@tempb}{\@tempb:\@tempc}{\expandafter \expandafter \csname
  mn@eprint@\@tempb\endcsname \expandafter{\@tempc}}}

\bibitem[\protect\citeauthoryear{{Ahn} et~al.,}{{Ahn} et~al.}{2012}]{AhnEA2012}
{Ahn} C.~P.,  et~al., 2012, \mn@doi [\apjs] {10.1088/0067-0049/203/2/21}, \href
  {https://ui.adsabs.harvard.edu/abs/2012ApJS..203...21A} {203, 21}

\bibitem[\protect\citeauthoryear{{Ahumada} et~al.,}{{Ahumada}
  et~al.}{2020}]{AhumadaEA2020}
{Ahumada} R.,  et~al., 2020, \mn@doi [\apjs] {10.3847/1538-4365/ab929e}, \href
  {https://ui.adsabs.harvard.edu/abs/2020ApJS..249....3A} {249, 3}

\bibitem[\protect\citeauthoryear{{Amaral}, {Vernstrom}  \& {Gaensler}}{{Amaral}
  et~al.}{2021}]{AmaralEA2021}
{Amaral} A.~D.,  {Vernstrom} T.,   {Gaensler} B.~M.,  2021, \mn@doi [\mnras]
  {10.1093/mnras/stab564}, \href
  {https://ui.adsabs.harvard.edu/abs/2021MNRAS.tmp..583A} {}

\bibitem[\protect\citeauthoryear{Banfield et~al.,}{Banfield
  et~al.}{2015}]{BanfieldEA2015}
Banfield J.~K.,  et~al., 2015, \mn@doi [Monthly Notices of the Royal
  Astronomical Society] {10.1093/mnras/stv1688}, 453, 2326

\bibitem[\protect\citeauthoryear{{Basu} \& {Roy}}{{Basu} \&
  {Roy}}{2013}]{BasuRoy2013}
{Basu} A.,  {Roy} S.,  2013, \mn@doi [\mnras] {10.1093/mnras/stt845}, \href
  {https://ui.adsabs.harvard.edu/abs/2013MNRAS.433.1675B} {433, 1675}

\bibitem[\protect\citeauthoryear{{Basu}, {Mao}, {Fletcher}, {Kanekar},
  {Shukurov}, {Schnitzeler}, {Vacca}  \& {Junklewitz}}{{Basu}
  et~al.}{2018}]{BasuEA2018}
{Basu} A.,  {Mao} S.~A.,  {Fletcher} A.,  {Kanekar} N.,  {Shukurov} A.,
  {Schnitzeler} D.,  {Vacca} V.,   {Junklewitz} H.,  2018, \mn@doi [\mnras]
  {10.1093/mnras/sty766}, \href
  {https://ui.adsabs.harvard.edu/abs/2018MNRAS.477.2528B} {477, 2528}

\bibitem[\protect\citeauthoryear{{Beck}}{{Beck}}{2007}]{BeckEA2007}
{Beck} R.,  2007, \mn@doi [\aap] {10.1051/0004-6361:20066988}, \href
  {https://ui.adsabs.harvard.edu/abs/2007A&A...470..539B} {470, 539}

\bibitem[\protect\citeauthoryear{{Beck}}{{Beck}}{2015}]{BeckEA2015}
{Beck} R.,  2015, \mn@doi [\aapr] {10.1007/s00159-015-0084-4}, \href
  {https://ui.adsabs.harvard.edu/abs/2015A&ARv..24....4B} {24, 4}

\bibitem[\protect\citeauthoryear{{Beck}, {Brandenburg}, {Moss}, {Shukurov}  \&
  {Sokoloff}}{{Beck} et~al.}{1996}]{Beck1996}
{Beck} R.,  {Brandenburg} A.,  {Moss} D.,  {Shukurov} A.,   {Sokoloff} D.,
  1996, \mn@doi [\araa] {10.1146/annurev.astro.34.1.155}, \href
  {http://adsabs.harvard.edu/abs/1996ARA%26A..34..155B} {34, 155}

\bibitem[\protect\citeauthoryear{{Becker}, {White}  \& {Helfand}}{{Becker}
  et~al.}{1995}]{BeckerEA1995}
{Becker} R.~H.,  {White} R.~L.,   {Helfand} D.~J.,  1995, \mn@doi [\apj]
  {10.1086/176166}, \href
  {https://ui.adsabs.harvard.edu/abs/1995ApJ...450..559B} {450, 559}

\bibitem[\protect\citeauthoryear{{Berkhuijsen} \& {M{\"u}ller}}{{Berkhuijsen}
  \& {M{\"u}ller}}{2008}]{BerkhuijsenM2008}
{Berkhuijsen} E.~M.,  {M{\"u}ller} P.,  2008, \mn@doi [\aap]
  {10.1051/0004-6361:200809675}, \href
  {https://ui.adsabs.harvard.edu/abs/2008A&A...490..179B} {490, 179}

\bibitem[\protect\citeauthoryear{{Bernet}, {Miniati}, {Lilly}, {Kronberg}  \&
  {Dessauges-Zavadsky}}{{Bernet} et~al.}{2008}]{Bernet08}
{Bernet} M.~L.,  {Miniati} F.,  {Lilly} S.~J.,  {Kronberg} P.~P.,
  {Dessauges-Zavadsky} M.,  2008, \mn@doi [\nat] {10.1038/nature07105}, \href
  {https://ui.adsabs.harvard.edu/#abs/2008Natur.454..302B} {454, 302}

\bibitem[\protect\citeauthoryear{{Bhat} \& {Subramanian}}{{Bhat} \&
  {Subramanian}}{2013}]{BhatEA2013}
{Bhat} P.,  {Subramanian} K.,  2013, \mn@doi [\mnras] {10.1093/mnras/sts516},
  \href {https://ui.adsabs.harvard.edu/abs/2013MNRAS.429.2469B} {429, 2469}

\bibitem[\protect\citeauthoryear{{Birnboim}, {Balberg}  \&
  {Teyssier}}{{Birnboim} et~al.}{2015}]{BirnboimEA2015}
{Birnboim} Y.,  {Balberg} S.,   {Teyssier} R.,  2015, \mn@doi [\mnras]
  {10.1093/mnras/stu2717}, \href
  {https://ui.adsabs.harvard.edu/abs/2015MNRAS.447.3678B} {447, 3678}

\bibitem[\protect\citeauthoryear{{Blackman}}{{Blackman}}{1998}]{BlackmanEA1998}
{Blackman} E.~G.,  1998, \mn@doi [\apjl] {10.1086/311238}, \href
  {https://ui.adsabs.harvard.edu/abs/1998ApJ...496L..17B} {496, L17}

\bibitem[\protect\citeauthoryear{{Brandenburg} \& {Subramanian}}{{Brandenburg}
  \& {Subramanian}}{2005}]{BS2005}
{Brandenburg} A.,  {Subramanian} K.,  2005, \mn@doi [\physrep]
  {10.1016/j.physrep.2005.06.005}, \href
  {http://adsabs.harvard.edu/abs/2005PhR...417....1B} {417, 1}

\bibitem[\protect\citeauthoryear{{Burn}}{{Burn}}{1966}]{BurnEA1966}
{Burn} B.~J.,  1966, \mn@doi [\mnras] {10.1093/mnras/133.1.67}, \href
  {https://ui.adsabs.harvard.edu/abs/1966MNRAS.133...67B} {133, 67}

\bibitem[\protect\citeauthoryear{{Cesarsky}}{{Cesarsky}}{1980}]{Cesarsky1980}
{Cesarsky} C.~J.,  1980, \mn@doi [\araa] {10.1146/annurev.aa.18.090180.001445},
  \href {http://adsabs.harvard.edu/abs/1980ARA%26A..18..289C} {18, 289}

\bibitem[\protect\citeauthoryear{{Condon}, {Cotton}, {Greisen}, {Yin},
  {Perley}, {Taylor}  \& {Broderick}}{{Condon} et~al.}{1998}]{CondonEA1998}
{Condon} J.~J.,  {Cotton} W.~D.,  {Greisen} E.~W.,  {Yin} Q.~F.,  {Perley}
  R.~A.,  {Taylor} G.~B.,   {Broderick} J.~J.,  1998, \mn@doi [\aj]
  {10.1086/300337}, \href
  {https://ui.adsabs.harvard.edu/abs/1998AJ....115.1693C} {115, 1693}

\bibitem[\protect\citeauthoryear{{Crocker}, {Bureau}, {Young}  \&
  {Combes}}{{Crocker} et~al.}{2011}]{CrockerEA2011}
{Crocker} A.~F.,  {Bureau} M.,  {Young} L.~M.,   {Combes} F.,  2011, \mn@doi
  [\mnras] {10.1111/j.1365-2966.2010.17537.x}, \href
  {https://ui.adsabs.harvard.edu/abs/2011MNRAS.410.1197C} {410, 1197}

\bibitem[\protect\citeauthoryear{{Dabhade}, {Combes}, {Salom{\'e}}, {Bagchi}
  \& {Mahato}}{{Dabhade} et~al.}{2020}]{DabhadeEA2020}
{Dabhade} P.,  {Combes} F.,  {Salom{\'e}} P.,  {Bagchi} J.,   {Mahato} M.,
  2020, \mn@doi [\aap] {10.1051/0004-6361/202038676}, \href
  {https://ui.adsabs.harvard.edu/abs/2020A&A...643A.111D} {643, A111}

\bibitem[\protect\citeauthoryear{{Dawson} et~al.,}{{Dawson}
  et~al.}{2016}]{DawsonEA2016}
{Dawson} K.~S.,  et~al., 2016, \mn@doi [\aj] {10.3847/0004-6256/151/2/44},
  \href {https://ui.adsabs.harvard.edu/abs/2016AJ....151...44D} {151, 44}

\bibitem[\protect\citeauthoryear{{Duric}}{{Duric}}{1988}]{Duric1988}
{Duric} N.,  1988, \mn@doi [\ssr] {10.1007/BF00183130}, \href
  {https://ui.adsabs.harvard.edu/abs/1988SSRv...48...73D} {48, 73}

\bibitem[\protect\citeauthoryear{{Evirgen}, {Gent}, {Shukurov}, {Fletcher}  \&
  {Bushby}}{{Evirgen} et~al.}{2017}]{EGSFB17}
{Evirgen} C.~C.,  {Gent} F.~A.,  {Shukurov} A.,  {Fletcher} A.,   {Bushby} P.,
  2017, \mn@doi [\mnras] {10.1093/mnrasl/slw196}, \href
  {http://adsabs.harvard.edu/abs/2017MNRAS.464L.105E} {464, L105}

\bibitem[\protect\citeauthoryear{{Farnes}, {Gaensler}  \& {Carretti}}{{Farnes}
  et~al.}{2014a}]{FarnesEA2014}
{Farnes} J.~S.,  {Gaensler} B.~M.,   {Carretti} E.,  2014a, \mn@doi [\apjs]
  {10.1088/0067-0049/212/1/15}, \href
  {https://ui.adsabs.harvard.edu/abs/2014ApJS..212...15F} {212, 15}

\bibitem[\protect\citeauthoryear{{Farnes}, {O'Sullivan}, {Corrigan}  \&
  {Gaensler}}{{Farnes} et~al.}{2014b}]{FarnesEA2014mg2}
{Farnes} J.~S.,  {O'Sullivan} S.~P.,  {Corrigan} M.~E.,   {Gaensler} B.~M.,
  2014b, \mn@doi [\apj] {10.1088/0004-637X/795/1/63}, \href
  {https://ui.adsabs.harvard.edu/abs/2014ApJ...795...63F} {795, 63}

\bibitem[\protect\citeauthoryear{{Federrath}}{{Federrath}}{2016}]{Fed16}
{Federrath} C.,  2016, \mn@doi [Journal of Plasma Physics]
  {10.1017/S0022377816001069}, \href
  {https://ui.adsabs.harvard.edu/#abs/2016JPlPh..82f5301F} {82, 535820601}

\bibitem[\protect\citeauthoryear{{Federrath}, {Chabrier}, {Schober},
  {Banerjee}, {Klessen}  \& {Schleicher}}{{Federrath} et~al.}{2011}]{Fed11}
{Federrath} C.,  {Chabrier} G.,  {Schober} J.,  {Banerjee} R.,  {Klessen}
  R.~S.,   {Schleicher} D.~R.~G.,  2011, \mn@doi [\prl]
  {10.1103/PhysRevLett.107.114504}, \href
  {https://ui.adsabs.harvard.edu/#abs/2011PhRvL.107k4504F} {107, 114504}

\bibitem[\protect\citeauthoryear{{Federrath}, {Schober}, {Bovino}  \&
  {Schleicher}}{{Federrath} et~al.}{2014}]{Fed14}
{Federrath} C.,  {Schober} J.,  {Bovino} S.,   {Schleicher} D. R.~G.,  2014,
  \mn@doi [\apj] {10.1088/2041-8205/797/2/L19}, \href
  {https://ui.adsabs.harvard.edu/#abs/2014ApJ...797L..19F} {797, L19}

\bibitem[\protect\citeauthoryear{{Felten}}{{Felten}}{1996}]{FeltenEA1996}
{Felten} J.~E.,  1996, in {Trimble} V.,  {Reisenegger} A.,  eds,  Astronomical
  Society of the Pacific Conference Series Vol. 88, Clusters, Lensing, and the
  Future of the Universe. p.~271

\bibitem[\protect\citeauthoryear{{Fletcher}}{{Fletcher}}{2010}]{Fletcher10}
{Fletcher} A.,  2010, in {Kothes} R.,  {Landecker} T.~L.,   {Willis} A.~G.,
  eds, ~ Vol. 438, The Dynamic Interstellar Medium: A Celebration of the
  Canadian Galactic Plane Survey. p.~197 (\mn@eprint {arXiv} {1104.2427})

\bibitem[\protect\citeauthoryear{{Fletcher}, {Beck}, {Shukurov}, {Berkhuijsen}
  \& {Horellou}}{{Fletcher} et~al.}{2011}]{FletcherEA2011}
{Fletcher} A.,  {Beck} R.,  {Shukurov} A.,  {Berkhuijsen} E.~M.,   {Horellou}
  C.,  2011, \mn@doi [\mnras] {10.1111/j.1365-2966.2010.18065.x}, \href
  {https://ui.adsabs.harvard.edu/abs/2011MNRAS.412.2396F} {412, 2396}

\bibitem[\protect\citeauthoryear{{Gaensler}, {Haverkorn}, {Staveley-Smith},
  {Dickey}, {McClure-Griffiths}, {Dickel}  \& {Wolleben}}{{Gaensler}
  et~al.}{2005}]{GaenslerEA2005}
{Gaensler} B.~M.,  {Haverkorn} M.,  {Staveley-Smith} L.,  {Dickey} J.~M.,
  {McClure-Griffiths} N.~M.,  {Dickel} J.~R.,   {Wolleben} M.,  2005, \mn@doi
  [Science] {10.1126/science.1108832}, \href
  {https://ui.adsabs.harvard.edu/abs/2005Sci...307.1610G} {307, 1610}

\bibitem[\protect\citeauthoryear{{Gaensler}, {Madsen}, {Chatterjee}  \&
  {Mao}}{{Gaensler} et~al.}{2008}]{GaenslerEA2008}
{Gaensler} B.~M.,  {Madsen} G.~J.,  {Chatterjee} S.,   {Mao} S.~A.,  2008,
  \mn@doi [\pasa] {10.1071/AS08004}, \href
  {https://ui.adsabs.harvard.edu/abs/2008PASA...25..184G} {25, 184}

\bibitem[\protect\citeauthoryear{{Garrington} \& {Conway}}{{Garrington} \&
  {Conway}}{1991}]{GarringtonEA1991(int)}
{Garrington} S.~T.,  {Conway} R.~G.,  1991, \mn@doi [\mnras]
  {10.1093/mnras/250.1.198}, \href
  {https://ui.adsabs.harvard.edu/abs/1991MNRAS.250..198G} {250, 198}

\bibitem[\protect\citeauthoryear{{Garrington}, {Leahy}, {Conway}  \&
  {Laing}}{{Garrington} et~al.}{1988}]{GarringtonEA1988}
{Garrington} S.~T.,  {Leahy} J.~P.,  {Conway} R.~G.,   {Laing} R.~A.,  1988,
  \mn@doi [\nat] {10.1038/331147a0}, \href
  {https://ui.adsabs.harvard.edu/abs/1988Natur.331..147G} {331, 147}

\bibitem[\protect\citeauthoryear{Gerhard}{Gerhard}{2010}]{Gerhard_2010}
Gerhard O.,  2010, \mn@doi [Galaxies and their Masks]
  {10.1007/978-1-4419-7317-7_29}, p. 339–346

\bibitem[\protect\citeauthoryear{{Grasso} \& {Rubinstein}}{{Grasso} \&
  {Rubinstein}}{2001}]{GrassoEA2001}
{Grasso} D.,  {Rubinstein} H.~R.,  2001, \mn@doi [\physrep]
  {10.1016/S0370-1573(00)00110-1}, \href
  {https://ui.adsabs.harvard.edu/abs/2001PhR...348..163G} {348, 163}

\bibitem[\protect\citeauthoryear{{Hamilton}, {Turnshek}  \&
  {Casertano}}{{Hamilton} et~al.}{2000}]{HamiltonEA200}
{Hamilton} T.~S.,  {Turnshek} D.~A.,   {Casertano} S.,  2000, in American
  Astronomical Society Meeting Abstracts. p. 109.02

\bibitem[\protect\citeauthoryear{{Hammond}, {Robishaw}  \&
  {Gaensler}}{{Hammond} et~al.}{2012}]{HammondEA2012}
{Hammond} A.~M.,  {Robishaw} T.,   {Gaensler} B.~M.,  2012, arXiv e-prints,
  \href {https://ui.adsabs.harvard.edu/abs/2012arXiv1209.1438H} {p.
  arXiv:1209.1438}

\bibitem[\protect\citeauthoryear{{Herzenberg}}{{Herzenberg}}{1958}]{Herzenberg1958}
{Herzenberg} A.,  1958, \mn@doi [Philosophical Transactions of the Royal
  Society of London Series A] {10.1098/rsta.1958.0007}, \href
  {https://ui.adsabs.harvard.edu/abs/1958RSPTA.250..543H} {250, 543}

\bibitem[\protect\citeauthoryear{{Ho}}{{Ho}}{2009}]{HoEA2009}
{Ho} L.~C.,  2009, \mn@doi [\apj] {10.1088/0004-637X/699/1/626}, \href
  {https://ui.adsabs.harvard.edu/abs/2009ApJ...699..626H} {699, 626}

\bibitem[\protect\citeauthoryear{{Hopkins} et~al.,}{{Hopkins}
  et~al.}{2018}]{HopkinsEA2018}
{Hopkins} P.~F.,  et~al., 2018, \mn@doi [\mnras] {10.1093/mnras/sty1690}, \href
  {https://ui.adsabs.harvard.edu/abs/2018MNRAS.480..800H} {480, 800}

\bibitem[\protect\citeauthoryear{{Hutschenreuter} et~al.,}{{Hutschenreuter}
  et~al.}{2021}]{Faraday2020}
{Hutschenreuter} S.,  et~al., 2021, arXiv e-prints, \href
  {https://ui.adsabs.harvard.edu/abs/2021arXiv210201709H} {p. arXiv:2102.01709}

\bibitem[\protect\citeauthoryear{Jiang, Ciotti, Ostriker  \& Spitkovsky}{Jiang
  et~al.}{2010}]{Jiang_2010}
Jiang Y.-F.,  Ciotti L.,  Ostriker J.~P.,   Spitkovsky A.,  2010, \mn@doi [The
  Astrophysical Journal] {10.1088/0004-637x/711/1/125}, 711, 125

\bibitem[\protect\citeauthoryear{{Kazantsev}}{{Kazantsev}}{1968}]{Kazantsev1968}
{Kazantsev} A.~P.,  1968, Soviet Journal of Experimental and Theoretical
  Physics, \href {http://adsabs.harvard.edu/abs/1968JETP...26.1031K} {26, 1031}

\bibitem[\protect\citeauthoryear{{Kierdorf} et~al.,}{{Kierdorf}
  et~al.}{2020}]{KierdorfEA2020}
{Kierdorf} M.,  et~al., 2020, \mn@doi [\aap] {10.1051/0004-6361/202037847},
  \href {https://ui.adsabs.harvard.edu/abs/2020A&A...642A.118K} {642, A118}

\bibitem[\protect\citeauthoryear{{Klein} \& {Fletcher}}{{Klein} \&
  {Fletcher}}{2015}]{KleinEA2015}
{Klein} U.,  {Fletcher} A.,  2015, {Galactic and Intergalactic Magnetic Fields}

\bibitem[\protect\citeauthoryear{{Krumholz} \& {Federrath}}{{Krumholz} \&
  {Federrath}}{2019}]{KrumholzF2019}
{Krumholz} M.~R.,  {Federrath} C.,  2019, \mn@doi [Frontiers in Astronomy and
  Space Sciences] {10.3389/fspas.2019.00007}, \href
  {https://ui.adsabs.harvard.edu/abs/2019FrASS...6....7K} {6, 7}

\bibitem[\protect\citeauthoryear{{Krumholz}, {Burkhart}, {Forbes}  \&
  {Crocker}}{{Krumholz} et~al.}{2018}]{KrumholzEA2018}
{Krumholz} M.~R.,  {Burkhart} B.,  {Forbes} J.~C.,   {Crocker} R.~M.,  2018,
  \mn@doi [\mnras] {10.1093/mnras/sty852}, \href
  {https://ui.adsabs.harvard.edu/abs/2018MNRAS.477.2716K} {477, 2716}

\bibitem[\protect\citeauthoryear{{Laing}}{{Laing}}{1988}]{LaingEA1988}
{Laing} R.~A.,  1988, \mn@doi [\nat] {10.1038/331149a0}, \href
  {https://ui.adsabs.harvard.edu/abs/1988Natur.331..149L} {331, 149}

\bibitem[\protect\citeauthoryear{{Laing}, {Bridle}, {Parma}  \&
  {Murgia}}{{Laing} et~al.}{2008a}]{Laing2008}
{Laing} R.~A.,  {Bridle} A.~H.,  {Parma} P.,   {Murgia} M.,  2008a, \mn@doi
  [\mnras] {10.1111/j.1365-2966.2008.13895.x}, \href
  {http://adsabs.harvard.edu/abs/2008MNRAS.391..521L} {391, 521}

\bibitem[\protect\citeauthoryear{{Laing}, {Bridle}, {Parma}  \&
  {Murgia}}{{Laing} et~al.}{2008b}]{LaingEA2008}
{Laing} R.~A.,  {Bridle} A.~H.,  {Parma} P.,   {Murgia} M.,  2008b, \mn@doi
  [\mnras] {10.1111/j.1365-2966.2008.13895.x}, \href
  {https://ui.adsabs.harvard.edu/abs/2008MNRAS.391..521L} {391, 521}

\bibitem[\protect\citeauthoryear{{Lan} \& {Prochaska}}{{Lan} \&
  {Prochaska}}{2020}]{LanP2020}
{Lan} T.-W.,  {Prochaska} J.~X.,  2020, \mn@doi [\mnras]
  {10.1093/mnras/staa1750}, \href
  {https://ui.adsabs.harvard.edu/abs/2020MNRAS.496.3142L} {496, 3142}

\bibitem[\protect\citeauthoryear{{Larmor}}{{Larmor}}{1919}]{Larmor1919}
{Larmor} J.,  1919, Reports of the British Association, 87, 159

\bibitem[\protect\citeauthoryear{{Li}, {Li}, {Bryan}, {Ostriker}  \&
  {Quataert}}{{Li} et~al.}{2020a}]{Li2020_SN}
{Li} M.,  {Li} Y.,  {Bryan} G.~L.,  {Ostriker} E.~C.,   {Quataert} E.,  2020a,
  \mn@doi [\apj] {10.3847/1538-4357/ab86b4}, \href
  {https://ui.adsabs.harvard.edu/abs/2020ApJ...894...44L} {894, 44}

\bibitem[\protect\citeauthoryear{{Li}, {Li}, {Bryan}, {Ostriker}  \&
  {Quataert}}{{Li} et~al.}{2020b}]{Li2020_SN2}
{Li} M.,  {Li} Y.,  {Bryan} G.~L.,  {Ostriker} E.~C.,   {Quataert} E.,  2020b,
  \mn@doi [\apj] {10.3847/1538-4357/ab9c22}, \href
  {https://ui.adsabs.harvard.edu/abs/2020ApJ...898...23L} {898, 23}

\bibitem[\protect\citeauthoryear{{Lilly} \& {Prestage}}{{Lilly} \&
  {Prestage}}{1987}]{LillyEA1987}
{Lilly} S.~J.,  {Prestage} R.~M.,  1987, \mn@doi [\mnras]
  {10.1093/mnras/225.3.531}, \href
  {https://ui.adsabs.harvard.edu/abs/1987MNRAS.225..531L} {225, 531}

\bibitem[\protect\citeauthoryear{{Lintott} et~al.,}{{Lintott}
  et~al.}{2008}]{LintottEA2008}
{Lintott} C.~J.,  et~al., 2008, \mn@doi [\mnras]
  {10.1111/j.1365-2966.2008.13689.x}, \href
  {https://ui.adsabs.harvard.edu/abs/2008MNRAS.389.1179L} {389, 1179}

\bibitem[\protect\citeauthoryear{{Lintott} et~al.,}{{Lintott}
  et~al.}{2011}]{LintottEA2011}
{Lintott} C.,  et~al., 2011, \mn@doi [\mnras]
  {10.1111/j.1365-2966.2010.17432.x}, \href
  {https://ui.adsabs.harvard.edu/abs/2011MNRAS.410..166L} {410, 166}

\bibitem[\protect\citeauthoryear{{Ma}, {Mao}, {Stil}, {Basu}, {West}, {Heiles},
  {Hill}  \& {Betti}}{{Ma} et~al.}{2019}]{MaEA2019}
{Ma} Y.~K.,  {Mao} S.~A.,  {Stil} J.,  {Basu} A.,  {West} J.,  {Heiles} C.,
  {Hill} A.~S.,   {Betti} S.~K.,  2019, \mn@doi [\mnras]
  {10.1093/mnras/stz1328}, \href
  {https://ui.adsabs.harvard.edu/abs/2019MNRAS.487.3454M} {487, 3454}

\bibitem[\protect\citeauthoryear{{Mac Low} \& {Klessen}}{{Mac Low} \&
  {Klessen}}{2004}]{MacLowK2004}
{Mac Low} M.-M.,  {Klessen} R.~S.,  2004, \mn@doi [Reviews of Modern Physics]
  {10.1103/RevModPhys.76.125}, \href
  {https://ui.adsabs.harvard.edu/abs/2004RvMP...76..125M} {76, 125}

\bibitem[\protect\citeauthoryear{{Majewski} et~al.,}{{Majewski}
  et~al.}{2017}]{MajewskiEA2017}
{Majewski} S.~R.,  et~al., 2017, \mn@doi [\aj] {10.3847/1538-3881/aa784d},
  \href {https://ui.adsabs.harvard.edu/abs/2017AJ....154...94M} {154, 94}

\bibitem[\protect\citeauthoryear{{Malarecki}, {Staveley-Smith}, {Saripalli},
  {Subrahmanyan}, {Jones}, {Duffy}  \& {Rioja}}{{Malarecki}
  et~al.}{2013}]{MalareckiEA2013}
{Malarecki} J.~M.,  {Staveley-Smith} L.,  {Saripalli} L.,  {Subrahmanyan} R.,
  {Jones} D.~H.,  {Duffy} A.~R.,   {Rioja} M.,  2013, \mn@doi [\mnras]
  {10.1093/mnras/stt471}, \href
  {https://ui.adsabs.harvard.edu/abs/2013MNRAS.432..200M} {432, 200}

\bibitem[\protect\citeauthoryear{{Malik}, {Chand}  \& {Seshadri}}{{Malik}
  et~al.}{2020}]{MalikEA2020}
{Malik} S.,  {Chand} H.,   {Seshadri} T.~R.,  2020, \mn@doi [\apj]
  {10.3847/1538-4357/ab6bd5}, \href
  {https://ui.adsabs.harvard.edu/abs/2020ApJ...890..132M} {890, 132}

\bibitem[\protect\citeauthoryear{Mathews \& Brighenti}{Mathews \&
  Brighenti}{1997}]{MathewsEA1997}
Mathews W.~G.,  Brighenti F.,  1997, \mn@doi [The Astrophysical Journal]
  {10.1086/304728}, 488, 595

\bibitem[\protect\citeauthoryear{{Mathews} \& {Brighenti}}{{Mathews} \&
  {Brighenti}}{2003}]{MathewsEA2003}
{Mathews} W.~G.,  {Brighenti} F.,  2003, \mn@doi [\araa]
  {10.1146/annurev.astro.41.090401.094542}, \href
  {https://ui.adsabs.harvard.edu/abs/2003ARA&A..41..191M} {41, 191}

\bibitem[\protect\citeauthoryear{{Matthaeus}, {Qin}, {Bieber}  \&
  {Zank}}{{Matthaeus} et~al.}{2003}]{MatthaeusEA2003}
{Matthaeus} W.~H.,  {Qin} G.,  {Bieber} J.~W.,   {Zank} G.~P.,  2003, \mn@doi
  [\apjl] {10.1086/376613}, \href
  {http://adsabs.harvard.edu/abs/2003ApJ...590L..53M} {590, L53}

\bibitem[\protect\citeauthoryear{{Moss} \& {Shukurov}}{{Moss} \&
  {Shukurov}}{1996}]{MShu96}
{Moss} D.,  {Shukurov} A.,  1996, \mn@doi [\mnras] {10.1093/mnras/279.1.229},
  \href {http://adsabs.harvard.edu/abs/1996MNRAS.279..229M} {279, 229}

\bibitem[\protect\citeauthoryear{{Nelson} et~al.,}{{Nelson}
  et~al.}{2020}]{NelsonEA2020}
{Nelson} D.,  et~al., 2020, \mn@doi [\mnras] {10.1093/mnras/staa2419}, \href
  {https://ui.adsabs.harvard.edu/abs/2020MNRAS.498.2391N} {498, 2391}

\bibitem[\protect\citeauthoryear{{Nyland} et~al.,}{{Nyland}
  et~al.}{2017}]{Nyland2017}
{Nyland} K.,  et~al., 2017, \mn@doi [\mnras] {10.1093/mnras/stw2385}, \href
  {https://ui.adsabs.harvard.edu/#abs/2017MNRAS.464.1029N} {464, 1029}

\bibitem[\protect\citeauthoryear{{Oppermann} et~al.,}{{Oppermann}
  et~al.}{2012}]{OppermannEA2012}
{Oppermann} N.,  et~al., 2012, \mn@doi [\aap] {10.1051/0004-6361/201118526},
  \href {https://ui.adsabs.harvard.edu/abs/2012A&A...542A..93O} {542, A93}

\bibitem[\protect\citeauthoryear{{Oppermann} et~al.,}{{Oppermann}
  et~al.}{2015}]{OppermannEA2015}
{Oppermann} N.,  et~al., 2015, \mn@doi [\aap] {10.1051/0004-6361/201423995},
  \href {https://ui.adsabs.harvard.edu/abs/2015A&A...575A.118O} {575, A118}

\bibitem[\protect\citeauthoryear{{Owen} \& {Laing}}{{Owen} \&
  {Laing}}{1989}]{OwenEA1989}
{Owen} F.~N.,  {Laing} R.~A.,  1989, \mn@doi [\mnras]
  {10.1093/mnras/238.2.357}, \href
  {https://ui.adsabs.harvard.edu/abs/1989MNRAS.238..357O} {238, 357}

\bibitem[\protect\citeauthoryear{{Pakmor} et~al.,}{{Pakmor}
  et~al.}{2020}]{PakmorEA2020}
{Pakmor} R.,  et~al., 2020, \mn@doi [\mnras] {10.1093/mnras/staa2530}, \href
  {https://ui.adsabs.harvard.edu/abs/2020MNRAS.498.3125P} {498, 3125}

\bibitem[\protect\citeauthoryear{{Putman}, {Peek}  \& {Joung}}{{Putman}
  et~al.}{2012}]{Putman2012CGM}
{Putman} M.~E.,  {Peek} J.~E.~G.,   {Joung} M.~R.,  2012, \mn@doi [\araa]
  {10.1146/annurev-astro-081811-125612}, \href
  {https://ui.adsabs.harvard.edu/abs/2012ARA&A..50..491P} {50, 491}

\bibitem[\protect\citeauthoryear{{Raymond}}{{Raymond}}{1992}]{Raymond1992}
{Raymond} J.~C.,  1992, \mn@doi [\apj] {10.1086/170892}, \href
  {https://ui.adsabs.harvard.edu/abs/1992ApJ...384..502R} {384, 502}

\bibitem[\protect\citeauthoryear{{Rincon}}{{Rincon}}{2019}]{Rincon19}
{Rincon} F.,  2019, \mn@doi [Journal of Plasma Physics]
  {10.1017/S0022377819000539}, \href
  {https://ui.adsabs.harvard.edu/abs/2019JPlPh..85d2001R} {85, 205850401}

\bibitem[\protect\citeauthoryear{{Ruzmaikin}, {Sokoloff}  \&
  {Shukurov}}{{Ruzmaikin} et~al.}{1988}]{RSS88}
{Ruzmaikin} A.~A.,  {Sokoloff} D.~D.,   {Shukurov} A.~M.,  eds, 1988, {Magnetic
  fields of galaxies}  Astrophysics and Space Science Library Vol. 133,
  \mn@doi{10.1007/978-94-009-2835-0.
}

\bibitem[\protect\citeauthoryear{{Sarala} \& {Jain}}{{Sarala} \&
  {Jain}}{2001}]{SaralaEA2001}
{Sarala} S.,  {Jain} P.,  2001, \mn@doi [\mnras]
  {10.1046/j.1365-8711.2001.04932.x}, \href
  {https://ui.adsabs.harvard.edu/abs/2001MNRAS.328..623S} {328, 623}

\bibitem[\protect\citeauthoryear{{Sarazin} \& {White}}{{Sarazin} \&
  {White}}{1988}]{SarazinEA1988_King}
{Sarazin} C.~L.,  {White} Raymond~E. I.,  1988, \mn@doi [\apj]
  {10.1086/166540}, \href
  {https://ui.adsabs.harvard.edu/abs/1988ApJ...331..102S} {331, 102}

\bibitem[\protect\citeauthoryear{Saripalli}{Saripalli}{2012}]{Saripalli_2012}
Saripalli L.,  2012, \mn@doi [The Astronomical Journal]
  {10.1088/0004-6256/144/3/85}, 144, 85

\bibitem[\protect\citeauthoryear{{Schekochihin}, {Cowley}, {Taylor}, {Maron}
  \& {McWilliams}}{{Schekochihin} et~al.}{2004}]{SCTMM04}
{Schekochihin} A.~A.,  {Cowley} S.~C.,  {Taylor} S.~F.,  {Maron} J.~L.,
  {McWilliams} J.~C.,  2004, \mn@doi [\apj] {10.1086/422547}, \href
  {http://adsabs.harvard.edu/abs/2004ApJ...612..276S} {612, 276}

\bibitem[\protect\citeauthoryear{{Seta}}{{Seta}}{2019}]{Seta2019}
{Seta} A.,  2019, PhD thesis, Newcastle University, Newcastle Upon Tyne, UK,
  \url {http://theses.ncl.ac.uk/jspui/handle/10443/4685}

\bibitem[\protect\citeauthoryear{{Seta} \& {Federrath}}{{Seta} \&
  {Federrath}}{2020}]{SetaF2020}
{Seta} A.,  {Federrath} C.,  2020, \mn@doi [\mnras] {10.1093/mnras/staa2978},
  \href {https://ui.adsabs.harvard.edu/abs/2020MNRAS.499.2076S} {499, 2076}

\bibitem[\protect\citeauthoryear{{Seta} \& {Federrath}}{{Seta} \&
  {Federrath}}{2021}]{SetaF2021}
{Seta} A.,  {Federrath} C.,  2021, \mn@doi [\mnras] {10.1093/mnras/stab128},
  \href {https://ui.adsabs.harvard.edu/abs/2021MNRAS.502.2220S} {502, 2220}

\bibitem[\protect\citeauthoryear{{Seta}, {Shukurov}, {Wood}, {Bushby}  \&
  {Snodin}}{{Seta} et~al.}{2018}]{SSWBS18}
{Seta} A.,  {Shukurov} A.,  {Wood} T.~S.,  {Bushby} P.~J.,   {Snodin} A.~P.,
  2018, \mn@doi [\mnras] {10.1093/mnras/stx2606}, \href
  {http://adsabs.harvard.edu/abs/2018MNRAS.473.4544S} {473, 4544}

\bibitem[\protect\citeauthoryear{{Seta}, {Bushby}, {Shukurov}  \&
  {Wood}}{{Seta} et~al.}{2020}]{SetaEA2020}
{Seta} A.,  {Bushby} P.~J.,  {Shukurov} A.,   {Wood} T.~S.,  2020, \mn@doi
  [Physical Review Fluids] {10.1103/PhysRevFluids.5.043702}, \href
  {https://ui.adsabs.harvard.edu/abs/2020PhRvF...5d3702S} {5, 043702}

\bibitem[\protect\citeauthoryear{{Seta}, {Rodrigues}, {Federrath}  \&
  {Hales}}{{Seta} et~al.}{2021}]{SetaEA2020II}
{Seta} A.,  {Rodrigues} L. F.~S.,  {Federrath} C.,   {Hales} C.~A.,  2021,
  \mn@doi [\apj] {10.3847/1538-4357/abd2bb}, \href
  {https://ui.adsabs.harvard.edu/abs/2021ApJ...907....2S} {907, 2}

\bibitem[\protect\citeauthoryear{{Shukurov} \& {Sokoloff}}{{Shukurov} \&
  {Sokoloff}}{2007}]{SS2008}
{Shukurov} A.,  {Sokoloff} D.,  2007, in {Cardin} P.,  {Cugliandolo} L.~F.,
  eds, ~ Vol. 88, Les Houches, Session LXXXVIII, Dynamos. Amsterdam: Elsevier,
  pp 251--299, \mn@doi{10.1016/S0924-8099(08)80008-X}, \url
  {http://dx.doi.org/10.1016/S0924-8099(08)80008-X}

\bibitem[\protect\citeauthoryear{{Shukurov}, {Snodin}, {Seta}, {Bushby}  \&
  {Wood}}{{Shukurov} et~al.}{2017}]{SSSBW17}
{Shukurov} A.,  {Snodin} A.~P.,  {Seta} A.,  {Bushby} P.~J.,   {Wood} T.~S.,
  2017, \mn@doi [\apjl] {10.3847/2041-8213/aa6aa6}, \href
  {http://adsabs.harvard.edu/abs/2017ApJ...839L..16S} {839, L16}

\bibitem[\protect\citeauthoryear{{Sokoloff}, {Bykov}, {Shukurov},
  {Berkhuijsen}, {Beck}  \& {Poezd}}{{Sokoloff} et~al.}{1998}]{Sokoloff1998}
{Sokoloff} D.~D.,  {Bykov} A.~A.,  {Shukurov} A.,  {Berkhuijsen} E.~M.,  {Beck}
  R.,   {Poezd} A.~D.,  1998, \mn@doi [\mnras]
  {10.1046/j.1365-8711.1998.01782.x}, \href
  {http://adsabs.harvard.edu/abs/1998MNRAS.299..189S} {299, 189}

\bibitem[\protect\citeauthoryear{{Tabatabaei}, {Krause}, {Fletcher}  \&
  {Beck}}{{Tabatabaei} et~al.}{2008}]{TabatabaeiEA2008}
{Tabatabaei} F.~S.,  {Krause} M.,  {Fletcher} A.,   {Beck} R.,  2008, \mn@doi
  [\aap] {10.1051/0004-6361:200810590}, \href
  {https://ui.adsabs.harvard.edu/abs/2008A&A...490.1005T} {490, 1005}

\bibitem[\protect\citeauthoryear{{Taylor} \& {Cordes}}{{Taylor} \&
  {Cordes}}{1993}]{TaylorEA1993}
{Taylor} J.~H.,  {Cordes} J.~M.,  1993, \mn@doi [\apj] {10.1086/172870}, \href
  {https://ui.adsabs.harvard.edu/abs/1993ApJ...411..674T} {411, 674}

\bibitem[\protect\citeauthoryear{{Taylor}, {Stil}  \& {Sunstrum}}{{Taylor}
  et~al.}{2009}]{TaylorEA2009}
{Taylor} A.~R.,  {Stil} J.~M.,   {Sunstrum} C.,  2009, \mn@doi [\apj]
  {10.1088/0004-637X/702/2/1230}, \href
  {https://ui.adsabs.harvard.edu/abs/2009ApJ...702.1230T} {702, 1230}

\bibitem[\protect\citeauthoryear{{Thakar}, {Szalay}, {Fekete}  \&
  {Gray}}{{Thakar} et~al.}{2008}]{ThakarEA2008}
{Thakar} A.~R.,  {Szalay} A.,  {Fekete} G.,   {Gray} J.,  2008, \mn@doi
  [Computing in Science and Engineering] {10.1109/MCSE.2008.15}, \href
  {https://ui.adsabs.harvard.edu/abs/2008CSE....10...30T} {10, 30}

\bibitem[\protect\citeauthoryear{{Tumlinson}, {Peeples}  \& {Werk}}{{Tumlinson}
  et~al.}{2017}]{TumlinsonEA2017}
{Tumlinson} J.,  {Peeples} M.~S.,   {Werk} J.~K.,  2017, \mn@doi [\araa]
  {10.1146/annurev-astro-091916-055240}, \href
  {https://ui.adsabs.harvard.edu/abs/2017ARA&A..55..389T} {55, 389}

\bibitem[\protect\citeauthoryear{{Uson}, {Boughn}  \& {Kuhn}}{{Uson}
  et~al.}{1990}]{UsonEA1990}
{Uson} J.~M.,  {Boughn} S.~P.,   {Kuhn} J.~R.,  1990, \mn@doi [Science]
  {10.1126/science.250.4980.539}, \href
  {https://ui.adsabs.harvard.edu/abs/1990Sci...250..539U} {250, 539}

\bibitem[\protect\citeauthoryear{{Vernstrom}, {Gaensler}, {Rudnick}  \&
  {Andernach}}{{Vernstrom} et~al.}{2019}]{VernstromEA2019}
{Vernstrom} T.,  {Gaensler} B.~M.,  {Rudnick} L.,   {Andernach} H.,  2019,
  \mn@doi [\apj] {10.3847/1538-4357/ab1f83}, \href
  {https://ui.adsabs.harvard.edu/abs/2019ApJ...878...92V} {878, 92}

\bibitem[\protect\citeauthoryear{{Wright} et~al.,}{{Wright}
  et~al.}{2010}]{WrightEA2010}
{Wright} E.~L.,  et~al., 2010, \mn@doi [\aj] {10.1088/0004-6256/140/6/1868},
  \href {https://ui.adsabs.harvard.edu/abs/2010AJ....140.1868W} {140, 1868}

\bibitem[\protect\citeauthoryear{{Yao}, {Manchester}  \& {Wang}}{{Yao}
  et~al.}{2017}]{YaoEA2017}
{Yao} J.~M.,  {Manchester} R.~N.,   {Wang} N.,  2017, \mn@doi [\apj]
  {10.3847/1538-4357/835/1/29}, \href
  {https://ui.adsabs.harvard.edu/abs/2017ApJ...835...29Y} {835, 29}

\bibitem[\protect\citeauthoryear{{de Gouveia Dal Pino}}{{de Gouveia Dal
  Pino}}{2006}]{ElisabeteEA2006}
{de Gouveia Dal Pino} E.~M.,  2006, in {Herrera-Vel{\'a}zquez} J. J.~E.,  ed.,
  American Institute of Physics Conference Series Vol. 875, Plasma and Fusion
  Science: 16th IAEA Technical Meeting on Research using Small Fusion Devices.
  pp 289--295 (\mn@eprint {arXiv} {astro-ph/0603065}),
  \mn@doi{10.1063/1.2405951}

\bibitem[\protect\citeauthoryear{{van de Voort}, {Bieri}, {Pakmor},
  {G{\'o}mez}, {Grand}  \& {Marinacci}}{{van de Voort}
  et~al.}{2021}]{VoortEA2020}
{van de Voort} F.,  {Bieri} R.,  {Pakmor} R.,  {G{\'o}mez} F.~A.,  {Grand} R.
  J.~J.,   {Marinacci} F.,  2021, \mn@doi [\mnras] {10.1093/mnras/staa3938},
  \href {https://ui.adsabs.harvard.edu/abs/2021MNRAS.501.4888V} {501, 4888}

\makeatother
\end{thebibliography}



\appendix


\section{Background RM Method Simulation}
\label{sec:bg_sim}
In this section, we explore the dependence of $\sigmarrm$ on $\ningal$ via simple numerical simulations. We create an ideal dataset of $50000$ background sources with their $\RM$ distribution following the three (Gaussian and non-Gaussian) probability density functions (G, SH, and LB) discussed in \Sec{sec:sdss_analysis} (check \Fig{fig:three_cases} for their application on the SDSS data). The parameters of these distributions are set such that the standard deviation of the rotation measure is approximately $20 \rad/\m^2$ for all three of them. This is motivated by the fact that the intercepts of slopes when applied to the SDSS data (see \rev{\Fig{fig:gen_line} \revb{and} \Fig{fig:spi_line}}) lie in the range $15~\text{--}~24 \rad/\m^{2}$. \Tab{table:init_table} contains the statistical properties, such as the mean $\mean$, standard deviation $\stddev$, skewness $\skewness$, and kurtosis $\kurtosis$, of the background $\RM$ distribution for all three cases.

\begin{table}
\centering
\begin{tabular}{ccccc}
\hline
Distributions & $\mean$ & $\sigmarrm$ & $\skewness$ & $\kurtosis$  \\
\hline
G & -0.07 & 20.04 & -0.01 & -0.02  \\
SH & -0.09 & 19.98 & -0.04 & 2.71 \\
LB & -0.09 & 20.05 & -0.05 & 3.17 \\
\hline
\end{tabular}
\caption{Parameters of the initial background $\RM$ distributions. We have fixed the $\sigmarrm$ to be $20~\rad/\m^2$, and the $\mean$ and $\skewness$ to be 0. The slight variations from the fixed values in second decimal places are a result of the finite number of elements in a sample.}
\label{table:init_table}
\end{table}

\begin{figure}

\begin{subfigure}{0.45\textwidth}
\includegraphics[width=1\columnwidth,height=7cm]{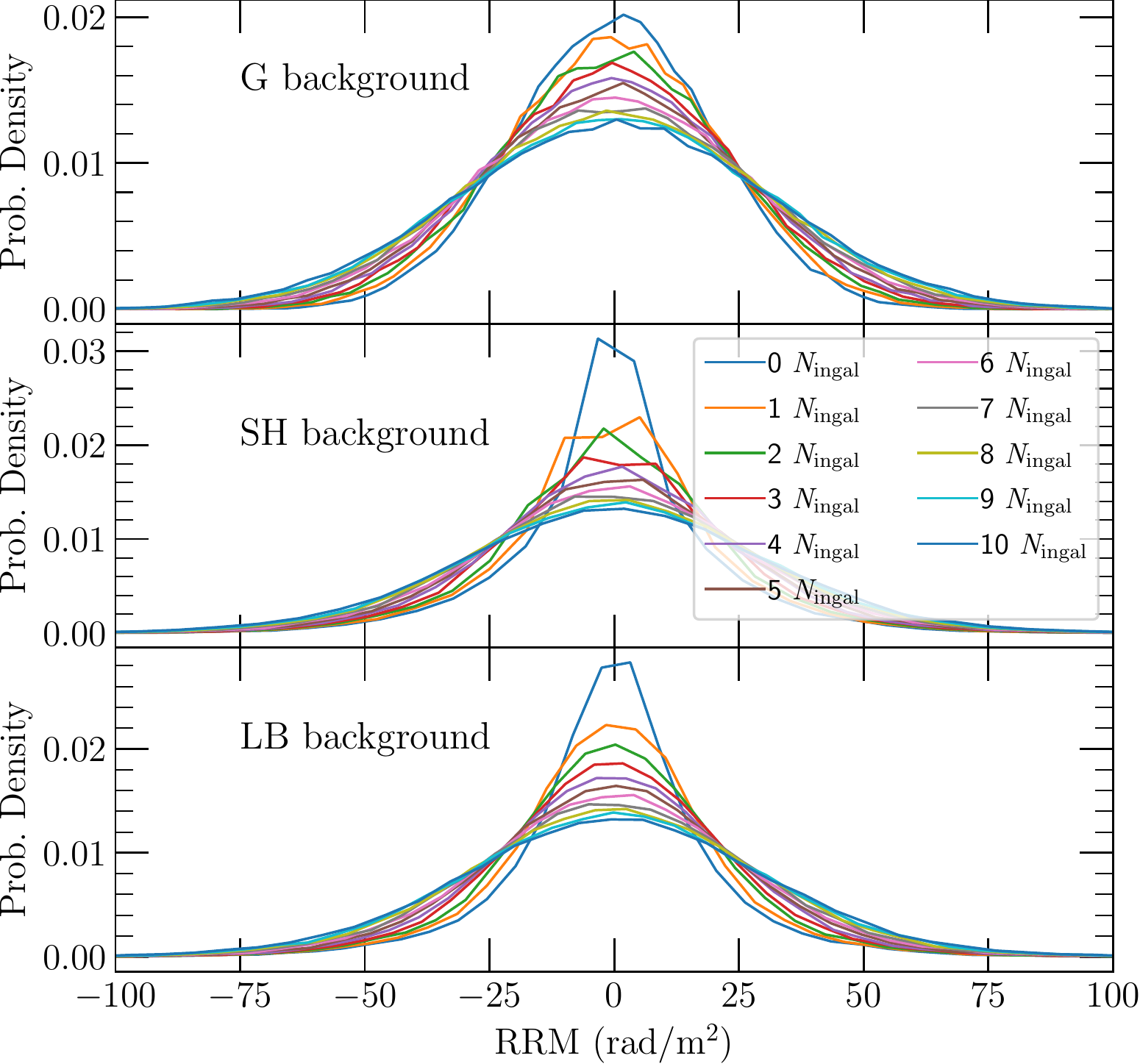} 
\caption{The PDF distributions with $|\RRM| < ~100~\rad/\m^2$ having either G, SH, LB distributions as initial background with random $\RMIngal$ contributions between $4~\text{--}~11 \rad/\m^2$ attributed to every intervening galaxy for a background source.}
\label{fig:subim1}
\end{subfigure}

\begin{subfigure}{0.45\textwidth}
\includegraphics[width=1\columnwidth]{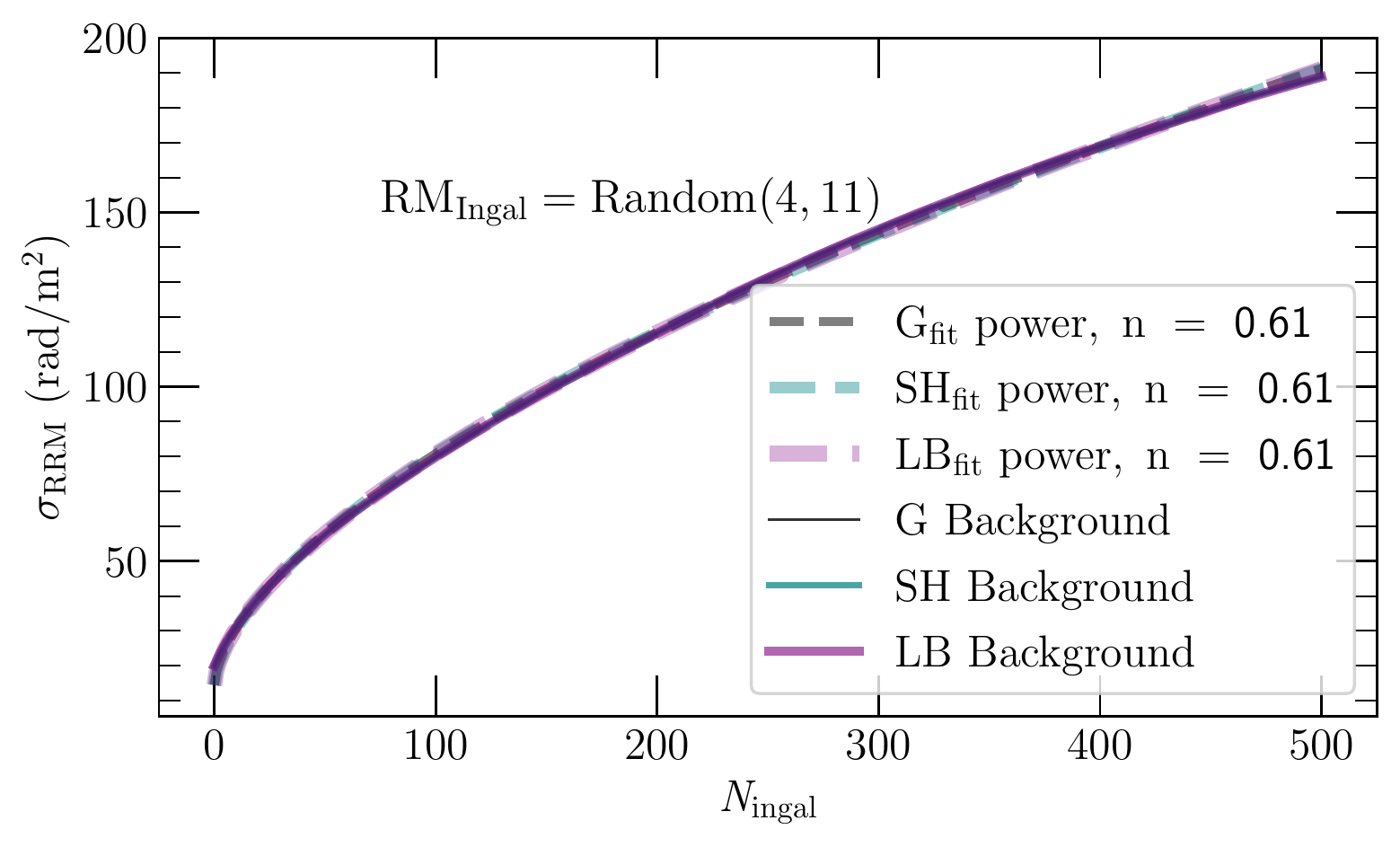}
\caption{For $\RMIngal$ between 4 and 11 $\rad/\m^2$, $\sigmarrm$ VS $\ningal$ reveals that the properties of the curves are independent of the background source distribution, as all the six curves (the simulated data, and its curve-fits according to \Eq{eq:dependence}) perfectly overlap over each other.}
\label{fig:overlap}
\end{subfigure}

\begin{subfigure}{0.45\textwidth}
\includegraphics[width=1\columnwidth]{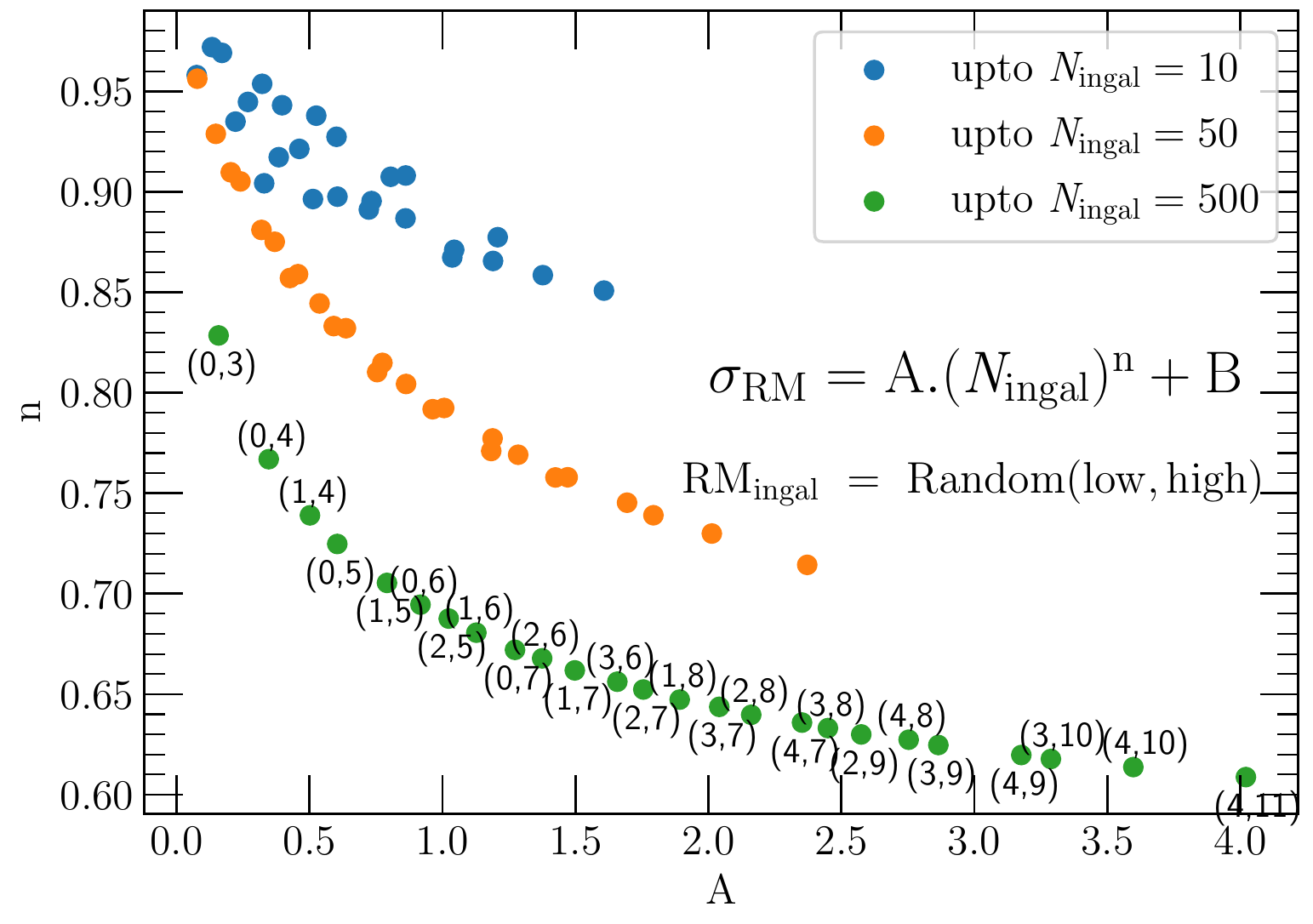}
\caption{Variation of A and n from \Eq{eq:dependence} with varying $\RMIngal$. Random(low, high) selects random $\RMIngal$ between $\rm low$ and $\rm high$ and annotated brackets represent these values. Due to congestion, the brackets are not annotated below the blue and orange dots, but they occur in the same order as the green dots. For low $\ningal$ fit, the power (n) $\approx 0.9$, implying an almost linear dependence.}
\label{fig:pow_slo}
\end{subfigure}

\label{fig:image2}
\end{figure}

To simulate the contribution of intervening galaxies, a random $\RM$ of the intervening galaxy, $\RMIngal$, in the range $4~\text{--}~11 \rad/\m^2$ (either positive or negative) is assigned randomly to every intervening galaxy. The probability density functions (PDFs) of $\RRM$ for 50000 sources for $\ningal$ ranging from $1$ to $10$, considering all three different background distributions, are shown in \Fig{fig:subim1}. The dispersion ($\sigmarrm$) increases as the no. of intervening galaxies ($\ningal$) increases (as the curve spreads more at higher $\ningal$). 

We assume the dependence of $\sigmarrm$ on $\ningal$ of the form
\begin{ceqn}
\begin{align}\label{eq:dependence}
    \sigmarrm (\ningal) = \mathrm{A}(\ningal)^\mathrm{n}+\mathrm{B},
\end{align}
\end{ceqn}
where $\rm A, B, n$ are to be determined. We fit the simulated data for upto 500 $\ningal$ with $\RMIngal$ randomly selected in the range $4~\text{--}~11 \rad/\m^2$ (sign also selected randomly). The fitting parameters $\rm (A, B, n)$ of \Eq{eq:dependence} do not vary with the background source distribution (decided by G, SH, and LB functions), as evident from \Fig{fig:overlap} as $\rm n=0.61$ for all the fits. \Eq{eq:dependence} is an excellent fit as seen from the overlapping curves of \Fig{fig:overlap} for ${\rm n=0.61}$, this is close to the expectation that $\sigmarrm \propto \sqrt{\ningal}$ \citep{LanP2020}. We emphasize this is only true for a very large number of background sources and intervening galaxies. 

As \Eq{eq:dependence} does not vary with the background source distributions, we take a Gaussian background of $\RM$ sources and calculate the fitting parameters from \Eq{eq:dependence} for different $\RMIngal$ ranges as shown in \Fig{fig:pow_slo}. In a realistic scenario $\ningal=500$ is a rarity, hence we also compute \Eq{eq:dependence} for different $\ningal$. There is an increasing trend of power (n) when $\ningal$ reduces. For $\ningal$ up to 10, the power (n) is $\approx$0.9, implying that \Eq{eq:dependence} is almost a linear equation. From the actual SDSS data, we only receive data upto $\ningal=6$ (see \Fig{fig:gen_line}), thus $\sigmarrm$ dependence on $\ningal$ can be safely assumed as linear for low $\ningal$. 

\bsp	
\label{lastpage}
\end{document}


\begin{landscape}
\begin{longtable}{|c|c|c|c|c|c|c|c|c|c|c|c|}
\hline
\caption{This table is a representative of the given and extracted values from the data of Vernstrom et al. (2019) for depolarization by galaxies (physical separation between the lobes less than $500$ kpc). The columns are as follows: column1: Vernstrom et al. (2019) source identifier number, column2: redshift of the source, column3: path length or projected physical separation between the lobes, column4: diameter of the halo of the host galaxy of the radio lobes (of Vernstrom et al. (2019)) identified in SDSS by cross-matching, column5: ratio of the polarization values of the jet and counterjet, column6: calculated standard deviation of rotation measure in elliptical's CGM/halo, column7: small-scale magnetic fields with their one-sigma uncertainties for the uniform $n_{\rm e}$ case, column8: small-scale magnetic fields with their one-sigma uncertainties at the center of ellipticals for the King profile $n_{\rm e}$ case, column9: large-scale magnetic fields, column10: ratio of small-to-large scale magnetic fields for the uniform $n_{\rm e}$ case, column11: ratio of small-to-large scale magnetic fields for the King profile case. The dashes represent the lack of values for the corresponding parameter due to a failure in meeting the error conditions.  }
\\ \hline
V19 no. & $z$ & \multicolumn{1}{p{0.9cm}}{\centering $L$ \\ $\rm (kpc)$} & \multicolumn{1}{|p{1.7cm}|}{\centering $\rm SDSS$ \\ $\rm identified$ \\ $\rm host~size$ $\rm (kpc)$} & $p_{\rm j}/p_{\rm cj}$ & \multicolumn{1}{p{1.5cm}}{\centering $\sigma_{\rm RM}$ \\ $\rm (rad/m^2)$} & \multicolumn{1}{|p{1.8cm}}{\centering $b_{\rm rms}$ \\ $(\mu{\rm G})$} & \multicolumn{1}{|p{1.8cm}}{\centering $b_{\rm rms}(0)$ \\ $(\mu{\rm G})$} & \multicolumn{1}{|p{2cm}|}{\centering $\overline{B_{0 \parallel}}$ \\ $(\mu{\rm G})$} & $b_{\rm rms} / |\overline{B_{0 \parallel}}|$ & $b_{\rm rms}(0) / |\overline{B_{0 \parallel}}|$ \\
\hline
38 & 0.097 & 174 & - & 1.29 & 6.53 & 0.18 $\pm$ 0.03 & 0.82 $\pm$ 0.15 & - & - & - \\
45 & 0.3 & 404 & - & 1.36 & 5.07 & - & - & - & - & - \\
52 & 0.156 & 310 & - & 3.54 & 13.0 & 0.24 $\pm$ 0.04 & 2.12 $\pm$ 0.34 & - & - & - \\
63 & 0.106 & 199 & - & 1.08 & 3.47 & - & - & - & - & - \\
88 & 0.108 & 238 & - & 1.06 & 2.98 & 0.07 $\pm$ 0.01 & 0.43 $\pm$ 0.07 & - & - & - \\
142 & 0.098 & 208 & 293 & 1.53 & 8.34 & 0.2 $\pm$ 0.02 & 1.14 $\pm$ 0.11 & - & - & - \\
229 & 0.123 & 481 & - & 1.34 & 6.59 & 0.1 $\pm$ 0.03 & 1.31 $\pm$ 0.36 & 0.0022 $\pm$ 0.0008 & 46.91 $\pm$ 30.28 & - \\
231 & 0.123 & 240 & - & 1.9 & 9.81 & 0.21 $\pm$ 0.03 & 1.43 $\pm$ 0.17 & - & - & - \\
237 & 0.112 & 345 & 395 & 1.04 & 2.35 & - & - & - & - & - \\
254 & 0.302 & 406 & - & 1.57 & 6.13 & 0.08 $\pm$ 0.01 & 1.13 $\pm$ 0.14 & - & - & - \\
337 & 0.123 & 240 & - & 2.2 & 10.86 & 0.24 $\pm$ 0.05 & 1.59 $\pm$ 0.33 & - & - & - \\
437 & 0.03 & 334 & - & 2.72 & 14.56 & 0.32 $\pm$ 0.01 & 2.46 $\pm$ 0.05 & 0.004 $\pm$ 0.0009 & 80.15 $\pm$ 20.59 & - \\
438 & 0.03 & 384 & - & 1.18 & 5.9 & 0.12 $\pm$ 0.02 & 1.06 $\pm$ 0.19 & 0.0061 $\pm$ 0.0007 & 19.97 $\pm$ 5.99 & 174.67 $\pm$ 52.39 \\
470 & 0.211 & 354 & - & 1.04 & 2.02 & - & - & - & - & - \\
521 & 0.3 & 431 & - & 1.04 & 1.82 & - & - & - & - & - \\
744 & 0.021 & 82 & - & 1.47 & 9.22 & 0.42 $\pm$ 0.08 & 0.84 $\pm$ 0.16 & - & - & - \\
786 & 0.087 & 157 & - & 9.6 & 19.67 & 0.57 $\pm$ 0.01 & 2.38 $\pm$ 0.03 & 0.0198 $\pm$ 0.007 & 28.57 $\pm$ 10.42 & 119.96 $\pm$ 43.73 \\
1132 & 0.25 & 355 & - & 1.38 & 5.64 & 0.08 $\pm$ 0.03 & 0.98 $\pm$ 0.34 & -0.0025 $\pm$ 0.0008 & 32.74 $\pm$ 22.19 & - \\
1224 & 0.127 & 220 & - & 1.88 & 9.67 & - & - & - & - & - \\
1231 & 0.055 & 129 & - & 1.94 & 11.29 & 0.38 $\pm$ 0.03 & 1.25 $\pm$ 0.1 & - & - & - \\
1291 & 0.046 & 98 & - & 1.38 & 7.99 & 0.31 $\pm$ 0.12 & 0.78 $\pm$ 0.29 & - & - & - \\
1426 & 0.119 & 272 & - & 1.67 & 8.82 & 0.18 $\pm$ 0.09 & 1.36 $\pm$ 0.64 & 0.0021 $\pm$ 0.0007 & 86.83 $\pm$ 71.32 & - \\
1507 & 0.06 & 112 & 177 & 1.03 & 2.47 & - & - & - & - & - \\
1577 & 0.322 & 453 & - & 1.01 & 1.04 & - & - & - & - & - \\
1654 & 0.133 & 257 & 2000 & 1.34 & 6.51 & 0.14 $\pm$ 0.02 & 0.98 $\pm$ 0.13 & - & - & - \\
1691 & 0.097 & 206 & 207 & 4.24 & 15.43 & 0.38 $\pm$ 0.05 & 2.1 $\pm$ 0.26 & 0.0232 $\pm$ 0.0096 & 16.43 $\pm$ 8.82 & 90.54 $\pm$ 48.62 \\
1706 & 0.283 & 466 & - & 1.24 & 4.36 & - & - & - & - & - \\
1792 & 0.055 & 142 & - & 1.1 & 4.31 & - & - & - & - & - \\
2035 & 0.201 & 341 & - & 1.42 & 6.35 & 0.1 $\pm$ 0.05 & 1.08 $\pm$ 0.51 & - & - & - \\
2080 & 0.029 & 70 & 2000 & 1.49 & 9.23 & 0.44 $\pm$ 0.19 & 0.78 $\pm$ 0.33 & - & - & - \\
2143 & 0.128 & 235 & 321 & 1.22 & 5.37 & - & - & - & - & - \\
2442 & 0.03 & 80 & 783 & 1.11 & 4.65 & - & - & - & - & - \\
2508 & 0.099 & 166 & - & 1.19 & 5.3 & - & - & - & - & - \\
2633 & 0.032 & 58 & 1357 & 2.27 & 13.13 & 0.69 $\pm$ 0.18 & 1.02 $\pm$ 0.27 & - & - & - \\
2643 & 0.111 & 244 & 383 & 1.17 & 4.92 & - & - & - & - & - \\
2663 & 0.299 & 457 & - & 1.39 & 5.23 & 0.06 $\pm$ 0.01 & 1.01 $\pm$ 0.21 & 0.0056 $\pm$ 0.0026 & 10.96 $\pm$ 7.42 & 179.61 $\pm$ 121.56 \\
2894 & 0.26 & 487 & - & 1.31 & 5.05 & 0.06 $\pm$ 0.02 & 1.01 $\pm$ 0.27 & - & - & - \\
2924 & 0.007 & 53 & 264 & 1.15 & 5.66 & 0.33 $\pm$ 0.08 & 0.42 $\pm$ 0.1 & -0.0467 $\pm$ 0.0087 & 7.01 $\pm$ 3.02 & 9.08 $\pm$ 3.91 \\
2925 & 0.007 & 62 & 264 & 2.41 & 14.28 & 0.77 $\pm$ 0.03 & 1.14 $\pm$ 0.04 & 0.046 $\pm$ 0.0053 & 16.63 $\pm$ 2.51 & 24.84 $\pm$ 3.74 \\
2926 & 0.007 & 43 & 264 & 1.89 & 12.14 & 0.78 $\pm$ 0.02 & 0.83 $\pm$ 0.02 & -0.0846 $\pm$ 0.0097 & 9.26 $\pm$ 1.29 & 9.77 $\pm$ 1.37 \\
2928 & 0.007 & 52 & 264 & 1.11 & 4.95 & 0.29 $\pm$ 0.04 & 0.37 $\pm$ 0.05 & 0.0329 $\pm$ 0.0048 & 8.77 $\pm$ 2.46 & 11.19 $\pm$ 3.13 \\
2968 & 0.003 & 6 & - & 2.25 & 13.82 & - & 0.46 $\pm$ 0.02 & - & - & - \\
3049 & 0.33 & 489 & - & 2.1 & 7.53 & 0.08 $\pm$ 0.01 & 1.51 $\pm$ 0.21 & - & - & - \\
3123 & 0.078 & 232 & 2000 & 1.52 & 8.59 & 0.21 $\pm$ 0.05 & 1.23 $\pm$ 0.3 & 0.0098 $\pm$ 0.0048 & 21.21 $\pm$ 15.45 & 126.1 $\pm$ 91.87 \\
3143 & 0.097 & 293 & - & 1.8 & 9.84 & 0.2 $\pm$ 0.06 & 1.57 $\pm$ 0.48 & - & - & - \\
3158 & 0.341 & 500 & - & 1.04 & 1.69 & - & - & - & - & - \\
3182 & 0.177 & 362 & - & 1.15 & 4.22 & - & - & - & - & - \\
3235 & 0.036 & 65 & 2000 & 4.58 & 17.76 & 0.88 $\pm$ 0.05 & 1.45 $\pm$ 0.08 & - & - & - \\
3308 & 0.044 & 120 & - & 1.37 & 7.91 & 0.28 $\pm$ 0.13 & 0.85 $\pm$ 0.38 & 0.0161 $\pm$ 0.0032 & 17.54 $\pm$ 11.41 & 52.74 $\pm$ 34.31 \\
3353 & 0.216 & 381 & - & 1.51 & 6.7 & 0.1 $\pm$ 0.04 & 1.2 $\pm$ 0.46 & -0.0029 $\pm$ 0.0009 & 33.96 $\pm$ 23.04 & - \\
3377 & 0.013 & 151 & - & 1.73 & 11.13 & 0.38 $\pm$ 0.05 & 1.32 $\pm$ 0.16 & 0.0108 $\pm$ 0.0039 & 35.06 $\pm$ 16.89 & 122.72 $\pm$ 59.13 \\
3378 & 0.013 & 183 & - & 2.64 & 14.83 & 0.46 $\pm$ 0.03 & 1.92 $\pm$ 0.12 & 0.0177 $\pm$ 0.0031 & 25.73 $\pm$ 6.08 & 108.01 $\pm$ 25.53 \\
3380 & 0.013 & 34 & - & 1.53 & 9.8 & 0.7 $\pm$ 0.05 & 0.6 $\pm$ 0.04 & 0.0482 $\pm$ 0.0045 & 14.6 $\pm$ 2.39 & 12.51 $\pm$ 2.05 \\
3384 & 0.013 & 289 & - & 1.28 & 7.53 & 0.18 $\pm$ 0.02 & 1.19 $\pm$ 0.13 & 0.0065 $\pm$ 0.001 & 28.21 $\pm$ 7.41 & 182.14 $\pm$ 47.88 \\
3458 & 0.129 & 209 & - & 1.21 & 5.27 & - & - & - & - & - \\
3632 & 0.156 & 245 & - & 2.21 & 10.31 & 0.21 $\pm$ 0.05 & 1.52 $\pm$ 0.38 & -0.0203 $\pm$ 0.0063 & 10.35 $\pm$ 5.78 & 74.54 $\pm$ 41.6 \\
3646 & 0.075 & 267 & - & 1.01 & 1.51 & - & - & - & - & - \\
3647 & 0.075 & 189 & - & 1.07 & 3.44 & - & - & -0.0092 $\pm$ 0.002 & - & - \\
3751 & 0.106 & 188 & 226 & 1.55 & 8.37 & 0.21 $\pm$ 0.04 & 1.09 $\pm$ 0.22 & - & - & - \\
3892 & 0.095 & 160 & - & 1.43 & 7.74 & 0.22 $\pm$ 0.1 & 0.94 $\pm$ 0.42 & - & - & - \\
3912 & 0.06 & 140 & - & 1.16 & 5.27 & 0.17 $\pm$ 0.02 & 0.61 $\pm$ 0.09 & - & - & - \\
4197 & 0.311 & 470 & - & 1.47 & 5.58 & 0.06 $\pm$ 0.02 & 1.09 $\pm$ 0.34 & - & - & - \\
4311 & 0.162 & 253 & 2000 & 1.04 & 2.35 & - & - & - & - & - \\
4350 & 0.072 & 166 & - & 1.09 & 3.98 & 0.11 $\pm$ 0.01 & 0.49 $\pm$ 0.06 & - & - & - \\
4359 & 0.106 & 176 & 423 & 1.01 & 0.99 & - & - & - & - & - \\
4446 & 0.207 & 328 & 367 & 2.27 & 9.6 & 0.16 $\pm$ 0.03 & 1.61 $\pm$ 0.36 & - & - & - \\
4451 & 0.242 & 369 & - & 1.66 & 7.12 & 0.1 $\pm$ 0.02 & 1.26 $\pm$ 0.22 & -0.0111 $\pm$ 0.0046 & 9.26 $\pm$ 5.51 & 113.44 $\pm$ 67.47 \\
4494 & 0.186 & 283 & - & 3.62 & 12.47 & 0.23 $\pm$ 0.01 & 1.95 $\pm$ 0.05 & - & - & - \\
4519 & 0.1 & 245 & - & 1.55 & 8.43 & 0.19 $\pm$ 0.04 & 1.24 $\pm$ 0.24 & - & - & - \\
4789 & 0.062 & 115 & - & 1.33 & 7.35 & 0.26 $\pm$ 0.04 & 0.78 $\pm$ 0.13 & - & - & - \\
4801 & 0.407 & 493 & - & 1.28 & 3.9 & - & - & 0.0058 $\pm$ 0.0012 & - & - \\
4838 & 0.049 & 104 & 851 & 1.61 & 9.7 & 0.37 $\pm$ 0.07 & 0.98 $\pm$ 0.17 & - & - & - \\
4862 & 0.2 & 359 & - & 1.86 & 8.46 & 0.13 $\pm$ 0.01 & 1.47 $\pm$ 0.14 & 0.0014 $\pm$ 0.0004 & 92.21 $\pm$ 34.21 & - \\
4881 & 0.059 & 276 & - & 1.18 & 5.63 & - & - & - & - & - \\
4883 & 0.059 & 365 & - & 3.65 & 15.67 & 0.31 $\pm$ 0.02 & 2.75 $\pm$ 0.15 & - & - & - \\
4936 & 0.033 & 64 & - & 1.39 & 8.34 & 0.42 $\pm$ 0.07 & 0.68 $\pm$ 0.11 & 0.0143 $\pm$ 0.0068 & 29.29 $\pm$ 18.74 & 47.31 $\pm$ 30.27 \\
4949 & 0.157 & 361 & - & 1.61 & 7.96 & 0.13 $\pm$ 0.03 & 1.39 $\pm$ 0.3 & -0.0101 $\pm$ 0.0038 & 13.22 $\pm$ 7.92 & 137.56 $\pm$ 82.39 \\
4978 & 0.026 & 152 & 275 & 1.49 & 9.28 & 0.31 $\pm$ 0.03 & 1.1 $\pm$ 0.11 & - & - & - \\
4993 & 0.11 & 485 & - & 1.21 & 5.47 & - & - & - & - & - \\
5008 & 0.244 & 441 & - & 2.0 & 8.31 & 0.11 $\pm$ 0.01 & 1.59 $\pm$ 0.14 & - & - & - \\
5191 & 0.206 & 368 & - & 1.1 & 3.28 & - & - & - & - & - \\
5296 & 0.319 & 450 & - & 1.56 & 5.94 & 0.07 $\pm$ 0.03 & 1.14 $\pm$ 0.5 & - & - & - \\
5324 & 0.13 & 280 & - & 1.52 & 7.83 & 0.16 $\pm$ 0.07 & 1.22 $\pm$ 0.55 & - & - & - \\
5402 & 0.295 & 453 & - & 1.11 & 3.03 & - & - & - & - & - \\

\hline
\end{longtable}
\end{landscape}